\documentclass[reqno,11pt]{amsart}
\usepackage{hyperref}
\usepackage{graphicx}
\usepackage{amscd}
\usepackage{slashed}
\usepackage[mathscr]{eucal}
\numberwithin{equation}{section}
\allowdisplaybreaks[1]

\newcommand{\SetFigFont}[3]{}

\title[A Lorentzian Quantum Geometry]{A Lorentzian Quantum Geometry}

\author[F.\ Finster]{Felix Finster}
\thanks{Supported in part by the Deutsche Forschungsgemeinschaft.}
\address{Fakult\"at f\"ur Mathematik \\ Universit\"at Regensburg \\ D-93040 Regensburg \\ Germany}
\email{finster@ur.de}
\author[A.\ Grotz]{Andreas Grotz \\ \\ July 2011 / April 2013}
\address{Department of Mathematics \\ Harvard University \\ Cambridge, MA 02138 \\ USA}
\email{agrotz@math.harvard.edu}

\newtheorem{Def}{Definition}[section]
\newtheorem{Thm}[Def]{Theorem}
\newtheorem{Prp}[Def]{Proposition}
\newtheorem{Lemma}[Def]{Lemma}

\newtheorem{Corollary}[Def]{Corollary}

\newcommand{\beq}{\begin{equation}}
\newcommand{\eeq}{\end{equation}}
\newcommand{\Proof}{\begin{proof}}
\newcommand{\QED}{\end{proof} \noindent}

\newcommand{\la}{\langle}
\newcommand{\ra}{\rangle}
\newcommand{\bra}{\,<\!\!}
\newcommand{\ket}{\!\!>\,}
\newcommand{\Sl}{\mbox{$\prec \!\!$ \nolinebreak}}
\newcommand{\Sr}{\mbox{\nolinebreak $\succ$}}

\newcommand{\C}{\mathbb{C}}
\newcommand{\R}{\mathbb{R}}
\newcommand{\1}{\mbox{\rm 1 \hspace{-1.05 em} 1}}
\newcommand{\Z}{\mathbb{Z}}
\newcommand{\N}{\mathbb{N}}
\newcommand{\Pdd}{\mbox{$\partial$ \hspace{-1.2 em} $/$}}
\newcommand{\slsh}{\mbox{ \hspace{-1.13 em} $/$}}
\renewcommand{\H}{\mathscr{H}}
\newcommand{\U}{{\rm{U}}}
\newcommand{\SU}{{\rm{SU}}}
\newcommand{\SO}{{\rm{SO}}}
\newcommand{\T}{\mathscr{T}}

\newcommand{\G}{\mathscr{G}}
\newcommand{\e}{{\mathfrak{e}}}
\newcommand{\f}{{\mathfrak{f}}}
\newcommand{\xis}{\slashed{\xi}}
\newcommand{\bep}{\begin{pmatrix}}
\newcommand{\enp}{\end{pmatrix}}
\newcommand{\bca}{\begin{cases}}
\newcommand{\eca}{\end{cases}}
\newcommand{\notcaus}{\mbox{\,$\not$ \hspace{-0.7 em} $\prec$}\; }
\newcommand{\nn}{\mathcal{N}}

\newcommand{\dd}{\mathcal{D}}
\newcommand{\rr}{\mathcal{R}}
\renewcommand{\O}{\mathscr{O}}

\newcommand{\Lin}{\text{\rm{L}}}
\newcommand{\F}{{\mathscr{F}}}
\newcommand{\E}{{\mathfrak{E}}}
\newcommand{\I}{{\mathcal{I}}}

\newcommand{\ie}{\iota^{\,\varepsilon}}
\newcommand{\nablaLC}{\nabla^\text{\tiny{\tt{LC}}}}
\newcommand{\DLC}{D^\text{\tiny{\tt{LC}}}}
\newcommand{\cliff}{\!\cdot\!}

\DeclareMathOperator{\Rea}{Re}
\DeclareMathOperator{\Ima}{Im}
\DeclareMathOperator{\Ric}{Ric}

\DeclareMathOperator{\Tr}{Tr}
\DeclareMathOperator{\tr}{tr}
\DeclareMathOperator{\Symm}{\mbox{\rm{Symm}}}

\DeclareMathOperator{\grad}{grad}
\DeclareMathOperator{\supp}{supp}
\DeclareMathOperator{\dvg}{div}
\DeclareMathOperator{\vleck}{\mathcal{V}}
\DeclareMathOperator{\s}{\text{\rm{scal}}}
\DeclareMathOperator{\Texp}{Texp}

\begin{document}
\maketitle

\begin{abstract}
We propose a formulation of a Lorentzian quantum geometry based on the framework
of causal fermion systems. After giving the general definition of causal fermion systems,
we deduce space-time as a topological space with an underlying causal structure.
Restricting attention to systems of spin dimension two, we derive the objects of
our quantum geometry: the spin space, the tangent space endowed with a Lorentzian metric,
connection and curvature. In order to get the correspondence to differential geometry,
we construct examples of causal fermion systems by regularizing Dirac sea configurations
in Minkowski space and on a globally hyperbolic Lorentzian manifold.
When removing the regularization, the objects of our quantum geometry
reduce precisely to the common objects of Lorentzian spin geometry, up to
higher order curvature corrections.
\end{abstract}

\tableofcontents

\section{Introduction}
General relativity is formulated in the language of Lorentzian geometry. Likewise, quantum
field theory is commonly set up in Minkowski space or on a Lorentzian manifold.
However, the ultraviolet divergences of quantum field theory and the problems in quantizing
gravity indicate that on the microscopic scale, a smooth manifold structure might no longer
be the appropriate model of space-time. Instead, a ``classical'' Lorentzian manifold should be
replaced by a ``quantum space-time''.
On the macroscopic scale, this quantum space-time should go over to
a Lorentzian manifold, whereas on the microscopic scale it should allow for a
more general structure. Consequently, the notions of Lorentzian geometry 
(like metric, connection and curvature) should be extended to
a corresponding ``quantum geometry''.

Although different proposals have been made,
there is no consensus on what the mathematical framework of quantum geometry should be.
Maybe the mathematically most advanced approach is Connes' non-commutative geometry~\cite{connes},
where the geometry is encoded in the spectral triple~$({\mathcal{A}}, \dd, \H)$ consisting of
an algebra~${\mathcal{A}}$ of operators on the Hilbert space~$\H$ and a
generalized Dirac operator~$\dd$. The correspondence to differential geometry
is obtained by choosing the algebra as the commutative algebra of functions on a manifold,
and~$\dd$ as the classical Dirac operator, giving back the setting of spin geometry.
By choosing~${\mathcal{A}}$ as a non-commutative algebra, one can extend the
notions of differential geometry to a much broader setting.
One disadvantage of non-commutative geometry is that it is mostly worked out in the Euclidean
setting (however, for the connection to the Lorentzian case see~\cite{strohmaier, paschke+verch}).
Moreover, it is not clear whether the spectral triple really gives a proper description of
quantum effects on the microscopic scale.
Other prominent approaches are canonical quantum gravity (see~\cite{kiefer}),
string theory (see~\cite{becker+schwarz}) and loop quantum gravity (see~\cite{thiemann}); for other interesting
ideas see~\cite{sorkin, QFTconf}.

In this paper, we present a framework for a quantum geometry
which is naturally adapted to the Lorentzian setting.
The physical motivation is coming from the fermio\-nic projector approach~\cite{PFP}.
We here begin with the more general formulation in the framework of {\em{causal fermion systems}}.
We give general definitions of geometric objects like the tangent space, spinors, connection
and curvature. It is shown that in a suitable limit, these objects reduce to the
corresponding objects of differential geometry on a globally hyperbolic Lorentzian manifold.
But our framework is more general, as it allows
to also describe space-times with a non-trivial microstructure
(like discrete space-times, space-time lattices or regularized space-times).
In this way, the notions of Lorentzian geometry are extended to a much broader
context, potentially including an appropriate model of the physical quantum space-time.

More specifically, in Section~\ref{secdiracsys} we introduce the general framework
of causal fermion systems and define notions of {\em{spinors}} as well as a {\em{causal structure}}.
In Section~\ref{secconstruct}, we proceed by constructing the objects of
our Lorentzian quantum geometry: We first define the {\em{tangent space}} endowed with
a Minkowski metric. Then we construct a {\em{spin connection}} relating spin spaces at different
space-time points. Similarly, a corresponding {\em{metric connection}} relates tangent spaces
at different space-time points. These connections give rise to corresponding notions
of {\em{curvature}}. We also find a distinguished {\em{time direction}} and
discuss the connection to causal sets.

In the following Sections~\ref{secvac} and~\ref{secglobhyp}, we explain how the objects
of our quantum geometry correspond to the common objects of differential geometry in
Minkowski space or on a Lorentzian manifold: In Section~\ref{secvac} we construct
a class of causal fermion systems by considering a Dirac sea configuration and introducing
an ultraviolet regularization. We show that if the ultraviolet regularization is removed,
we get back the topological, causal and metric structure of Minkowski space, whereas the
connections and curvature become trivial. 
In Section~\ref{secglobhyp} we consider causal fermion systems constructed from a globally
hyperbolic space-time. Removing the regularization, we recover the topological,
causal and metric structure of the Lorentzian manifold. The spin connection 
and the metric connection go over to the spin and Levi-Civita connections on the manifold,
respectively, up to higher order curvature corrections.

\section{Causal Fermion Systems of Spin Dimension Two}\label{secdiracsys}

\subsection{The General Framework of Causal Fermion Systems}
We begin with the general definition of causal fermion systems
(see~\cite{lrev, srev} for the physical motivation and~\cite[Section~1]{rrev}
for more details on the abstract framework).

\begin{Def} \label{def1} {\em{
Given a complex Hilbert space~$(\H, \la .|. \ra_\H)$ (the {\em{particle space}})
and a parameter~$n \in \N$ (the {\em{spin dimension}}), we let~$\F \subset \Lin(\H)$ be the set of all
self-adjoint operators on~$\H$ of finite rank, which (counting with multiplicities) have
at most~$n$ positive and at most~$n$ negative eigenvalues. On~$\F$ we are given
a positive measure~$\rho$ (defined on a $\sigma$-algebra of subsets of~$\F$), the so-called
{\em{universal measure}}. We refer to~$(\H, \F, \rho)$ as a {\em{causal fermion system in the
particle representation}}.
}}
\end{Def} \noindent
On~$\F$ we consider the topology induced by the
operator norm
\beq \label{onorm}
\|A\| := \sup \{ \|A u \|_\H \text{ with } \| u \|_\H = 1 \}\:.
\eeq
A vector~$\psi \in \H$ has the interpretation as an occupied fermionic state of our system.
The name ``universal measure'' is motivated by the fact that~$\rho$ describes a space-time ``universe''.
More precisely, we define {\em{space-time}}~$M$ as the support of the universal measure,
$M := \supp \rho$; it is a closed subset of~$\F$.
The induced measure~$\mu := \rho|_M$ on~$M$ allows us compute the volume of regions of space-time.
The interesting point in the above definition is that by considering the spectral properties of
the operator products~$x y$, we get relations between the space-time points~$x, y \in M$.
The goal of this article is to analyze these relations in detail.
The first relation is a notion of causality, which was also the motivation for the name ``causal'' fermion system.
\begin{Def} \label{def2} (causal structure)
{\em{ For any~$x, y \in \F$, the product~$x y$ is an operator
of rank at most~$2n$. We denote its non-trivial eigenvalues (counting with algebraic multiplicities)
by~$\lambda^{xy}_1, \ldots, \lambda^{xy}_{2n}$. The points~$x$ and~$y$ are
called {\em{timelike}} separated if the~$\lambda^{xy}_j$ are all real. They are said to be
{\em{spacelike}} separated if the~$\lambda^{xy}_j$ are complex
and all have the same absolute value.
In all other cases, the points~$x$ and~$y$ are said to be {\em{lightlike}} separated. }}
\end{Def} \noindent
Restricting the causal structure of~$\F$ to~$M$, we get causal relations in space-time.

In order to put the above definition into the context of previous work, it is
useful to introduce the inclusion map~$F : M \hookrightarrow \F$.
Slightly changing our point of view, we can now take the space-time~$(M, \mu)$ and the
mapping~$F : M \rightarrow \F$ as the
starting point. Identifying~$M$ with $F(M) \subset \F$ and
constructing the measure~$\rho$ on~$\F$ as the push-forward,
\beq \label{push}
\rho = F_* \mu \::\: \Omega \mapsto \rho(\Omega) := \mu(F^{-1}(\Omega)) \:,
\eeq
we get back to the setting of Definition~\ref{def1}.
If we assume that~$\H$ is finite dimensional and that the total volume~$\mu(M)$ is finite,
we thus recover the framework used in~\cite[Section~2]{continuum} for the formulation of so-called
{\em{causal variational principles}}.
Interpreting~$F(x)$ as local correlation matrices, one can
construct the corresponding fermion system formulated on an indefinite inner product space
(see~\cite[Sections~3.2 and~3.3]{continuum}).
In this setting, the dimension~$f$ of~$\H$ is interpreted as the number of particles,
whereas~$\mu(M)$ is the total volume of space-time.
If we assume furthermore that~$\rho$ is a finite counting
measure, we get into the framework of {\em{fermion systems in discrete space-time}}
as considered in~\cite{discrete, osymm}.
Thus Definition~\ref{def1} is compatible with previous papers, but it is slightly more general in
that we allow for an infinite number of particles and an infinite space-time volume.
These generalizations are useful for describing the infinite volume limit of the
systems analyzed in~\cite[Section~2]{continuum}.

\subsection{The Spin Space and the Euclidean Operator}
For every~$x \in \F$, we define the {\em{spin space}}~$S_x$ by
\beq \label{Sxdef}
S_x = x(\H)\:;
\eeq
it is a subspace of~$\H$ of dimension
at most~$2n$. On~$S_x$ we introduce the {\em{spin scalar product}} $\Sl .|. \Sr_x$ by
\beq \label{ssp}
\Sl u | v \Sr_x = -\la u | x v \ra_\H \qquad \text{(for all $u,v \in S_x$)}\:;
\eeq
it is an indefinite inner product of signature~$(p,q)$ with~$p,q \leq n$.
A {\em{wave function}}~$\psi$ is defined as a $\rho$-measurable function
which to every~$x \in M$ associates a vector of the corresponding spin space,
\beq \label{wavefunction}
\psi \::\: M \rightarrow \H \qquad \text{with} \qquad \psi(x) \in S_x \quad \text{for all~$x \in M$}\:.
\eeq
Thus the number of components of the wave functions at the space-time point~$x$
is given by~$p+q$. Having four-component Dirac spinors in mind, we are led to
the case of {\em{spin dimension two}}. Moreover, we impose that~$S_x$ has maximal rank.
\begin{Def} \label{defregular}
{\em{ Let~$(\H, \F, \rho)$ be a fermion system of spin dimension two.
A space-time point~$x \in M$ is called {\em{regular}} if~$S_x$ has dimension four.
}} \end{Def} \noindent
We remark that for points that are not regular, one could extend the spin space to
a four-dimensional vector space (see~\cite[Section~3.3]{continuum} for a similar construction).
However, the construction of the spin connection in Section~\ref{sec33}
only works for regular points. With this in mind, it seems preferable to always restrict attention
to regular points.

For a regular point~$x$, the operator~$(-x)$ on~$\H$ has two positive and two negative
eigenvalues. We denote its positive and negative spectral subspaces by~$S_x^+$ and~$S_x^-$,
respectively. In view of~\eqref{ssp}, these subspaces are also orthogonal with respect to the
spin scalar product,
\[ S_x = S_x^+ \oplus S_x^- \:. \]
We introduce the {\em{Euclidean operator}}~$\E_x$ by
\[ \E_x = -x^{-1} : S_x \rightarrow S_x\:. \]
It is obviously invariant on the subspaces~$S_x^\pm$. It is useful because it
allows us to recover the scalar product of~$\H$ from the spin scalar product,
\beq \label{Eop}
\la u, v \ra_\H |_{S_x \times S_x} = \Sl u | \E_x v \Sr_x \:.
\eeq
Often, the precise eigenvalues of~$x$ and~$\E_x$ will not be relevant; we only need
to be concerned about their signs. To this end, we
introduce the {\em{Euclidean sign operator}}~$s_x$ as a symmetric operator on~$S_x$
whose eigenspaces corresponding to the eigenvalues~$\pm 1$ are the spaces~$S_x^+$
and~$S_x^-$, respectively.

In order to relate two space-time points~$x, y \in M$ we define the {\em{kernel of the fermionic
operator}}~$P(x,y)$ by
\beq \label{Pxydef}
P(x,y) = \pi_x \,y \::\: S_y \rightarrow S_x \:,
\eeq
where~$\pi_x$ is the orthogonal projection onto the subspace~$S_x \subset \H$.
The calculation
\begin{align*}
\Sl P(x,y) \,\psi(y) \,|\, \psi(x) \Sr_x &= - \la (\pi_x \,y \,\psi(y)) \,|\, x \,\phi(x) \ra_\H \\
&= - \la \psi(y) \,|\, y x \,\phi(x) \ra_\H = \Sl \psi(y) \,|\,  P(y,x) \,\psi(x) \Sr_y
\end{align*}
shows that this kernel is {\em{symmetric}} in the sense that
\[ P(x,y)^* = P(y,x)\:, \]
where the star denotes the adjoint with respect to the spin scalar product.
The {\em{closed chain}} is defined as the product
\beq \label{Axydef}
A_{xy} = P(x,y)\, P(y,x) \::\: S_x \rightarrow S_x\:.
\eeq
It is obviously symmetric with respect to the spin scalar product,
\beq \label{Axysymm}
A_{xy}^* = A_{xy}\:.
\eeq
Moreover, as it is an endomorphism of~$S_x$, we can compute its eigenvalues.
The calculation~$A_{xy} = (\pi_x y)(\pi_y x) = \pi_x\, yx$ shows that these eigenvalues
coincide precisely with the non-trivial eigenvalues~$\lambda^{xy}_1, \ldots, \lambda^{xy}_4$
of the operator~$xy$ as considered in Definition~\ref{def2}. In this way, the kernel of the fermionic
operator encodes the causal structure of~$M$. Considering the closed chain has the advantage
that instead of working in the high- or even infinite-dimensional
Hilbert space~$\H$, it suffices to consider a symmetric operator on the four-dimensional vector space~$S_x$.
Then the appearance of complex eigenvalues in Definition~\ref{def2} can be understood from
the fact that the spectrum of symmetric operators in indefinite inner product spaces need not be
real, as complex conjugate pairs may appear (for details see~\cite{GLR}).

\subsection{The Connection to Dirac Spinors, Preparatory Considerations}
From the physical point of view, the appearance of indefinite inner products shows that we are
dealing with a {\em{relativistic}} system. In general terms, this can be understood from the fact
that the isometry group of an indefinite inner product space is non-compact, allowing for the
possibility that it may contain the Lorentz group.

More specifically, we have the context of Dirac spinors on a Lorentzian manifold~$(M, g)$ in
mind. In this case, the spinor bundle~$SM$ is a vector bundle, whose fibre~$(S_xM, \Sl .|. \Sr)$ is
a four-dimensional complex vector space endowed with an inner product of signature~$(2,2)$.
The connection to causal fermion systems is obtained by identifying this vector space
with~$(S_x, \Sl .|. \Sr_x)$
as defined by~\eqref{Sxdef} and~\eqref{ssp}. But clearly, in the context of Lorentzian spin geometry
one has many more structures. In particular, the Clifford multiplication associates to every tangent vector
$u \in T_xM$ a symmetric linear operator on~$S_xM$. Choosing a local frame and trivialization of the
bundle, the Clifford multiplication can also be expressed in terms of Dirac matrices~$\gamma^j(x)$,
which satisfy the anti-communication relations
\beq \label{anticommute}
\{ \gamma^i, \gamma^j \} = 2\, g^{ij}\,\1\:.
\eeq
Furthermore, on the spinor bundle one can introduce the {\em{spinorial Levi-Civita connection}} $\nablaLC$,
which induces on the tangent bundle an associated {\em{metric connection}}.

The goal of the present paper is to construct objects for general causal fermion systems
which correspond to the tangent space, the spin connection and the metric connection
in Lorentzian spin geometry and generalize these notions to the setting of a ``Lorentzian quantum
geometry.'' The key for constructing the tangent space is to observe that~$T_xM$ can be identified
with the subspace of the symmetric operators on~$S_xM$ spanned by the Dirac matrices.
The problem is that the anti-commutation relations~\eqref{anticommute} are not sufficient
to distinguish this subspace, as there are many different representations of these anti-commutation
relations. We refer to such a representation as a {\em{Clifford subspace}}. Thus in order to get a
connection to the setting of spin geometry, we need to distinguish a specific Clifford subspace.
The simplest idea for constructing the spin connection would be to use a polar decomposition of~$P(x,y)$.
Thus decomposing~$P(x,y)$ as
\[ P(x,y) = U(x) \, \rho(x,y) \,U(y)^{-1} \]
with a positive operator~$\rho(x,y)$ and unitary operators~$U(x)$ and~$U(y)$, we would like
to introduce the spin connection as the unitary mapping
\beq \label{idea}
D_{x,y} = U(x)\, U(y)^{-1} \::\: S_y \rightarrow S_x \:.
\eeq
The problem with this idea is that it is not clear how this spin connection should give rise
to a corresponding metric connection. Moreover, one already sees in the simple example
of a regularized Dirac sea vacuum (see Section~\ref{secvac}) that in Minkowski space
this spin connection does not reduce to the trivial connection.
Thus the main difficulty is to modify~\eqref{idea} such as to obtain a spin connection which induces
a metric connection and becomes trivial in Minkowski space. This
difficulty is indeed closely related to the problem of distinguishing a specific Clifford subspace.

The key for resolving these problems will be to use the Euclidean operator~$\E_x$ 
in a specific way. In order to explain the physical significance of this operator,
we point out that, apart from the Lorentzian point of view discussed above,
we can also go over to the {\em{Euclidean framework}} by considering instead of the spin
scalar product the scalar product on~$\H$. In view of the identity~\eqref{Eop},
the transition to the Euclidean framework can be described by the Euclidean operator,
which motivates its name. The physical picture is that the causal fermion systems of Definition~\ref{def1}
involve a regularization which breaks the Lorentz symmetry. This fact becomes apparent
in the Euclidean operator, which allows us to introduce a scalar product on spinors~\eqref{Eop}
which violates Lorentz invariance. The subtle point in the constructions in this paper is
to use the Euclidean sign operator to distinguish certain Clifford subspaces, but in such a way
that the Lorentz invariance of the resulting objects is preserved. The connection between the
Euclidean operator and the regularization will become clearer in the examples in
Sections~\ref{secvac} and~\ref{secglobhyp}.

We finally give a construction which will not be needed later on, but which is nevertheless
useful to get a closer connection to Dirac spinors in relativistic quantum mechanics.
To this end, we consider wave functions~$\psi, \phi$ of the form~\eqref{wavefunction} which are square
integrable. Setting
\beq \label{stip}
\bra \psi | \phi \ket = \int_M \Sl \psi(x) | \phi(x) \Sr_x\: d\mu(x) \:,
\eeq
the vector space of wave functions becomes an indefinite inner product space.
Interpreting~$P(x,y)$ as an integral kernel, we can introduce the {\em{fermionic operator}} by
\[ (P \psi)(x,y) = \int_M P(x,y)\, \psi(y)\: d\mu(y)\:. \]
Additionally imposing the idempotence condition~$P^2=P$, we obtain the {\em{fermionic projector}}
as considered in~\cite{PFP, discrete}. In this context, the inner product~\eqref{stip} reduces to the integral over Minkowski space~$\int_M \overline{\psi(x)} \phi(x)\, d^4x$, where~$\overline{\psi} \phi$
is the Lorentz invariant inner product on Dirac spinors.

\section{Construction of a Lorentzian Quantum Geometry}\label{secconstruct}

\subsection{Clifford Extensions and the Tangent Space}
We proceed with constructions in the spin space~$(S_x, \Sl .|. \Sr)$ at a fixed space-time point~$x \in M$.
We denote the set of symmetric linear endomorphisms of~$S_x$ by~$\Symm(S_x)$; it is
a $16$-dimensional real vector space.

We want to introduce the Dirac matrices, but without specifying a particular representation.
Since we do not want to prescribe the dimension of the resulting space-time, it is preferable
to work with the maximal number of five generators (for the minimal dimensions of
Clifford representations see for example~\cite{baum}).

\begin{Def} \label{defcliffsubspace} A five-dimensional subspace~$K \subset \Symm(S_x)$ is called
a {\bf{Clifford subspace}} if the following conditions hold:
\begin{itemize}
\item[(i)] For any~$u, v \in K$, the anti-commutator~$\{ u,v \} \equiv u v + v u$ is a multiple
of the identity on~$S_x$.
\item[(ii)] The bilinear form~$\la .,. \ra$ on~$K$ defined by
\beq \label{anticommute2}
\frac{1}{2} \left\{ u,v \right\} = \la u,v \ra \, \1 \qquad {\text{for all~$u,v \in K$}}
\eeq
is non-degenerate.
\end{itemize}
The set of all Clifford subspaces~$(K, \la .,. \ra$) is denoted by~$\T$.
\end{Def} \noindent

Our next lemma characterizes the possible signatures of Clifford subspaces.
\begin{Lemma} \label{lemma0}
The inner product~$\la .,. \ra$ on a Clifford subspace has either
the signature~$(1,4)$ or the signature~$(3,2)$. In the first (second) case, the inner product
\beq \label{iprod}
\Sl .\,| u \,.\, \Sr_x \::\: S_x \times S_x \rightarrow \C
\eeq
is definite (respectively indefinite) for every vector~$u \in K$ with~$\la u,u \ra > 0$.\end{Lemma}
\Proof
Taking the trace of~\eqref{anticommute2}, one sees that the inner product on~$K$
can be extended to all of~$\Symm(S_x)$ by
\[ \la .,. \ra \::\: \Symm(S_x) \times \Symm(S_x) \rightarrow \C ,\;
(A, B) \mapsto \frac{1}{4}\:\Tr(A B)\:. \]
A direct calculation shows that this inner product has signature~$(8,8)$
(it is convenient to work in the basis of~$\Symm(S_x)$ given by the
matrices~$(\1, \gamma^i, i\gamma^5, \gamma^5 \gamma^i, \sigma^{jk})$ in the usual Dirac representation;
see~\cite[Section~2.4]{bjorken}).

Since~$\la .,. \ra$ is assumed to be non-degenerate, it has signature~$(p, 5-p)$ with
a parameter~$p \in \{0,\ldots, 5\}$. 
We choose a basis~$e_0, \ldots, e_4$ of~$K$ where the bilinear form is diagonal,
\beq \label{ac1}
\left\{ e_j, e_k \right\} = 2 s_j \:\delta_{jk}\:\1 \qquad \text{with} \qquad 
s_0, \ldots, s_{p-1} =1 \text{ and }
s_{p}, \ldots, s_4 =-1\:.
\eeq
These basis vectors generate a Clifford algebra. Using the uniqueness results on Clifford
representations~\cite[Theorem~5.7]{lawson+michelsohn}, we find that in a suitable
basis of~$S_x$, the operators~$e_j$ have the basis representations
\beq \label{Clifford}
e_0 = c_0 \begin{pmatrix} \1 & 0 \\ 0 & -\1 \end{pmatrix} , \qquad
e_\alpha = c_\alpha \begin{pmatrix} 0 & i \sigma_\alpha \cr
-i \sigma_\alpha & 0 \end{pmatrix} , \qquad
e_4 = c_4 \begin{pmatrix} 0 & \1 \\ \1 & 0 \end{pmatrix}
\eeq
with coefficients
\[ c_0, \ldots, c_{p-1} \in \{1, -1\} \:,\qquad c_p, \ldots, c_{4} \in \{i,  -i \}\:. \]
Here~$\alpha \in \{1,2,3\}$, and~$\sigma^\alpha$ are the three Pauli matrices
\[ \sigma^1 = \begin{pmatrix} 0 & 1 \\ 1 & 0 \end{pmatrix}\:,\quad
\sigma^2 = \begin{pmatrix} 0 & -i \\ i & 0 \end{pmatrix}\:,\quad
\sigma^3 = \begin{pmatrix} 1 & 0 \\ 0 & -1 \end{pmatrix} . \]
In particular, one sees that the~$e_j$ are all trace-free.
We next introduce the ten bilinear operators
\[ \sigma_{jk} := i e_j e_k  \qquad \text{with} \qquad 1 \leq j < k \leq 5\:. \]
Taking the trace and using that~$e_j$ and~$e_k$ anti-commute, one sees that the bilinear
operators are also trace-free. Furthermore, using the anti-commutation relations~\eqref{ac1},
one finds that
\[  \la \sigma_{jk}, \sigma_{lm} \ra = s_j s_k \:\delta_{jl} \delta_{km}\:. \]
Thus the operators~$\{ \1, e_j, \sigma_{jk}\}$ form a pseudo-orthonormal basis
of $\Symm(S_x)$.

In the cases~$p=0$ and~$p=5$, the operators~$\sigma_{jk}$ would span a ten-dimensional
definite subspace of~$\Symm(S_x)$,
in contradiction to the above observation that~$\Symm(S_x)$ has signature~$(8,8)$.
Similarly, in the cases~$p=2$ and~$p=4$, the signature of~$\Symm(S_x)$ would be
equal to~$(7,9)$ and~$(11,5)$, again giving a contradiction.
We conclude that the possible signatures of~$K$ are~$(1,4)$ and~$(3,2)$.

We represent the spin scalar product in the spinor basis of~\eqref{Clifford}
with a signature matrix~$S$,
\[ \Sl .|. \Sr_x = \la .| S \,. \ra_{\C^4}\:. \]
Let us compute~$S$. In the case of signature~$(1,4)$, the fact that the operators~$e_j$
are symmetric gives rise to the conditions
\beq \label{cond14}
[S, e_0] = 0 \qquad \text{and} \qquad \{S, e_j \} = 0 \quad \text{for~$j=1,\ldots, 4$}\: .
\eeq
A short calculation yields~$S = \lambda e_0$ for~$\lambda \in \R \setminus \{ 0 \}$.
This implies that the bilinear form~$\Sl .| e_0\,. \Sr_x$ is definite. Moreover, a direct calculation
shows that~\eqref{iprod} is definite for any vector~$u \in K$ with~$\la u,u \ra > 0$.

In the case of signature~$(3,2)$, we obtain similar to~\eqref{cond14} the conditions
\[ [S, e_j] = 0 \quad \text{for~$j=0,1,2$} \qquad \text{and} \qquad
\{S, e_j \} = 0 \quad \text{for~$j=3, 4$}\: . \]
It follows that~$S= i \lambda e_3 e_4$. Another direct calculation yields that
the bilinear form~\eqref{iprod} is indefinite for any~$u \in K$ with~$\la u,u \ra > 0$.
\QED

We shall always restrict attention to Clifford subspaces of signature~$(1,4)$.
This is motivated physically because the Clifford subspaces of signature~$(3,2)$
only have two spatial dimensions, so that by dimensional reduction we cannot
get to Lorentzian signature~$(1,3)$. 
Alternatively, this can be understood from the analogy to Dirac spinors, where the inner
product~$\overline{\psi} u^j \gamma_j \phi$ is definite for any timelike vector~$u$.
Finally, for the Clifford subspaces of signature~$(3,2)$
the constructions following Definition~\ref{defclifford} would not work.

From now on, we implicitly assume that all Clifford subspaces have signature~$(1,4)$.
We next show that such a Clifford subspace is uniquely determined by a two-dimen\-sio\-nal
subspace of signature~$(1,1)$.
\begin{Lemma} \label{lemmaextend}
Assume that~$L \subset K$ is a two-dimensional subspace of a Clifford subspace~$K$,
such that the inner product~$\la .,. \ra|_{L \times L}$ has signature~$(1,1)$.
Then for every Clifford subspace~$\tilde{K}$ the following implication holds:
\[ L \subset \tilde{K}  \quad \Longrightarrow \quad \tilde{K} = K\:. \]
\end{Lemma}
\Proof We choose a pseudo-orthonormal basis of~$L$, which we denote by~$(e_0, e_4)$.
Since~$e_0^2=\1$, the spectrum of~$e_0$ is contained in the set~$\{\pm 1\}$.
The calculation~$e_0 (e_0 \pm \1) = \1 \pm e_0 = \pm (e_0 \pm \1)$ shows
that the corresponding invariant subspaces are indeed eigenspaces.
Moreover, as the bilinear form~$\Sl .|e_0 . \Sr_x$ is definite,
the eigenspaces are also definite.
Thus we may choose a pseudo-orthonormal eigenvector basis~$(\f_1, \ldots, \f_4)$ in which
\[ e_0 = \pm \begin{pmatrix} \1 & 0 \\ 0 & -\1 \end{pmatrix}\:. \]

We next consider the operator~$e_4$. Using that it anti-commutes with~$e_0$,
is symmetric and that~$(e_4)^2=-\1$, one easily sees that it has the matrix representation
\[ e_4 = \begin{pmatrix} 0 & -V \\ V^{-1} & 0 \end{pmatrix} 
\qquad \text{with} \qquad V \in \U(2)\:. \]
Thus after transforming the basis vectors~$\f_3$ and~$\f_4$ by
\beq \label{f34rep}
\begin{pmatrix} \f_3 \\ \f_4 \end{pmatrix} \rightarrow -i V \begin{pmatrix} \f_3 \\ \f_4 \end{pmatrix} \:,
\eeq
we can arrange that
\[  e_4 = i \begin{pmatrix} 0 & \1 \\ \1 & 0 \end{pmatrix} \:. \]

Now suppose that~$\tilde{K}$ extends~$L$ to a Clifford subspace.
We extend~$(e_0, e_4)$ to a pseudo-orthonormal basis~$(e_0, e_1, \ldots, e_4)$ of~$\tilde{K}$.
Using that the operators~$e_1, e_2$ and~$e_3$
anti-commute with~$e_0$ and~$e_4$ and are symmetric, we
see that each of these operators must be of the form
\beq \label{ealpha}
e_\alpha = \begin{pmatrix} 0 & A^\alpha \\ -A^\alpha & 0 \end{pmatrix}
\eeq
with Hermitian $2 \times 2$-matrices~$A_\alpha$. The anti-commutation relations~\eqref{anticommute2}
imply that the~$A_\alpha$ satisfy the anti-commutation relations of the Pauli matrices
\[ \left\{ A^\alpha, A^\beta \right\} = 2 \delta^{\alpha \beta}\:. \]
The general representation of these relations is obtained from the Pauli matrices by
an~$\SU(2)$-transformation and possible sign flips,
\[ A^\alpha = \pm U \sigma^\alpha U^{-1} \qquad \text{with} \qquad U \in \SU(2)\:. \]
Since~$U \sigma^\alpha U^{-1} = O^\alpha_\beta \sigma^\beta$ with~$O \in \SO(3)$,
we see that the~$A^\alpha$ are linear combinations of the Pauli matrices.
Hence the subspace spanned by the matrices~$e_1$, $e_2$ and~$e_3$ is uniquely
determined by~$L$. It follows that~$\tilde{K}=K$.
\QED

In the following corollary we choose a convenient matrix representation for a Clifford subspace.
\begin{Corollary} \label{corollary1}
For every pseudo-orthonormal basis~$(e_0, \ldots, e_4)$ of a
Clifford subspace~$K$, we can choose a pseudo-orthonormal basis~$(\f_1, \ldots, \f_4)$ of~$S_x$,
\beq \label{pseudoONB}
\Sl \f_\alpha | \f_\beta \Sr = s_\alpha\, \delta_{\alpha \beta} \qquad \text{with} \qquad
s_1=s_2=1 \text{ and } s_3=s_4=-1 \:,
\eeq
such that the operators~$e_i$ have the following matrix representations,
\beq \label{Dirac}
e_0 = \pm \begin{pmatrix} \1 & 0 \\ 0 & -\1 \end{pmatrix} , \qquad
e_\alpha = \pm \begin{pmatrix} 0 & \sigma_\alpha \cr
-\sigma_\alpha & 0 \end{pmatrix} , \qquad
e_4 = i \begin{pmatrix} 0 & \1 \\ \1 & 0 \end{pmatrix} .
\eeq
\end{Corollary} \noindent
\Proof As in the proof of Lemma~\ref{lemmaextend}, we can choose a pseudo-orthonormal
basis $(\f_1, \ldots, \f_4)$ of~$S_x$ satisfying~\eqref{pseudoONB} such that~$e_0$
and~$e_4$ have the desired representation. Moreover, in this basis
the operators~$e_1$, $e_2$ and~$e_3$ are of the form~\eqref{ealpha}.
Hence by the transformation of the spin basis
\[ \begin{pmatrix} \f_1 \\ \f_2 \end{pmatrix} \rightarrow U^{-1} \begin{pmatrix} \f_1 \\ \f_2 \end{pmatrix} ,
\qquad
\begin{pmatrix} \f_3 \\ \f_4 \end{pmatrix} \rightarrow U^{-1} \begin{pmatrix} \f_3 \\ \f_4
\end{pmatrix} , \]
we obtain the desired representation~\eqref{Dirac}.
\QED

Our next step is to use the Euclidean sign operator to distinguish a specific subset of
Clifford subspaces. For later use, it is preferable to work instead of the Euclidean sign operator
with a more general class of operators defined as follows.
\begin{Def} \label{defsign}
An operator~$v \in \Symm(S_x)$ is called a {\bf{sign operator}} if~$v^2 = \1$
and if the inner product~$\Sl .|v \,. \Sr_x\::\: S_x \times S_x \rightarrow \C$ is positive definite.
\end{Def} \noindent
Clearly, the Euclidean sign operator~$s_x$ is an example of a sign operator.

Since a sign operator~$v$ is symmetric with respect to the
positive definite inner product~$\Sl .|v \,. \Sr$, it can be diagonalized.
Again using that the inner product~$\Sl .|v \,. \Sr$ is positive, one finds that
the eigenvectors corresponding to the eigenvalues $+1$ and~$-1$ are positive 
and negative definite, respectively. Thus we may choose a pseudo-orthonormal basis~\eqref{pseudoONB}
in which~$v$ has the matrix representation~$v = \text{diag}(1,1,-1,-1)$.
In this spin basis, $v$ is represented by the matrix~$\gamma^0$
(in the usual Dirac representation).
Thus by adding the spatial Dirac matrices, we can extend~$v$ to a Clifford subspace. We now form the
set of all such extensions.
\begin{Def} \label{defclifford}
For a given sign operator~$v$, the set of {\bf{Clifford extensions}} $\T^v$
is defined as the set of all Clifford subspaces containing~$v$,
\[ \T^v = \{K {\text{ Clifford subspace with }} v \in K \}\:. \]
\end{Def}

After these preparations, we want to study how different Clifford subspaces or
Clifford extensions can be related to each other by unitary transformations.
We denote the group of unitary endomorphisms of~$S_x$ by~$\U(S_x)$;
it is isomorphic to the group~$\U(2,2)$. Thus for given~$K, \tilde{K} \in \T$ (or~$\T^v$),
we want to determine the unitary operators~$U \in \U(S_x)$ such that
\beq \label{tKK}
\tilde{K} = U K U^{-1}\:.
\eeq
Clearly, the subgroup~$\exp(i \R \1) \simeq \U(1)$ is irrelevant for this problem, because
in~\eqref{tKK} phase transformations drop out. For this reason, it is useful to divide out this group by
setting
\beq \label{gaugedef}
\G(S_x) = \U(S_x)/\exp(i \R \1) \:.
\eeq
We refer to~$\G$ as the {\em{gauge group}}
(this name is motivated by the formulation of spinors in curved space-time as
a gauge theory; see~\cite{U22}). It is a $15$-dimensional non-compact
Lie group whose corresponding Lie algebra is formed of all trace-free elements of~$\Symm(S_x)$.
It is locally isomorphic to the group~$\SU(2,2)$ of $\U(2,2)$-matrices with determinant one.
However, we point out that~$\G$ is {\em{not}} isomorphic to~$\SU(2,2)$,
because the four-element subgroup~$\Z_4 := \exp(i \pi \Z \1/2) \subset \SU(2,2)$ 
is to be identified with the neutral element in~$\G$. In other words, the groups are isomorphic only
after dividing out this discrete subgroup, $\G \simeq \SU(2,2)/\Z_4$.

\begin{Corollary} \label{cor1} For any two Clifford subspaces~$K, \tilde{K} \in \T$, there is a
gauge transformation~$U \in \G$ such that~\eqref{tKK} holds.
\end{Corollary}
\Proof We choose spin bases~$(\f_\alpha)$ and similarly~$(\tilde{\f}_\alpha)$ as in
Corollary~\ref{corollary1} and let~$U$ be the unitary transformation describing the basis transformation.
\QED

Next, we consider the subgroups of~$\G$ which leave the sign operator~$v$ and possibly
a Clifford subspace~$K \in \T^v$ invariant:
\beq \label{stabilizer}
\begin{split}
\G_v &= \left\{ U \in \G \text{ with } U v U^{-1} = v \right\} \\
\G_{v, K} &= \left\{ U \in \G \text{ with } U v U^{-1} = v \text{ and } U K U^{-1} = K \right\} .
\end{split}
\eeq
We refer to these groups as the {\bf{stabilizer subgroups}} of~$v$ and~$(v, K)$, respectively.

\begin{Lemma} \label{cor2}
For any Clifford extension~$K \in \T^v$, the stabilizer subgroups are related by
\[ \G_v = \exp(i \R v) \times \G_{v,K} \:. \]
Furthermore,
\[ \G_{v,K} \simeq (\SU(2) \times \SU(2))/\Z_2 \simeq \SO(4) \:, \]
where the group~$\SO(4)$ acts on any pseudo-orthonormal basis~$(v, e_1, \ldots, e_4)$
of~$K$ by
\beq \label{SO4}
e_i \mapsto \sum_{j=1}^4 O^j_i \,e_j \:,\qquad O \in \SO(4)\:.
\eeq
\end{Lemma}
\Proof The elements of~$\G_v$ are represented by unitary operators which commute with~$v$.
Thus choosing a spin frame where
\beq \label{vrep}
v = \begin{pmatrix} \1 & 0 \\ 0 & -\1 \end{pmatrix} ,
\eeq
every~$U \in \G_v$ can be represented as
\[ U = \begin{pmatrix} V_1 & 0 \\ 0 & V_2 \end{pmatrix} \qquad \text{with} \qquad
V_{1,2} \in \U(2)\:. \]
Collecting phase factors, we can write
\[ U = e^{i \alpha} \begin{pmatrix} e^{i \beta} & 0 \\ 0 & e^{-i \beta} \end{pmatrix}
\begin{pmatrix} U_1 & 0 \\ 0 & U_2 \end{pmatrix} \qquad \text{with~$\alpha, \beta \in \R$
and~$U_{1,2} \in \SU(2)$}\:. \]
As the two matrices in this expression obviously commute, we obtain,
after dividing out a global phase,
\beq \label{Gfact}
\G_v \simeq \exp(i \R v) \times (\SU(2) \times \SU(2))/\Z_2 \:,
\eeq
where $\Z_2$ is the subgroup $\{\pm\1\}$ of $\SU(2) \times \SU(2)$.

Let us consider the group~$\SU(2) \times \SU(2)$ acting on the vectors of~$K$ by conjugation.
Obviously, $UvU^{-1} = v$. In order to compute~$U e_j U^{-1}$, we first apply the
identity
\[ e^{i \vec{u}_1 \vec{\sigma}} \: (i \rho\, \1 + \vec{w} \vec{\sigma})\:e^{-i \vec{u}_2 \vec{\sigma}} 
= i \rho'\,\1 + \vec{w}' \vec{\sigma} \:. \]
Taking the determinant of both sides, one sees that the 
vectors~$(\rho, \vec{w}), (\rho', \vec{w}') \in \R^4$ have the same Euclidean norm.
Thus the group~$\SU(2) \times \SU(2)$ describes
$\SO(4)$-trans\-for\-ma\-tions~\eqref{SO4}.
Counting dimensions, it follows
that~$\SU(2) \times \SU(2)$ is a covering of~$\SO(4)$.
Next it is easy to verify that the only elements of~$\SU(2) \times \SU(2)$ which
leave all~$\gamma^i$, $i=1,\ldots, 4$, invariant are multiples of the identity matrix.
We conclude that~$(\SU(2) \times \SU(2))/\Z_2 \simeq \SO(4)$
(this can be understood more abstractly from the fact that~$\SU(2) \times \SU(2) = \text{Spin}(4)$;
see for example~\cite[Chapter~1]{friedrich}).

To summarize, the factor~$\SU(2) \times \SU(2)$ in~\eqref{Gfact} leaves~$K$ invariant
and describes the transformations~\eqref{SO4}. However, the only elements of the group~$\exp(i \R v)$
which leave~$K$ invariant are multiples of the identity. This completes the proof.
\QED

Our method for introducing the tangent space is to form equivalence classes of Clifford extensions.
To this end, we introduce on~$\T^v$ the equivalence relation
\beq \label{equiv}
K \sim \tilde{K} \quad \Longleftrightarrow \quad {\text{there is~$U \in \exp(i \R v)$
with~$\tilde{K} = U K U^{-1}$}} \:.
\eeq
According to Corollary~\ref{cor1} and Lemma~\ref{cor2}, there is only one equivalence
class. In other words, for any~$K, \tilde{K} \in \T^v$
there is an operator~$U \in \exp(i \R v)$ such that~\eqref{tKK} holds. 
However, we point out that the operator~$U$ is not unique. Indeed, for two
choices~$U, U'$, the operator~$U^{-1} U'$ is an element
of~$\exp(i \R v) \cap \G_{v,K}$, meaning that~$U$ is unique only up to the transformations
\beq \label{Ufreedom}
U \rightarrow \pm U \qquad \text{and} \qquad U \rightarrow \pm i v \,U \:.
\eeq
The operator~$U$ gives rise to the so-called 
{\em{identification map}}
\beq \label{phidef}
\phi^v_{\tilde{K}, K} \::\: K \rightarrow \tilde{K} \,,\; w \mapsto U w U^{-1}\:.
\eeq
The freedom~\eqref{Ufreedom} implies that the mapping~$\phi^v_{\tilde{K}, K}$ is defined
only up to a {\em{parity transformation}}~$P^v$ which flips the sign of the orthogonal complement of~$v$,
\beq \label{parity}
\phi^v_{\tilde{K}, K} \rightarrow P^v\, \phi^v_{\tilde{K}, K} \qquad \text{with} \qquad
P^v w = - w + 2 \la w, v \ra\, v\:.
\eeq
As the identification map preserves the inner product~$\la .,. \ra$, the
quotient space~$\T^v / \sim$ is endowed with a Lorentzian metric.
We now take~$v$ as the Euclidean sign operator, which seems the most natural choice.
\begin{Def} \label{deftangent}
The {\bf{tangent space}}~$T_x$ is defined by
\[ T_x = \T^{s_x} / \exp(i \R s_x)\:. \]
It is endowed with an inner product~$\la .,. \ra$ of signature~$(1,4)$.
\end{Def}

We point out that, due to the freedom to perform the parity transformations~\eqref{parity},
the tangent space has no spatial orientation. In situations when a spatial orientation is needed,
one can fix the parity by distinguishing a class of representatives.
\begin{Def} \label{defparity}
A set of representatives~$\mathscr{U} \subset \T^{s_x}$ of the tangent space
is called {\bf{parity preserving}} if for any two~$K, \tilde{K} \in \mathscr{U}$,
the corresponding identification map~$\phi^{s_x}_{\tilde{K}, K}$ is of the form~\eqref{phidef}
with~$U = e^{i \beta s_x}$ and~$\beta \not \in \frac{\pi}{2} + \pi \Z$.
Then the {\bf{parity preserving identification map}} is defined by~\eqref{phidef} with
\beq \label{Ufix}
U = U^{s_x}_{\tilde{K}, K} := e^{i \beta s_x} \qquad \text{and} \qquad
\beta \in \Big( -\frac{\pi}{2}, \frac{\pi}{2} \Big) \:.
\eeq
\end{Def} \noindent
By identifying the elements of~$\mathscr{U}$ via the parity preserving identification maps,
one can give the tangent space a spatial orientation. In Section~\ref{secmetricconn}, we will
come back to this construction for a specific choice of~$\mathscr{U}$ induced by the spin connection.

\subsection{Synchronizing Generically Separated Sign Operators}
In this section, we will show that for two given sign operators~$v$ and~$\tilde{v}$
(again at a fixed space-time point~$x \in M$), under generic assumptions one can distinguish unique
Clifford extensions~$K \in \T^v$ and~$\tilde{K} \in \T^{\tilde{v}}$.
Moreover, we will construct the so-called synchronization map~$U^{\tilde{v}, v}$,
which transforms these two Clifford extensions into each other.

\begin{Def} \label{defgensep}
Two sign operators~$v, \tilde{v}$ are said to be {\bf{generically separated}} if
their commutator~$[v, \tilde{v}]$ has rank four.
\end{Def}

\begin{Lemma} \label{lemma3}
Assume that~$v$ and~$\tilde{v}$ are two generically separated sign operators.
Then there are unique Clifford extensions~$K \in \T^v$ and~$\tilde{K} \in \T^{\tilde{v}}$ 
and a unique vector~$\rho \in K \cap \tilde{K}$ with the following properties:
\begin{align}
&{\rm{(i)}} \hspace*{1cm} \{ v, \rho \} = 0 = \{ \tilde{v}, \rho \} \label{vrels} \\
&{\rm{(ii)}} \hspace*{1.15cm} \tilde{K} = e^{i \rho} \,K\, e^{-i \rho} \label{KKtrel} \\
&{\rm{(iii)}} \hspace*{0.7cm}
\text{If~$\{v, \tilde{v}\}$ is a multiple of the identity, then~$\rho=0$.} \label{special}
\end{align}
The operator~$\rho$ depends continuously on~$v$ and~$\tilde{v}$.
\end{Lemma}
\Proof Our first step is to choose a spin frame where~$v$ and~$\tilde{v}$ have a simple form.
Denoting the spectral projector of~$v$ corresponding to the eigenvalue one by~$E_+ =
(\1+v)/2$, we choose an orthonormal eigenvector basis~$(\f_1, \f_2)$ of
the operator~$E_+ \tilde{v} E_+$, i.e.\
\[ E_+ \tilde{v} E_+|_{E_+(S_x)} = \text{diag}(\nu_1, \nu_2)
\qquad \text{with~$\nu_1, \nu_2 \in \R$} \:. \]
Setting~$\f_3 = (\tilde{v}-\nu_1) \f_1$ and~$\f_4 = (\tilde{v}-\nu_2) \f_2$, these vectors are
clearly orthogonal to~$\f_1$ and~$\f_2$. They are both non-zero because otherwise the
commutator~$[\nu, \tilde{\nu}]$ would be singular. Next, being orthogonal to
the eigenspace of~$v$ corresponding to the eigenvalue one, they lie in the
eigenspace of~$v$ corresponding to the eigenvalue~$-1$, and are thus
both negative definite.
Moreover, the following calculation shows that they are orthogonal,
\begin{align*}
\Sl \f_3 | \f_4 \Sr &= \Sl (\tilde{v}-\nu_1) \f_1 | (\tilde{v}-\nu_2) \f_2 \Sr
= \Sl \f_1 |  (\tilde{v}-\nu_1)(\tilde{v}-\nu_2) \f_2 \Sr \\
&= \Sl \f_1 |  \left( 1 + \nu_1 \nu_2 - (\nu_1 + \nu_2) \tilde{v} \right) \f_2 \Sr = 0\:, 
\end{align*}
where in the last step we used that~$\f_2$ and~$\tilde{v} \f_2$ are orthogonal to~$\f_1$. 
The image of~$\f_3$ (and similarly~$\f_4$) is computed by
\[ \tilde{v} f_3 = \tilde{v}  (\tilde{v}-\nu_1) \f_1 = (1 - \nu_1 \tilde{v}) \f_1
= -\nu_1 \f_3 + (1-\nu_1^2)\, \f_1 \:. \]
We conclude that, after normalizing~$\f_3$ and~$\f_4$ by the replacement~$\f_i \rightarrow 
\f_i /\sqrt{-\Sl f_i | f_i \Sr}$, the matrix~$v$ is diagonal~\eqref{vrep},
whereas~$\tilde{v}$ is of the form
\beq \label{vtform}
\tilde{v} = \begin{pmatrix} \cosh \alpha & 0 & \sinh \alpha & 0 \\
0 & \cosh \beta & 0 & -\sinh \beta \\
-\sinh \alpha & 0 & -\cosh \alpha & 0 \\
0 & \sinh \beta & 0 & -\cosh \beta \end{pmatrix} \qquad \text{with~$\alpha, \beta>0$.}
\eeq

In the case~$\alpha = \beta$, the anti-commutator~$\{ v, \tilde{v}\}$ is a multiple of the
identity. Thus by assumption~(iii) we need to choose~$\rho=0$.
Then~$K=\tilde{K}$ must be the Clifford subspace spanned by the
matrices~$e_0, \ldots, e_4$ in~\eqref{Dirac}.

In the remaining case~$\alpha\neq\beta$, a short calculation shows that any operator~$\rho$
which anti-commutes with both~$v$ and~$\tilde{v}$ is a linear combination of the matrix~$e_4$
and the matrix~$i e_0 e_3$.
Since~$\rho$ should be an element of~$K$, its square must be a multiple of the identity.
This leaves us with the two cases
\beq \label{twocases}
\rho = \frac{\tau}{2}\: e_4 \qquad \text{or} \qquad \rho = \frac{\tau}{2}\: i e_0 e_3
\eeq
for a suitable real parameter~$\tau$. In the first case, we obtain
\[ e^{i \rho} v e^{-i \rho} = e^{2 i \rho} v = \begin{pmatrix} \1 \cosh \tau & \1 \sinh \tau \\
-\1 \sinh \tau & -\1 \cosh \tau \end{pmatrix} . \]
A straightforward calculation yields that the anti-commutator of this matrix with~$\tilde{v}$
is a multiple of the identity if and only if 
\[ \cosh(\alpha-\tau) = \cosh(\beta+\tau) \:, \]
determining~$\tau$ uniquely to~$\tau = (\alpha-\beta)/2$.
In the second case in~\eqref{twocases}, a similar calculation yields the
condition~$\cosh(\alpha-\tau) = \cosh(\beta-\tau)$, which has no solution.
We conclude that we must choose~$\rho$ as
\beq \label{rhoform}
\rho = \frac{\alpha-\beta}{4}\: e_4\:.
\eeq
In order to construct the corresponding Clifford subspaces~$K$ and~$\tilde{K}$,
we first replace~$\tilde{v}$ by the transformed operator~$e^{-i \rho} \tilde{v} e^{i \rho}$.
Then we are again in case~$\alpha=\beta>0$, where the unique Clifford subspace~$K$
is given by the span of the matrices~$e_0, \ldots, e_4$ in~\eqref{Dirac}.
Now we can use the formula in~(ii) to define~$\tilde{K}$; it follows by construction
that~$\tilde{v} \in \tilde{K}$.

In order to prove continuity, we first note that the constructions in the two cases~$\alpha=\beta$
and~$\alpha \neq \beta$ obviously depend continuously on~$v$ and~$\tilde{v}$.
Moreover, it is clear from~\eqref{rhoform} that~$\rho$ is continuous in the
limit~$\alpha - \beta \rightarrow 0$. This concludes the proof.
\QED

\begin{Def} \label{defsync}
For generically separated signature operators~$v, \tilde{v}$,
we denote the unique clifford extension~$K$ in Lemma~\ref{lemma3}
by~$K^{v, (\tilde{v})} \in \T^v$ and refer to it as the
{\bf{Clifford extension of~$v$ synchronized with~$\tilde{v}$}}.
Similarly, $K^{\tilde{v}, (v)} \in \T^{\tilde{v}}$ is
the Clifford extension of~$\tilde{v}$ synchronized with~$v$.
Moreover, we introduce the {\bf{synchronization map}}~$U^{\tilde{v}, v} := e^{i \rho} \in \U(S_x)$.
\end{Def} \noindent
According to Lemma~\ref{lemma3}, the synchronization map satisfies the relations
\[ U^{\tilde{v}, v} = (U^{v, \tilde{v}})^{-1}
\qquad \text{and} \qquad K^{\tilde{v}, (v)} = 
U^{\tilde{v}, v} K^{v, (\tilde{v})} U^{v, \tilde{v}} \:. \]

\subsection{The Spin Connection} \label{sec33}
For the constructions in this section we need a stronger version of Definition~\ref{def2}.

\begin{Def} \label{defproptl}
The space-time points~$x,y \in M$ are said to be {\bf{properly timelike}}
separated if the closed chain~$A_{xy}$ has a strictly positive spectrum and if the corresponding
eigenspaces are definite subspaces of~$S_x$.
\end{Def} \noindent
The condition that the eigenspaces should be definite ensures that~$A_{xy}$ is diagonalizable
(as one sees immediately by restricting~$A_{xy}$ to the orthogonal complement of
all eigenvectors). Let us verify that our definition is symmetric in~$x$ and~$y$:
Suppose that~$A_{xy} u = \lambda u$ with~$u \in S_x$ and~$\lambda \in \R \setminus \{0\}$.
Then the vector~$w := P(y,x) u \in S_y$ is an eigenvector of~$A_{yx}$ again to the
eigenvalue~$\lambda$,
\beq \label{eigenmap}
A_{yx} \,w = P(y,x) P(x,y) \,P(y,x) \, u 
= P(y,x) \,A_{xy} \,u = \lambda\, P(y,x)\, u = \lambda w \:.
\eeq
Moreover, the calculation
\beq \label{symm}
\begin{split}
\lambda \,\Sl u | u \Sr &= \Sl u | A_{xy} u \Sr = \Sl u \,|\, P(x,y)\, P(y,x) \,u \Sr \\
&= \Sl P(y,x) u \,|\, P(y,x) u \Sr = \Sl w | w \Sr
\end{split}
\eeq
shows that~$w$ is a definite vector if and only if~$u$ is.
We conclude that~$A_{yx}$ has the same eigenvalues as~$A_{xy}$
and again has definite eigenspaces.

According to~\eqref{symm}, the condition in Definition~\ref{defproptl} that the
spectrum of~$A_{xy}$ should be positive means that~$P(y,x)$ maps positive
and negative definite eigenvectors of~$A_{xy}$ to positive and negative
definite eigenvectors of~$A_{yx}$, respectively.
This property will be helpful in the subsequent constructions. But possibly this
condition could be weakened (for example, it seems likely that a spin connection
could also be constructed in the case that the eigenvalues of~$A_{xy}$ are all negative).
But in view of the fact that in the examples in Sections~\ref{secvac} and~\ref{secglobhyp},
the eigenvalues of~$A_{xy}$ are always positive in timelike directions, for our purposes
Definition~\ref{defproptl} is sufficiently general.

For given space-time points~$x, y \in M$, our goal is to use the form of~$P(x,y)$ and~$P(y,x)$
to construct the {\em{spin connection}} $D_{x,y} \in \U(S_y, S_x)$ as a unitary transformation
\beq \label{Dunit}
D_{x,y} : S_y \rightarrow S_x \qquad \text{and} \qquad
D_{y,x}=(D_{x,y})^{-1} = (D_{x,y})^*: S_x \rightarrow S_y\:,
\eeq
which should have the additional property that it gives rise to an isometry of the
corresponding tangent spaces.

We now give the general construction of the spin connection, first in specific bases and then
in an invariant way. At the end of this section, we will list all the
assumptions and properties of the resulting spin connection (see Theorem~\ref{thmspinconnect}).
The corresponding mapping of the tangent spaces will be constructed in Section~\ref{secmetricconn}.

Our first assumption is that the space-time points~$x$ and~$y$
should be properly timelike separated (see Definition~\ref{defproptl}).
Combining the positive definite eigenvectors of~$A_{xy}$,
we obtain a two-dimensional positive definite invariant subspace~$I_+$ of the operator~$A_{xy}$.
Similarly, there is a two-dimensional negative definite invariant subspace~$I_-$.
Since~$A_{xy}$ is symmetric, these invariant subspaces form an orthogonal decomposition,
$S_x = I_+ \oplus I_-$. We introduce the operator~$v_{xy} \in \Symm(S_x)$ as an
operator with the property that~$I_+$ and~$I_-$ are eigenspaces corresponding to the eigenvalues~$+1$
and~$-1$, respectively. Obviously, $v_{xy}$ is a sign operator (see Definition~\ref{defsign}).
Alternatively, it can be characterized in a basis-independent way as follows.
\begin{Def} \label{defdirsig}
The unique sign operator~$v_{xy} \in \Symm(S_x)$ which commutes with the operator~$A_{xy}$
is referred to as the {\bf{directional sign operator}} of~$A_{xy}$.
\end{Def}

We next assume that the Euclidean sign operator and the directional sign operator are
generically separated at both~$x$ and~$y$ (see Definition~\ref{defgensep}).
Then at the point~$x$, there is the unique Clifford extension~$K_{xy}:=K_x^{v_{xy}, (s_x)}
\in \T_x^{v_{xy}}$ of the directional sign operator synchronized with the Euclidean sign operator
(see Definition~\ref{defsync} and Definition~\ref{defclifford}, where for clarity we
added the base point~$x$ as a subscript). Similarly, at~$y$ we consider the Clifford
extension~$K_{yx} := K_y^{v_{yx}, (s_y)} \in \T_y^{v_{yx}}$.
In view of the later construction of the metric connection (see Section~\ref{secmetricconn}),
we need to impose that the spin connection should map these Clifford extensions into each other, i.e.\
\beq \label{Ucond}
K_{xy} = D_{x,y} \,K_{yx}\, D_{y,x} \:.
\eeq
To clarify our notation, we point out that by the subscript~$_{xy}$ we always denote an object at the
point~$x$, whereas the additional comma $_{x,y}$ denotes an operator which maps an object at~$y$
to an object at~$x$.
Moreover, it is natural to demand that
\beq
v_{xy} = D_{x,y}\, v_{yx}\, D_{y,x}\:. \label{Dc2}
\eeq

We now explain the construction of the spin connection in suitably chosen bases
of the Clifford subspaces and the spin spaces. We will then verify that this construction
does not depend on the choice of the bases. At the end of this section, we will
give a basis independent characterization of the spin connection.
In order to choose convenient bases at the point~$x$, 
we set~$e_0=v_{xy}$ and extend this vector to an pseudo-orthonormal
basis~$(e_0, \ldots, e_4)$ of~$K_{xy}$. We then choose the spinor basis of Corollary~\ref{corollary1}.
Similarly, at the point~$y$ we set~$e_0=v_{yx}$ and extend to a basis~$(e_0, \ldots, e_4)$ of~$K_{yx}$,
which we again represent in the form~\eqref{Dirac}.
Since~$v_{xy}$ and~$v_{yx}$ are sign operators, the inner
products~$\Sl .| v_{xy} \,. \Sr_x$ and~$\Sl .| v_{yx} \,. \Sr_y$
are positive definite, and thus these sign operators even have the representation~\eqref{vrep}.
In the chosen matrix representations, the condition~\eqref{Dc2} means that~$D_{x,y}$
is block diagonal. Moreover, in view of Lemma~\ref{cor2}, the conditions~\eqref{Ucond} imply
that~$D_{x,y}$ must be of the form
\beq \label{Dxyblock}
D_{x,y} = e^{i \vartheta_{xy}} \begin{pmatrix} D_{x,y}^+ & 0 \\
0 & D_{x,y}^- \end{pmatrix} \qquad \text{with $\vartheta_{xy} \in \R$ and~$D_{x,y}^\pm \in \SU(2)$}\:.
\eeq
Next, as observed in~\eqref{eigenmap} and~\eqref{symm},
$P(y,x)$ maps the eigenspaces of~$v_{xy}$ to the corresponding eigenspaces of~$v_{yx}$.
Thus in our spinor bases, the kernel of the fermionic operator has the form
\beq \label{Pform}
P(x,y) = \begin{pmatrix} P^+_{x,y} & 0 \\ 0 & P^-_{x,y} \end{pmatrix} , \quad
P(y,x) = \begin{pmatrix} P^+_{y,x} & 0 \\ 0 & P^-_{y,x} \end{pmatrix} \:,
\eeq
where~$P^\pm_{x,y}$ are invertible $2 \times 2$ matrices and~$P^\pm_{y,x} = (P^\pm_{x,y})^*$
(and the star simply denotes complex conjugation and transposition).

At this point, a polar decomposition of~$P^\pm_{x,y}$ is helpful.
Recall that any invertible $2 \times 2$-matrix~$X$ can be uniquely
decomposed in the form~$X = R V$ with a positive matrix~$R$ and a unitary
matrix~$V \in \U(2)$ (more precisely, one sets~$R = \sqrt{X^* X}$ and~$V = R^{-1} X $).
Since in~\eqref{Dxyblock} we are working with~$\SU(2)$-matrices, it is useful to extract from~$V$
a phase factor. Thus we write
\beq \label{polar}
P^s(x,y) = \:e^{i \vartheta_{xy}^s} R_{xy}^s\:V^s_{x,y} 
\eeq
with~$\vartheta_{xy}^s \in \R \!\!\mod 2 \pi$, $R^s_{xy} >0$ and~$V^s_{x,y} \in \SU(2)$, 
where~$s \in \{ +,- \}$.
Comparing~\eqref{polar} with~\eqref{Dxyblock}, the natural ansatz
for the spin connection is
\beq \label{Dxyansatz}
D_{x,y} = e^{\frac{i}{2}(\vartheta^+_{xy} + \vartheta^-_{xy})} \begin{pmatrix} V_{x,y}^+ & 0 \\
0 & V_{x,y}^- \end{pmatrix} .
\eeq

The construction so far suffers from the problem that the $\SU(2)$-matrices~$V^s_{x,y}$
in the polar decomposition~\eqref{polar} are determined only up to a sign, so that there
still is the freedom to perform the transformations
\beq \label{signfreedom}
V^s_{x,y} \rightarrow - V^s_{x,y} \:,\qquad
\vartheta_{xy}^s \rightarrow \vartheta_{xy}^s + \pi\:.
\eeq
If we flip the signs of both~$V^+_{x,y}$ and~$V^-_{x,y}$, then the factor~$e^{\frac{i}{2}(\vartheta^+_{x,y} + \vartheta^-_{x,y})}$ in~\eqref{Dxyansatz} also flips its sign,
so that~$D_{x,y}$ remains unchanged. The relative sign of~$V^+_{x,y}$ and~$V^-_{x,y}$,
however, does effect the ansatz~\eqref{Dxyansatz}. In order to fix the relative signs,
we need the following assumption, whose significance will be clarified
in Section~\ref{sectimedir} below.
\begin{Def} \label{deftimedir}
The space-time points~$x$ and~$y$ are said to be {\bf{time-directed}}
if the phases~$\vartheta_{xy}^\pm$ in~\eqref{polar} satisfy the condition
\[ \vartheta_{xy}^+ - \vartheta_{xy}^- \not \in \frac{\Z \pi}{2}\: . \]
\end{Def} \noindent
Then we can fix the relative signs by imposing that
\beq \label{phasefix}
\vartheta_{xy}^+ - \vartheta_{xy}^-  \in 
\Big(-\frac{3\pi}{2}, -\pi \Big) \cup
\Big(\pi, \frac{3\pi}{2} \Big)
\eeq
(the reason for this convention will become clear in Section~\ref{secvac42}).

We next consider the behavior under the transformations of bases.
At the point~$x$, the pseudo-orthonormal basis~$(v_{xy}=e_0, e_1, \ldots, e_4)$ of~$K_{xy}$
is unique up to $\SO(4)$-transformations of the basis vectors~$e_1, \ldots, e_4$.
According to Lemma~\ref{cor2}, this gives rise to a
$\U(1) \times \SU(2) \times \SU(2)$-freedom to transform the
spin basis~$\f_1, \ldots, \f_4$ (where~$\U(1)$ corresponds to
a phase transformation). At the point~$y$, we can independently perform
$\U(1) \times \SU(2) \times \SU(2)$-transformations of the spin basis.
This gives rise to the freedom to transform the kernel of the fermionic operator by
\beq \label{Ptrans}
P(x,y) \rightarrow U_x \,P(x,y)\, U_y^{-1} \qquad \text{and} \qquad
P(y,x) \rightarrow U_y \,P(y,x)\, U_x^{-1} \:,
\eeq
where
\beq \label{blockgauge}
U_z = e^{i \beta_z}
\begin{pmatrix} U_z^+ & 0 \\ 0 & U_z^- \end{pmatrix} \qquad \text{with~$\beta \in \R$
and~$U_z^\pm \in \SU(2)$}\:.
\eeq
The phase factors~$e^{\pm i \beta_z}$ shift the angles~$\vartheta^+_{xy}$
and~$\vartheta^-_{xy}$ by the same value, so that the difference of these angles
entering Definition~\ref{deftimedir} are not affected.
The $\SU(2)$-matrices~$U_z$ and~$U_z^{-1}$, on the other hand, modify the
polar decomposition~\eqref{polar} by
\[ V^s_{x,y} \rightarrow U^s_x \,V^s_{x,y} \,(U^s_y)^{-1} \qquad \text{and} \qquad
R_{xy}^s \rightarrow U^s_x \,R_{xy}^s \,(U^s_x)^{-1} \:. \]
The transformation law of the matrices~$V^s_{x,y}$ ensures that the ansatz~\eqref{Dxyansatz}
is indeed independent of the choice of bases.
We thus conclude that this ansatz indeed defines a spin connection.

The result of our construction is summarized as follows.
\begin{Def}\label{spinconable} Two space-time points~$x, y \in M$ are said to be {\bf{spin-connectable}}
if the following conditions hold:
\begin{itemize}
\item[(a)] The points~$x$ and~$y$ are properly timelike separated
(see Definition~\ref{defproptl}).
\item[(b)] The directional sign operator~$v_{xy}$ of~$A_{xy}$ is generically separated
from the Euclidean sign operator~$s_x$ (see Definitions~\ref{defdirsig} and~\ref{defgensep}).
Likewise, $v_{yx}$ is generically separated from~$s_y$.
\item[(c)] The points~$x$ and~$y$ are time-directed (see Definition~\ref{deftimedir}).
\end{itemize}
The {\bf{spin connection}}~$D$ is the set of spin-connectable pairs~$(x,y)$ together with the
corresponding maps~$D_{x,y} \in \U(S_y, S_x)$ which are uniquely
determined by~\eqref{Dxyansatz} and~\eqref{phasefix},
\[ D = \left\{ ((x,y), D_{x,y}) \text{ with $x,y$ spin-connectable} \right\} . \]
\end{Def}

We conclude this section by compiling properties of the spin connection
and by characterizing it in a basis independent way. To this end, we want to rewrite~\eqref{polar}
in a way which does not refer to our particular bases. First, using~\eqref{Pform} and~\eqref{polar},
we obtain for the closed chain
\beq \label{Axyframe}
A_{xy} = P(x,y)\: P(x,y)^* = \begin{pmatrix} (R^+_{xy})^2 & 0 \\ 0 & (R^-_{xy})^2 \end{pmatrix} \:.
\eeq
Taking the inverse and multiplying by~$P(x,y)$, the operators~$R^\pm_{xy}$ drop out,
\[ A_{xy}^{-\frac{1}{2}}\: P(x,y) = \begin{pmatrix} e^{i \vartheta^+_{xy}}\: V^+_{x,y} & 0 \\ 0 &
e^{i \vartheta^-_{xy}}\: V^-_{x,y} \end{pmatrix} \:. \]
Except for the relative phases on the diagonal, this 
coincides precisely with the definition of the spin connection~\eqref{Dxyansatz}.
Since in our chosen bases, the operator~$v_{xy}$ has the matrix representation~\eqref{vrep},
this relative phase can be removed by multiplying with the operator~$\exp(i \varphi_{xy} v_{xy})$,
where
\beq \label{phival}
\varphi_{xy} = -\frac{1}{2} \left( \vartheta^+_{xy} - \vartheta^-_{xy} \right) .
\eeq
Thus we can write the spin connection in the basis independent form
\beq \label{Dsecond}
D_{x,y} = e^{i \varphi_{xy}\, v_{xy}}\: A_{xy}^{-\frac{1}{2}}\: P(x,y) \:.
\eeq
Obviously, the value of~$\varphi_{xy}$ in~\eqref{phival} is also
determined without referring to our bases by using the condition~\eqref{Ucond}.
This makes it possible to reformulate our previous results in a manifestly invariant way.
\begin{Lemma} There is~$\varphi_{xy} \in \R$ such that~$D_{x,y}$ defined by~\eqref{Dsecond}
satisfies the conditions~\eqref{Dunit} and
\beq \label{DK}
(D_{x,y})^{-1} \,K_{xy}\, D_{x,y} = K_{yx}\:.
\eeq
The phase~$\varphi_{xy}$ is determined up to multiples of~$\frac{\pi}{2}$.
\end{Lemma}

\begin{Def} \label{deftimedir2}
The space-time points~$x$ and~$y$ are said to be {\bf{time-directed}}
if the phase~$\varphi_{xy}$ in~\eqref{Dsecond} satisfying~\eqref{DK}
is not a multiple of~$\frac{\pi}{4}$.
\end{Def} \noindent
We then uniquely determine~$\varphi_{xy}$ by the condition
\beq \label{phasefix2}
\varphi_{xy} \in \Big(-\frac{3\pi}{4}, -\frac{\pi}{2} \Big) \cup
\Big(\frac{\pi}{2}, \frac{3\pi}{4} \Big) \:.
\eeq

\begin{Thm} {\bf{(characterization of the spin connection)}} \label{thmspinconnect}
Assume that the points~$x, y$ are spin-connectable (see Definitions~\ref{spinconable}
and~\ref{deftimedir2}). Then the spin connection of Definition~\ref{spinconable}
is uniquely characterized by the following conditions:
\begin{itemize}
\item[(i)] $D_{x,y}$ is of the form~\eqref{Dsecond} with~$\varphi_{xy}$ in the range~\eqref{phasefix2}.
\item[(ii)] The relation~\eqref{Ucond} holds,
\[ D_{y,x} \,K_{xy}\, D_{x,y} = K_{yx} \:. \]
\end{itemize}
The spin connection has the properties
\begin{align}
D_{y,x} &= (D_{x,y})^{-1} = (D_{x,y})^* \label{unitary} \\
A_{xy} &= D_{x,y}\, A_{yx}\, D_{y,x} \label{ADrel} \\
v_{xy} &= D_{x,y}\, v_{yx}\, D_{y,x} \:. \label{Dc22}
\end{align}
\end{Thm}
\Proof The previous constructions show that the conditions~(i) and~(ii) give rise to a unique
unitary mapping~$D_{x,y} \in \U(S_y, S_x)$, which coincides with the spin connection
of Definition~\ref{spinconable}. Since~$\varphi_{xy}$ is uniquely fixed, it follows that
\[ \varphi_{yx} = - \varphi_{xy}\:, \]
and thus it is obvious from~\eqref{Dsecond} that the identity~$D_{y,x} = D_{x,y}^{-1}$ holds.

The identity~\eqref{ADrel} follows from the calculation
\begin{align*}
D_{x,y}\, A_{yx} &= \Big( e^{i \varphi_{xy}\, v_{xy}}\: A_{xy}^{-\frac{1}{2}} \, P(x,y) \Big) A_{yx}
= e^{i \varphi_{xy}\, v_{xy}}\: A_{xy}^{-\frac{1}{2}}\: A_{xy}\: P(x,y) \\
&= A_{xy} \Big( e^{i \varphi_{xy}\, v_{xy}}\: A_{xy}^{-\frac{1}{2}}\: P(x,y) \Big) = A_{xy}\, D_{x,y} \:,
\end{align*}
where we applied~\eqref{Dsecond} and used that the operators~$A_{xy}$ and~$v_{xy}$ commute.

The relations~\eqref{ADrel} and~\eqref{unitary} show that the operators~$A_{xy}$ and~$A_{yx}$
are mapped to each other by the unitary transformation~$D_{y,x}$.
As a consequence, these operators have the same spectrum, and~$D_{y,x}$ also maps the
corresponding eigenspaces to each other. This implies~\eqref{Dc22}
(note that this identity already appeared in our previous construction; see~\eqref{Dc2}).
\QED

\subsection{The Induced Metric Connection, Parity-Preserving Systems} \label{secmetricconn}
The spin connection induces a connection on the corresponding tangent spaces, as we now explain.
Suppose that~$x$ and~$y$ are two spin-connectable space-time points.
According to Lemma~\ref{lemma3}, the signature operators~$s_x$ and~$v_{xy}$
distinguish two Clifford subspaces at~$x$.
One of these Clifford subspaces was already used in the previous section;
we denoted it by~$K_{xy} := K_x^{v_{xy}, (s_x)}$ (see also Definition~\ref{defsync}).
Now we will also need the other Clifford subspace, which we denote
by~$K_x^{(y)} := K_x^{s_x, (v_{xy})}$. It is an element of~$\T^{s_x}$ and can therefore be regarded
as a representative of the tangent space. We denote the corresponding synchronization map
by~$U_{xy} = U^{v_{xy}, s_x}$, i.e.
\[ K_{xy} = U_{xy}\, K_x^{(y)}\, U_{xy}^{-1}\:. \]
Similarly, at the point~$y$ we represent the tangent space
by the Clifford subspace~$K_y^{(x)} := K_y^{s_y, (v_{yx})} \in \T^{s_y}$
and denote the synchronization map by~$U_{yx} = U^{v_{yx}, s_y}$.

Suppose that a tangent vector~$u_y \in T_y$ is given.
We can regard~$u_y$ as a vector in~$K_y^{(x)}$.
By applying the synchronization map, we obtain a vector in~$K_{yx}$,
\beq \label{metric1}
u_{yx} := U_{yx} \,u_y\, U_{yx}^{-1} \in K_{yx}\:.
\eeq
According to Theorem~\ref{thmspinconnect}~(ii), we can now ``parallel transport'' the vector to the
Clifford subspace~$K_{xy}$,
\beq \label{metric2}
u_{xy} := D_{x,y} \, u_{yx}\, D_{y,x} \in K_{xy}\:.
\eeq
Finally, we apply the inverse of the synchronization map to obtain the vector
\beq \label{metric3}
u_x := U_{xy}^{-1} \,u_{xy}\, U_{xy} \in K_x^{(y)} \:.
\eeq
As~$K_x^{(y)}$ is a representative of the tangent space~$T_x$ and all transformations were unitary,
we obtain an isometry from~$T_y$ to~$T_x$. 
\begin{Def} \label{defmetric}
The isometry between the tangent spaces defined by
\[ \nabla_{x,y} \::\: T_y \rightarrow T_x \,,\; u_y \mapsto u_x \]
is  referred to as the {\bf{metric connection}} corresponding to the spin connection~$D$.
\end{Def} \noindent
By construction, the metric connection satisfies the relation
\[ \nabla_{y,x} = (\nabla_{x,y})^{-1} \:. \]

We would like to introduce the notion that the metric connection preserves the spatial orientation.
This is not possible in general, because in view of~\eqref{parity} the tangent spaces themselves
have no spatial orientation. However, using the notions of Definition~\ref{defparity}
we can introduce a spatial orientation under additional assumptions.

\begin{Def}\label{def-parpreserv} A causal fermion system of spin dimension two is
said to be {\bf{parity preserving}} if for every point~$x \in M$, the set
\[ \mathscr{U}(x) := \{ K_x^{(y)} \text{with~$y$ spin-connectable to~$x$} \} \]
is parity preserving (see Definition~\ref{defparity}).
\end{Def} \noindent
Provided that this condition holds, the identification maps~$\phi^{s_x}_{\tilde{K}, K}$
with~$K, \tilde{K} \in \mathscr{U}(x)$ can be uniquely fixed by choosing them in the form~\eqref{phidef}
with~$U$ according to~\eqref{Ufix}. Denoting the corresponding equivalence relation
by~$\overset{\oplus}{\sim}$, we introduce the {\bf{space-oriented tangent space}}~$T_x^\oplus$ by
\[ T_x^\oplus = \mathscr{U}(x) / \overset{\oplus}{\sim} \:. \]
Considering the Clifford subspaces~$K_y^{(x)}$ and~$K_x^{(y)}$ as representatives
of~$T_y^\oplus$ and~$T_x^\oplus$, respectively, the above construction~\eqref{metric1}-\eqref{metric3}
gives rise to the {\bf{parity preserving metric connection}}
\[ \nabla_{x,y} \::\: T_y^\oplus \rightarrow T_x^\oplus\,,\; u_y \mapsto u_x \:. \]

\subsection{A Distinguished Direction of Time} \label{sectimedir}
For spin-connectable points we can distinguish a direction of time.
\begin{Def} \label{deftime} {\bf{(Time orientation of space-time)}}
Assume that the points~$x, y \in M$ are spin-connectable.
We say that~$y$ lies in the {\bf{future}} of~$x$ if the phase~$\varphi_{xy}$
as defined by~\eqref{Dsecond} and~\eqref{phasefix2} is positive.
Otherwise, $y$ is said to lie in the {\bf{past}} of~$x$.

We denote the points in the future of~$x$ by~$\I^\vee(x)$.
Likewise, the points in the past of~$y$ are denoted by~$\I^\wedge(x)$.
We also introduce the set
\[ \I(x) = \I^\vee(x) \cup \I^\wedge(x) \:; \]
it consists of all points which are spin-connectable to~$x$.
\end{Def} \noindent
Taking the adjoint of~\eqref{Dsecond} and using that~$D_{x,y}^* = D_{y,x}$,
one sees that~$\varphi_{xy} = -\varphi_{yx}$. Hence~$y$ lies in the future of~$x$ if and only if~$x$
lies in the past of~$y$. Moreover, as all the conditions in Definition~\ref{spinconable}
are stable under perturbations of~$y$ and the phase~$\varphi_{xy}$ is continuous in~$y$,
we know that~$\I^\vee(x)$ and~$\I^\wedge(x)$ are open subsets of~$M$.

On the tangent space, we can also introduce the notions of past and future, albeit in a
completely different way. We first give the definition and explain afterwards how the
different notions are related. Recall that, choosing a representative~$K \in \T^{s_x}$ of the tangent
space~$T_x$, every vector~$u \in T_x$ can be regarded as a vector in the Clifford subspace~$K$.
According to Lemma~\ref{lemma0}, the bilinear form~$\Sl .| u. \Sr_x$ on~$S_x$ is
definite if~$\la u,u \ra >0$. Using these facts, the following definition is
independent of the choice of the representatives.
\begin{Def} {\bf{(Time orientation of the tangent space)}} \label{deftimetangent}
\[ \text{A vector~$u \in T_x$ is called} \quad
\left\{\! \begin{array}{cl}
\text{timelike}  & \text{if~$\la u,u \ra > 0$} \\
\text{spacelike} & \text{if~$\la u,u \ra < 0$} \\
\text{lightlike} & \text{if~$\la u,u \ra = 0$}\:.
\end{array} \right. \]
We denote the timelike vectors by~$I_x \subset T_x$.
\[ \text{A vector~$u \in I_x$ is called} \quad
\left\{\! \begin{array}{cl}
\text{future-directed}  & \text{if~$\Sl .| u \,. \Sr_x > 0$} \\
\text{past-directed} & \text{if~$\Sl .| u \,. \Sr_x < 0$}\:.
\end{array} \right. \]
We denote the future-directed and past-directed vectors by~$I_x^\vee$ and~$I_x^\wedge$,
respectively.
\end{Def}

In order to clarify the connection between these definitions, we now construct a mapping
which to every point~$y \in \I(x)$ associates a timelike tangent vector $y_x \in I_x$,
such that the time orientation is preserved.
To this end, for given~$y \in \I(x)$ we consider the operator
\[ L_{xy} = -i D_{x,y}\, P(y,x) \::\: S_x \rightarrow S_x \]
and symmetrize it,
\[ M_{xy} = \frac{1}{2} \left( L_{xy} + L_{xy}^* \right) \in \Symm(S_x)\:. \]
The square of this operator need not be a multiple of the identity, and therefore
it cannot be regarded as a vector of a Clifford subspace. But we can take the
orthogonal projection $\text{pr}_{K_{xy}}$ of~$M_{xy}$ onto the Clifford
subspace~$K_{xy} \subset \Symm(S_x)$
(with respect to the inner product~$\la .,.  \ra$), giving us a vector in~$K_{xy}$.
Just as in~\eqref{metric3}, we can apply the synchronization map to obtain a
vector in~$K_x^{(y)}$, which then represents a vector of the tangent space~$T_x$.
We denote this vector by~$y_x$ and refer to it as the {\bf{time-directed tangent vector}} of~$y$ in~$T_x$,
\beq \label{yxdef}
y_x = U_{xy}^{-1} \,\text{pr}_{K_{xy}} (M_{xy})\, U_{xy} \in K_x^{(y)} \:.
\eeq
Moreover, it is useful to introduce the {\bf{directional tangent vector}} $\hat{y}_x$
of~$y$ in~$T_x$ by synchronizing the directional sign operator~$v_{xy}$,
\beq \label{hyxdef}
\hat{y}_x := U_{xy}^{-1} \,v_{xy}\, U_{xy} \in K_x^{(y)} \:.
\eeq
By definition of the sign operator, the inner product~$\Sl .| v_{xy} . \Sr_x$ is positive definite.
Since the synchronization map is unitary, it follows that the vector~$\hat{y}_x$ is
a future-directed unit vector in~$T_x$.

\begin{Prp} For any~$y \in \I(x)$, the time-directed tangent vector of~$y$ in~$T_x$
is timelike, $y_x \in I_x$. Moreover, the time orientation of the space-time points~$x, y \in M$
(see Definition~\ref{deftime}) agrees with the time orientation of~$y_x \in T_x$
(see Definition~\ref{deftimetangent}),
\[ y \in \I^\vee(x) \Longleftrightarrow y_x \in I^\vee_x \qquad \text{and} \qquad
y \in \I^\wedge(x) \Longleftrightarrow y_x \in I^\wedge_x \:. \]
Moreover,
\beq \label{yxsync}
y_x = \frac{1}{4}\, \sin(\varphi_{xy}) \Tr \Big( A_{xy}^{\frac{1}{2}} \Big)\: \hat{y}_x \:.
\eeq
\end{Prp}
\Proof From~\eqref{Dsecond} one sees that
\[ L_{xy} = -i e^{i \varphi_{xy}\, v_{xy}}\: A_{xy}^{\frac{1}{2}} \qquad \text{and} \qquad
M_{xy} = \sin(\varphi_{xy})\: v_{xy}\:  A_{xy}^{\frac{1}{2}}  \:. \]
We again choose the pseudo-orthonormal basis~$(e_0=v_{xy}, e_1, \ldots, e_4)$ of~$K_{xy}$
and the spinor basis of Corollary~\ref{corollary1}. Then~$v_{xy}$ has the form~\eqref{vrep},
whereas~$A_{xy}$ is block diagonal~\eqref{Axyframe}.
Since the matrices~$e_1, \ldots e_4$ vanish on the block diagonal, the
operators~$e_j M_{xy}$ are trace-free for~$j=1,\ldots, 4$. 
Hence the projection of~$M_{xy}$ is proportional to~$v_{xy}$,
\[ \text{pr}_{K_{xy}} (M_{xy}) = \frac{1}{4}\, \sin(\varphi_{xy}) \Tr \Big( A_{xy}^{\frac{1}{2}} \Big)
\: v_{xy}\:. \]
By synchronizing we obtain~\eqref{yxsync}.

The trace in~\eqref{yxsync} is positive because the operator~$A_{xy}$ has a strictly positive spectrum
(see Definition~\ref{defproptl}). Moreover, in view of~\eqref{phasefix2}
and Definition~\ref{deftime}, the factor~$\sin (\varphi_{xy})$ is positive if and only if
$y$ lies in the future of~$x$. Since~$\hat{y}_x$ is future-directed,
we conclude that~$y_x \in I^\vee_x$ if and only if~$y \in \I^\vee(x)$.
\QED

\subsection{Reduction of the Spatial Dimension}
We now explain how to reduce the dimension of the tangent space to four, with the
desired Lorentzian signature~$(1,3)$. 
\begin{Def}\label{def-chirsymm} A causal fermion system of spin dimension two
is called {\bf{chirally symmetric}} if to every~$x \in M$
we can associate a spacelike vector~$u(x) \in T_x$ which is orthogonal to all directional tangent vectors,
\[ \la u(x), \hat{y}_x \ra = 0 \qquad \text{for all~$y \in \I(x)$} \:, \]
and is parallel with respect to the metric connection, i.e.
\[ u(x) = \nabla_{x,y} \,u(y) \qquad \text{for all~$y \in \I(x)$} \:. \]
\end{Def}

\begin{Def} \label{defreduce}
For a chirally symmetric fermion system, we introduce the {\bf{reduced tangent
space}} $T_x^\text{red}$ by
\[ T_x^\text{red} = \langle u_x \rangle^\perp \subset T_x \:. \]
\end{Def} \noindent
Clearly, the reduced tangent space has dimension four and signature~$(1,3)$.
Moreover, the operator~$\nabla_{x,y}$ maps the reduced tangent spaces isometrically to each other.
The local operator~$e_5 := -i u/\sqrt{-u^2}$ takes the role of the {\bf{pseudoscalar matrix}}.

\subsection{Curvature and the Splice Maps}
We now introduce the curvature of the metric connection and the spin connection and
explain their relation.
Since our formalism should include discrete space-times, we cannot in general
work with an infinitesimal parallel transport. Instead, we must take two space-time points
and consider the spin or metric connection, which we defined in Sections~\ref{sec33}
and~\ref{secmetricconn} as mappings between the corresponding spin or tangent spaces.
By composing such mappings, we can form the analog of the parallel transport along a
polygonal line. Considering closed polygonal loops, we thus obtain the analog of
a holonomy. Since on a manifold, the curvature at~$x$ is immediately
obtained from the holonomy by considering the loops in a small neighborhood of~$x$,
this notion indeed generalizes the common notion of curvature to causal fermion systems.

We begin with the metric connection.
\begin{Def} \label{defcurvature}
Suppose that three points~$x, y, z \in M$ are pairwise spin-connectable. Then the
{\bf{metric curvature}}~$R$ is defined by
\beq \label{Rdef}
R(x,y,z) = \nabla_{x,y} \,\nabla_{y,z} \,\nabla_{z,x} \::\: T_x \rightarrow T_x\:.
\eeq
\end{Def} \noindent
Let us analyze this notion, for simplicity for parity-preserving systems.
According to~\eqref{metric1}-\eqref{metric3}, for a given tangent vector~$u_y \in K_y^{(x)}$
we have
\[ \nabla_{x,y} u_y = U u_y\, U^{-1} \in K_x^{(y)} \qquad \text{with} \qquad
U = U_{xy}^{-1} \, D_{x,y}\, U_{yx}\:. \]
Composing with~$\nabla_{z,x}$, we obtain
\[ \nabla_{z,x} \nabla_{x,y} u_y = U u_y U^{-1} \in K_z^{(x)} \]
with
\[ U = U_{zx}^{-1} \, D_{z,x}\,\Big( U_{xz}\, U^{s_x}_{K_x^{(z)}, K_x^{(y)}} \,
 U_{xy}^{-1} \Big)\, D_{x,y}\, U_{yx} \:, \]
where~$U^{s_x}_{K_x^{(z)}, K_x^{(y)}}$ is the unitary operator~\eqref{Ufix} of the
identification map.
We see that the composition of the metric connection can be written as the product
of spin connections, joined by the product of unitary operators in the brackets which
synchronize and identify suitable Clifford extensions. We give this operator product a convenient name.

\begin{Def}\label{def-cliffpara} The unitary mapping
\[ U_x^{(z|y)} = U_{xz}\, U^{s_x}_{K_{xz}, K_{xy}} \, U_{xy}^{-1} \in \U(S_x) \]
is referred to as the {\bf{splice map}}.
A causal fermion system of spin dimension two is called {\bf{Clifford-parallel}}
if all splice maps are trivial.
\end{Def} \noindent
Using the splice maps, the metric curvature can be written as
\[ R(x,y,z) \::\: K_x^{(z)} \rightarrow K_x^{(z)} \, , \; u_x \mapsto
U u_x U^{-1} \:, \]
where the unitary mapping~$U$ is given by
\beq \label{Ucompose}
U = U_{xz}^{-1}\;
\,U_x^{(z|y)}\, D_{x,y} \,U_y^{(x|z)}\,D_{y,z}\, U_z^{(y|x)} \,D_{z,x} \; U_{xz} \:.
\eeq
Thus two factors of the spin connection are always joined by an intermediate
splice map.

We now introduce the curvature of the spin connection. The most obvious way
is to simply replace the metric connection in~\eqref{Rdef} by the spin connection.
On the other hand, the formula~\eqref{Ucompose} suggests that it might be a good idea to
insert splice maps. As it is a-priori not clear which method is preferable, we define
both alternatives.
\begin{Def} Suppose that three points~$x, y, z \in M$ are pairwise spin-connectable. Then the
{\bf{unspliced spin curvature}}~$\mathfrak{R}^\text{us}$ is defined by
\beq \label{Rusdef}
{\mathfrak{R}}^\text{us}(x,y,z) = D_{x,y} \,D_{y,z} \,D_{z,x} \::\: S_x \rightarrow S_x\:.
\eeq
The (spliced) {\bf{spin curvature}} is introduced by
\beq \label{Rspliced}
{\mathfrak{R}}(x,y,z) = U_x^{(z|y)} \,D_{x,y} \,U_y^{(x|z)}\,
D_{y,z} \,U_z^{(y|x)}\, D_{z,x} \::\: S_x \rightarrow S_x\:.
\eeq
\end{Def} \noindent
Clearly, for Clifford-parallel systems, the spliced and unspliced spin curvatures
coincide. But if  the causal fermion system is not Clifford-parallel, the situation is more
involved. The spliced spin curvature and the metric curvature
are compatible in the sense that, after unitarily transforming 
to the Clifford subspace~$K_{xz}$, the following identity holds,
\[ U_{xz} \,R(x,y,z)\, U_{xz}^{-1} \::\: K_{xz} \rightarrow K_{xz} \,,\;
v \mapsto \mathfrak{R}(x,y,z) \,v\, \mathfrak{R}(x,y,z)^*\:. \]
Thus the metric curvature can be regarded as ``the square of the spliced spin curvature.''

We remark that the systems considered in Section~\ref{secvac} will all be Clifford parallel.
In the examples in Section~\ref{secglobhyp}, however, the systems
will {\em{not}} be Clifford parallel. In these examples, we shall see that it is indeed preferable to
work with the spliced spin curvature (for a detailed explanation see Section~\ref{secversus}).

We conclude this section with a construction which will be useful in Section~\ref{secglobhyp}.
In the causal fermion systems considered in these sections,
at every space-time point there is a distinguished representative of the tangent space,
making it possible to introduce the following notion.
\begin{Def} \label{deftangrep} We denote the set of five-dimensional subspaces of~$\Symm(\H)$
by ${\mathfrak{S}}_5(\H)$; it carries the topology induced by the operator norm~\eqref{onorm}.
A continuous mapping~$\mathfrak{K}$ which to very space-time point associates
a representative of the corresponding tangent space,
\[ \mathfrak{K} \::\: M \rightarrow {\mathfrak{S}}_5(\H) 
\qquad \text{with} \qquad
\mathfrak{K}(x) \in \T^{s_x} \quad \text{for all~$x \in M$}\:, \]
is referred to as a {\bf{representation map}} of the tangent spaces.
The system~$(\H, \F, \rho, \mathfrak{K})$ is referred to as a causal
fermion system with {\bf{distinguished representatives of the tangent spaces}}.
\end{Def} \noindent
If we have distinguished representatives of the tangent spaces, the spin connection
can be combined with synchronization and identification maps such that forming
compositions of this combination always gives rise to intermediate splice maps.
\begin{Def} \label{defsplicedD} Suppose that our fermion system is parity preserving and has
distinguished representatives of the tangent spaces. Introducing the splice maps~$U_.^{|.)}$
and~$U_.^{(.|}$ by
\[ U_x^{|y)} = U^{s_x}_{\mathfrak{K}(x) , K^{(y)}_x} \,U_{xy}^{-1} \qquad \text{and}
\qquad U_x^{(y|} = \big( U_x^{|y)} \big)^* = 
U_{xy} \,U^{s_x}_{K^{(y)}_x, \mathfrak{K}(x)} \:, \]
we define the {\bf{spliced spin connection}}~$D_{(.|.)}$ by
\beq \label{Dsplicedef}
D_{(x,y)} =  U_x^{|y)}\:D_{x,y}\: U_y^{(x|}
\::\: S_y \rightarrow S_x\:.
\eeq
\end{Def} \noindent
Our notation harmonizes with Definition~\ref{def-cliffpara} in that
\beq \label{Uharmony}
U_x^{(y|} \:U_x^{|z)} = U_x^{(y|z)}\:.
\eeq
Forming compositions and comparing with Definition~\ref{def-cliffpara}, one readily finds that
\[ D_{(x,y)} \,D_{(y,z)} = U_x^{|y)}\,
D_{x,y}\: U_y^{(x|z)} \,D_{y,z}\, U_z^{(x|} . \]
Proceeding iteratively, one sees that the spin curvature~\eqref{Rspliced} can be represented by
\[ {\mathfrak{R}}(x,y,z) = V \,D_{(x,y)} \, D_{(y,z)} \, D_{(z,x)} \, V^*
\qquad \text{with} \qquad
V = U_x^{(z|} \:. \]
Thus up to the unitary transformation~$V$, the spin curvature coincides
with the holonomy of the spliced spin connection.

\subsection{Causal Sets and Causal Neighborhoods}
The relation ``lies in the future of'' introduced in Definition~\ref{deftime}
reminds of the partial ordering on a causal set.
In order to explain the connection, we first recall the definition of a causal set
(for details see for example~\cite{sorkin}).

\begin{Def} \label{defcausalset}
A set~$C$ with a partial order relation~$\Sl$ is a {\bf{causal set}} if
the following conditions hold:
\begin{itemize}
\item[(i)] Irreflexivity: For all~$x \in C$, we have~$x \notcaus x$.
\item[(ii)] Transitivity: For all~$x, y, z \in C$, we have~$x \Sl y$ and~$y \Sl z$
implies~$x \Sl z$.
\item[(iii)] Local finiteness: For all~$x,z \in C$, the set~$\{y \in C \text{ with }
x \Sl y \Sl z\}$ is finite.
\end{itemize}
\end{Def} \noindent
Our relation ``lies in the future of'' agrees with~(i) because the sign operators~$s_x$
and~$v_{xx}$ coincide, and therefore every space-time point~$x$ is not spin-connectable
to itself. The condition~(iii) seems an appropriate assumption for causal fermion systems in discrete space-time
(in particular, it is trivial if~$M$ is a finite set). In the setting when space-time
is a general measure space~$(M, \mu)$, it is natural to replace~(iii) by the condition
that the set~$\{y \in C \text{ with } x \Sl y \Sl z\}$ should have finite measure.
The main difference between our setting and a causal set is that the relation
``lies in the future of'' is in general not transitive, so that~(ii) is violated.
However, it seems reasonable to weaken~(ii)
by a local condition of transitivity. We now give a possible definition.

\begin{Def} \label{locft} A subset~$U \subset M$ is called {\bf{future-transitive}} if for all
pairwise spin-connectable points~$x,y,z \in U$ the following implication holds:
\[ y \in \I^\vee(x) \text{ and } z \in \I^\vee(y) \quad \Longrightarrow \quad
z \in \I^\vee(x) \:. \]

A causal fermion system of spin dimension two is called {\bf{locally future-transitive}}
if every point~$x \in M$ has a neighborhood~$U$ which is future-transitive.
\end{Def} \noindent
This definition ensures that~$M$ locally includes the structure of a causal set.
As we shall see in the examples in Sections~\ref{secvac} and~\ref{secglobhyp},
Dirac sea configurations without regularization in Minkowski space or on
globally hyperbolic Lorentzian manifolds are indeed locally future-transitive.
However, it still needs to be investigated if Definition~\ref{locft}
applies to quantum space-times of physical interest.

\section{Example: The Regularized Dirac Sea Vacuum} \label{secvac}
As a first example, we now consider Dirac spinors in Minkowski space.
Taking~$\H$ as the space of all negative-energy solutions of the Dirac equation,
we construct a corresponding causal fermion system.
We show that the notions introduced in Section~\ref{secconstruct} give back the usual causal
and geometric structures of Minkowski space.

We first recall the basics and fix our notation (for details see for example~\cite{bjorken}
or~\cite[Chapter~1]{PFP}). Let~$(M, \la .,. \ra)$ be Minkowski space
(with the signature convention~$(+ - - -)$) and~$d\mu$ the standard volume measure
(thus~$d\mu = d^4x$ in a reference frame~$x= (x^0, \ldots, x^3)$).
Naturally identifying the spinor spaces at different space-time points and denoting
them by~$V=\C^4$, we write the free Dirac equation for particles of mass~$m>0$ as
\beq\label{deq}
(i\slashed{\partial} - m) \,\psi := (i \gamma^k \partial_k - m) \,\psi = 0\:, 
\eeq
where~$\gamma^k$ are the Dirac matrices in the Dirac representation,
and~$\psi : M \rightarrow V$ are four-component complex Dirac spinors.
The Dirac spinors are endowed with an inner product of signature~$(2,2)$,
which is usually written as~$\overline{\psi} \phi$, where~$\overline{\psi} = \psi^\dagger \gamma^0$
is the adjoint spinor. For notational consistency, we denote this inner product
on~$V$ by~$\Sl .|. \Sr$.
The free Dirac equation has plane wave solutions, which we denote
by~$\psi_{\vec{k}a \pm}$ with~$\vec{k} \in \R^3$ and $a \in \{1,2\}$.
They have the form
\beq \label{planewave}
\psi_{\vec{k}a \pm}(x) = \frac{1}{(2 \pi)^\frac{3}{2}} \:e^{\mp i \omega t + i \vec{k}\vec{x}}\:
\chi_{\vec{k}a \pm} ,
\eeq
where~$x=(t, \vec{x})$ and~$\omega:=\sqrt{|\vec{k}|^2 + m^2}$.
Here the spinor~$\chi_{\vec{k}a \pm}$ is a solution of the algebraic equation
\beq \label{Diralg}
(k\slsh - m) \chi_{\vec{k}a \pm} = 0 \:,
\eeq
where~$k\slsh = k^j \gamma_j$ and~$k=(\pm \omega, \vec{k}$).
Using the normalization convention
\[ \Sl \chi_{\vec{k}a \pm} | \chi_{\vec{k}a' \pm} \Sr = \pm \delta_{a,a'}\:, \]
the projector onto the two-dimensional solution space
of~\eqref{Diralg} can be written as
\beq \label{Dirproj}
\frac{\slashed{k}+m}{2m} = \pm \sum_{a=1,2}|\chi_{\vec{k}a \pm}\Sr\Sl\chi_{\vec{k}a \pm}| \:.
\eeq
The frequency~$\pm \omega$ of the plane wave~\eqref{planewave} is the energy of
the solution. More generally, by a negative-energy solution~$\psi$ of the Dirac equation
we mean a superposition of plane wave solutions of negative energy,
\beq \label{negen}
\psi(x) = \sum_{a=1,2} \int d^3k\: g_a(\vec{k})\: \psi_{\vec{k} a -}(x)\:.
\eeq
Dirac introduced the concept that in the vacuum all negative-energy states
should be occupied forming the so-called Dirac sea. Following this concept,
we want to introduce the Hilbert space~$(\H, \la .,. \ra_\H)$ as the space of all negative-energy
solutions, equipped with the usual scalar product obtained by integrating the probability density
\beq \label{pip}
\la \psi | \phi \ra_\H = 2 \pi \int_{t=\text{const}} \Sl \psi(t, \vec{x}) | \gamma^0 \phi(t, \vec{x}) \Sr\,
d\vec{x} \:.
\eeq
Note that the plane-wave solutions~$\psi_{\vec{k}a -}$ cannot be considered as vectors in~$\H$,
because the normalization integral~\eqref{pip} diverges.
But for the superposition~\eqref{negen}, the normalization integral is certainly finite
for test functions~$g_a(\vec{k}) \in C^\infty_0(\R^3)$, making it possible to
define~$(\H, \la .,. \ra_\H)$ as the completion of such wave functions.
Then due to current conservation, the integral in~\eqref{pip} is time independent.
For the plane-wave solutions, one can still make sense of the normalization integral in the
distributional sense. Namely, a short computation gives
\beq \label{psinorm}
\la \psi_{\vec{k}a -} | \psi_{\vec{k}' a' -} \ra_\H
=  \frac{2 \pi\omega}{m}\: \delta_{a,a'}\: \delta^3(\vec{k} - \vec{k}')\:.
\eeq
The completeness of the plane-wave solutions can be expressed by the
Plancherel formula
\beq \label{plancherel}
\psi(x) = \frac{m}{\pi}
\sum_{a=1,2} \int \frac{d^3k}{2 \omega}\: \psi_{\vec{k}a -}(x) \:\la \psi_{\vec{k}a -} | \psi \ra\:
\qquad \text{for all~$\psi \in \H$}\:.
\eeq

\subsection{Construction of the Causal Fermion System} \label{sec41}
In order to construct a causal fermion system of spin dimension two, to every~$x \in M$
we want to associate a self-adjoint operator~$F(x) \in L(\H)$, having at most two positive
and at most two negative eigenvalues.
By identifying~$x$ with~$F(x)$, we then get into the setting of Definition~\ref{def1}.
The idea is to define~$F(x)$ as an operator which describes the correlations of the wave functions
at the point~$x$,
\beq \label{Fdef}
\la \psi | F(x) \phi \ra_\H = - \Sl \psi(x) | \phi(x) \Sr \:.
\eeq
As the spin scalar product has signature~$(2,2)$,
this ansatz incorporates that~$F(x)$ should be a self-adjoint operator with
at most two positive and at most two negative eigenvalues.
Using the completeness relation~\eqref{plancherel}, $F(x)$ can be written in the explicit form
\beq \label{Fexplicit}
F(x) \,\phi = - \frac{m^2}{\pi^2} \sum_{a,a'=1,2}
\int \frac{d^3k}{2 \omega} \int \frac{d^3k'}{2 \omega'}\:
\psi_{\vec{k}a -} \Sl \psi_{\vec{k}a -}(x) | \psi_{\vec{k}' a' -}(x) \Sr\:
\la \psi_{\vec{k}' a' -} | \phi \ra_\H \:.
\eeq
Unfortunately, this simple method does not give rise to a well-defined operator~$F(x)$.
This is obvious in~\eqref{Fdef} because the wave functions~$\psi, \phi \in \H$
are in general not continuous and could even have a singularity at~$x$.
Alternatively, in~\eqref{Fexplicit} the momentum integrals will in general diverge.
This explains why we must introduce an ultraviolet regularization.
We do it in the simplest possible way by inserting convergence generating factors,
\beq \label{Feps}
\begin{split}
F^\varepsilon(x) \,\phi := - \frac{m^2}{\pi^2} \sum_{a,a'=1,2}
\int \frac{d^3k}{2 \omega}\: &e^{-\frac{\varepsilon \omega}{2}} 
\int \frac{d^3k'}{2 \omega'}\: e^{-\frac{\varepsilon \omega'}{2}} \\
&\times \psi_{\vec{k}a -} \Sl \psi_{\vec{k}a -}(x) | \psi_{\vec{k}' a' -}(x) \Sr\:
\la \psi_{\vec{k}' a' -} | \phi \ra_\H \:,
\end{split}
\eeq
where the parameter~$\varepsilon>0$ is the length scale of the regularization.
Note that this regularization is spherically symmetric, but the Lorentz invariance is broken.
Moreover, the operator~$F^\varepsilon(x)$ is no longer a local operator, meaning that
space-time is ``smeared out'' on the scale~$\varepsilon$.

In order to show that~$F^\varepsilon$ defines a causal fermion system,
we need to compute the eigenvalues of~$F^\varepsilon(x)$.
To this end, it is helpful to write~$F^\varepsilon$ similar to a Gram matrix as
\beq \label{Fepsdef}
F^\varepsilon(x) = -\iota^\varepsilon_x\, (\iota^\varepsilon_x)^* \:,
\eeq
where~$\iota_x$ is the operator
\beq \label{iotaepsdef}
\iota^\varepsilon_x \::\: V \rightarrow \H \,,\;
u \mapsto -\frac{m}{\pi} \sum_{a=1,2}
\int \frac{d^3k}{2 \omega}\: e^{-\frac{\varepsilon \omega}{2}} \:
\psi_{\vec{k}a -} \Sl \psi_{\vec{k}a -}(x) | u \Sr\:,
\eeq
and the star denotes the adjoint with respect to the corresponding inner products~$\Sl .|. \Sr$
and~$\la .,. \ra_\H$.
From this decomposition, one sees right away that~$F^\varepsilon(x)$ has 
at most two positive and at most two negative eigenvalues. Moreover, these eigenvalues
coincide with those of the operator~$-(\iota^\varepsilon_x)^* \iota^\varepsilon_x \::\:
V \rightarrow V$, which can be computed as follows:
\begin{align}
-(\iota^\varepsilon_x)^* \iota^\varepsilon_x \,u &= -\frac{m^2}{\pi^2}\!\!\!
\sum_{a,a'=1,2} \iint \frac{d^3k\, d^3k'}{4 \omega \omega'}\: e^{-\frac{\varepsilon (\omega+\omega')}{2}} 
\:\psi_{\vec{k}a -}(x)\,
\la \psi_{\vec{k}a -} | \psi_{\vec{k}' a' -}\ra_\H \:\Sl \psi_{\vec{k}' a' -}(x) | u \Sr \nonumber \\
&\!\!\overset{\eqref{psinorm}}{=}
-\frac{m}{\pi} \sum_{a=1,2} \int \frac{d^3k}{2 \omega}\: e^{-\varepsilon \omega} 
\:\psi_{\vec{k}a -}(x)\: \:\Sl \psi_{\vec{k} a -}(x) | u \Sr \label{iis0} \\
&\!\!\overset{\eqref{Dirproj}}{=} \frac{m}{\pi} \int \frac{d^3k}{2 \omega}\: e^{-\varepsilon \omega} 
\:\frac{k\slsh+m}{2m} \, u 
=  \frac{m}{\pi}  \int \frac{d^3k}{2 \omega}\: e^{-\varepsilon \omega} 
\:\frac{-\omega \gamma^0+m}{2m} \, u \:, \label{iis}
\end{align}
where in the last step we used the spherical symmetry.
\begin{Prp} \label{prp41} For any~$\varepsilon>0$, the
operator~$F^\varepsilon(x) : \H \rightarrow \H$ has rank four
and has two positive and two negative eigenvalues. The mapping
\[ F \::\: M \rightarrow \F \,,\; x \mapsto F^\varepsilon(x) \]
is injective. Identifying~$x$ with~$F^\varepsilon(x)$ and introducing the measure~$\rho^\varepsilon
= F^\varepsilon_* \mu$ on~$\F$ as the push-forward~\eqref{push},
the resulting tupel~$(\H, \F, \rho^\varepsilon)$ is a causal fermion system of spin dimension
two. Every space-time point is regular (see Definition~\ref{defregular}).

More specifically, the non-trivial eigenvalues~$\nu_1, \ldots, \nu_4$ of the operator~$F^\varepsilon(x)$ are
\[ \begin{split}
\nu^\varepsilon_1 &= \nu^\varepsilon_2 =
\int \frac{d^3k}{4 \pi \omega}\: e^{-\varepsilon \omega} 
\left( -\omega+m \right) < 0 \\
\nu^\varepsilon_3 &= \nu^\varepsilon_4 =
\int \frac{d^3k}{4 \pi \omega}\: e^{-\varepsilon \omega} 
\;\:\left( \omega+m \right) \,\, >0 \:.
\end{split} \]
The corresponding eigenvectors~$\f^\varepsilon_1, \ldots, \f^\varepsilon_4$ are given by
\beq \label{falpha}
\f^\varepsilon_\alpha(x) = \frac{1}{\nu^\varepsilon_\alpha}\: \iota^\varepsilon_x(\e_\alpha) \:,
\eeq
where~$(\e_\alpha)$ denotes the canonical basis of~$V=\C^4$.
\end{Prp}
\Proof It is obvious from~\eqref{iis} that~$\e_\alpha$ is an eigenvector basis of the
operator~$-(\iota^\varepsilon_x)^* \iota^\varepsilon_x$,
\beq \label{eeigen}
-(\iota^\varepsilon_x)^* \iota^\varepsilon_x \, \e_\alpha = \nu_\alpha\, \e_\alpha \:.
\eeq
Next, the calculation
\[ F^\varepsilon(x)\: (\iota^\varepsilon_x \e_\alpha)
= \iota^\varepsilon_x\: \big( - (\iota^\varepsilon_x)^* \iota^\varepsilon_x \big) \e_\alpha
= \nu_\alpha^\varepsilon\, (\iota^\varepsilon_x\,  \e_\alpha) \]
shows that the vectors~$\f_\alpha^\varepsilon$ are eigenvectors of~$F^\varepsilon(x)$
corresponding to the same eigenvalues (our normalization convention will be explained
in~\eqref{nconvent} below).

To prove the injectivity of~$F^\varepsilon$, assume that~$F^\varepsilon(x) = F^\varepsilon(y)$.
We consider the expectation
value~$\la \psi | (F^\varepsilon(x) - F^\varepsilon(y)) \phi \ra_\H$.
Since this expectation value vanishes for all~$\phi$ and~$\psi$, we conclude
from~\eqref{Feps} that
\[ \Sl \psi_{\vec{k}a -}(x) | \psi_{\vec{k}' a' -}(x) \Sr = \Sl \psi_{\vec{k}a -}(y) | \psi_{\vec{k}' a' -}(y) \Sr \]
for all~$a,a' \in \{1,2\}$ and~$\vec{k}, \vec{k}' \in \R^3$.
Using~\eqref{planewave}, the left and right side of this equation are plane waves
of the form~$e^{i (k-k') x}$ and~$e^{i (k-k') y}$, respectively.
We conclude that~$x=y$.
\QED

We now introduce for every~$x \in M$ the spin space~$(S^\varepsilon_x, \Sl .|. \Sr_x)$
by~\eqref{Sxdef} and~\eqref{ssp}. By construction, the eigenvectors~$\f^\varepsilon_\alpha(x)$
in~\eqref{falpha} form a basis of~$S^\varepsilon_x$. Moreover, this basis is
pseudo-orthonormal, as the following calculation shows:
\begin{align}
\Sl \f_\alpha^\varepsilon(x)|\f_\beta^\varepsilon(x) \Sr_x
&= - \la \f_\alpha^\varepsilon(x) | F^\varepsilon(x)\, \f_\beta^\varepsilon(x) \ra_\H 
= - \nu^\varepsilon_\beta \: \la \f_\alpha^\varepsilon(x) | \f_\beta^\varepsilon(x) \ra_\H  \nonumber \\
&= -\frac{1}{\nu^\varepsilon_\alpha} \:
\la \iota^\varepsilon_x \e_\alpha | \iota^\varepsilon_x \e_\beta \ra_\H
= -\frac{1}{\nu^\varepsilon_\alpha} \:
\Sl \e_\alpha | (\iota^\varepsilon_x)^* \iota^\varepsilon_x \e_\beta \Sr \nonumber \\
&\!\!\!\overset{\eqref{eeigen}}{=} \frac{\nu^\varepsilon_\beta}{\nu^\varepsilon_\alpha} \:
\Sl \e_\alpha | \e_\beta \Sr = s_\alpha\: \delta_{\alpha \beta} \:, \label{nconvent}
\end{align}
where we again used the notation of Corollary~\ref{corollary1}.
It is useful to always identify the inner product space~$(V, \Sl.|.\Sr)$
(and thus also the spinor space~$S_xM$; see before~\eqref{deq})
with the spin space~$(S^\varepsilon_x, \Sl.|.\Sr_x)$ 
via the isometry~$\mathfrak{J}^\varepsilon_x$
given by
\beq \label{Jdef}
{\mathfrak{J}}^\varepsilon_x \::\:
S_xM \simeq V \rightarrow S^\varepsilon_x \,,\; \e_\alpha \mapsto \f_\alpha^\varepsilon(x)\:.
\eeq
Then, as the~$\f^\varepsilon(x)$ form an eigenvector basis of~$F^\varepsilon(x)$,
the Euclidean operator takes the form
\beq \label{Euklid}
s_x = \gamma^0\:.
\eeq
Moreover, we obtain a convenient matrix representation of the kernel of fermionic operator~\eqref{Pxydef},
which again under the identification of~$x$ with~$F^\varepsilon(x)$ we now write as
\beq \label{PepsF}
P^\varepsilon(x,y) = \pi_{F^\varepsilon(x)} \,F^\varepsilon(y) \:.
\eeq

\begin{Lemma} \label{lemmapeps}
In the spinor basis~$(\e_\alpha)$ given by~\eqref{Jdef}, the kernel of the fermionic operator
takes the form
\beq \label{Pepsdef}
P^\varepsilon(x,y) 
= \int \frac{d^4k}{(2 \pi)^4}\: e^{-\varepsilon |k^0|}\:
(k\slsh+m)\: \delta(k^2-m^2)\: \Theta(-k^0)\: e^{-ik(x-y)} \:.
\eeq
\end{Lemma}
\Proof Using~\eqref{ssp}, we find that~$\Sl .| \pi_x y . \Sr_x = -\la .| xy . \ra_\H$.
Thus, applying Proposition~\ref{prp41}, we find
\beq \label{Pecalc} \begin{split}
\Sl &\f^\varepsilon_\alpha(x) |  P^\varepsilon(x,y)\, \f^\varepsilon_\beta(y) \Sr_x \\
&= -\la \f^\varepsilon_\alpha(x) | F^\varepsilon(x) \,F^\varepsilon(y)\, \f^\varepsilon_\beta(y) \ra_\H 
= -\la F^\varepsilon(x)\, \f^\varepsilon_\alpha(x) | F^\varepsilon(y)\, \f^\varepsilon_\beta(y) \ra_\H \\
&= -\nu^\varepsilon_\alpha \nu^\varepsilon_\beta\: 
\la \f^\varepsilon_\alpha(x) | \f^\varepsilon_\beta(y) \ra_\H
= -\la \iota^\varepsilon_x \e_\alpha | \iota^\varepsilon_y \e_\beta \ra_\H
= -\Sl \e_\alpha | (\iota^\varepsilon_x)^* \iota^\varepsilon_y \e_\beta \Sr\:.
\end{split} \eeq
Identifying~$\f^\varepsilon_\alpha(x)$ and~$\f^\varepsilon_\alpha(y)$ with~$\e_\alpha$, we conclude that
the kernel of the fermionic operator has the representation
\begin{align*}
P^\varepsilon(x,y) &= -(\iota^\varepsilon_x)^* \iota^\varepsilon_y =
-\frac{m}{\pi} \sum_{a=1,2} \int \frac{d^3k}{2 \omega}\: e^{-\varepsilon \omega} 
\:|\psi_{\vec{k}a -}(x) \Sr \Sl \psi_{\vec{k} a -}(y) | \\
&\!\!\overset{\eqref{planewave}}{=} -2m \sum_{a=1,2} \int \frac{d^3k}{2 \omega\, (2 \pi)^4}\: e^{-ik(x-y)}
e^{-\varepsilon \omega} \:|\chi_{\vec{k}a -}(x)\Sr \Sl \chi_{\vec{k} a -}(y)| \\
&\!\!\overset{\eqref{Dirproj}}{=} \int \frac{d^3k}{2 \omega\, (2 \pi)^4}\: e^{-ik(x-y)}
e^{-\varepsilon \omega} \:(k\slsh + m) \:,
\end{align*}
where again~$k=(- \omega, \vec{k}$). Carrying out the $k^0$-integration in~\eqref{Pepsdef}
gives the result.
\QED \noindent
We point out that in the limit~$\varepsilon \searrow 0$ when the regularization is removed,
$P^\varepsilon(x,y)$ converges to the Lorentz invariant distribution
\beq \label{Pdef}
P(x,y)  = \int \frac{d^4k}{(2 \pi)^4}\:
(k\slsh+m)\: \delta(k^2-m^2)\: \Theta(-k^0)\: e^{-ik(x-y)} \:. 
\eeq
This distribution is supported on the lower mass shell and thus describes the
Dirac sea vacuum where all negative-energy solutions are occupied.
It is the starting point of the fermionic projector approach (see~\cite{PFP, srev}).

With the spin space~$(S_x^\varepsilon, \Sl .|. \Sr_x)$, the Euclidean operator~\eqref{Euklid} and the
kernel of the fermionic operator~\eqref{Pepsdef},
we have introduced all the objects needed for the constructions in Section~\ref{secconstruct}.
Before analyzing the resulting geometric structures in detail, we conclude this subsection
by computing the Fourier integral in~\eqref{Pepsdef}
and discussing the resulting formulas. Setting
\beq \label{xidef}
\xi=y-x
\eeq
and~$t=\xi^0$, $r=|\vec{\xi}|$, $p=|\vec{k}|$, we obtain
\begin{align}
	P^\varepsilon(x,y)&=(i\slashed{\partial}_x+m)\int \frac{d^4k}{(2\pi)^4}
	\delta(k^2-m^2)\:\Theta(-k^0) \:e^{ik \xi} \:e^{-\varepsilon|k^0|} \nonumber \\
	&=(i\slashed{\partial}_x+m)\int \frac{d^3k}{(2\pi)^4}\:\frac{1}{2\sqrt{\vec{k}^2+m^2}}
	\: e^{-i\sqrt{\vec{k}^2+m^2}\, t-i \vec{k}\vec{\xi}}\:e^{-\varepsilon \sqrt{\vec{k}^2+m^2}} \nonumber \\
	&=(i\slashed{\partial}_x+m)\int_0^\infty \frac{dp}{2(2\pi)^3} \int_{-1}^1d\cos\theta
	\: \frac{p^2}{\sqrt{p^2+m^2}} \:e^{-(\varepsilon+it) \sqrt{p^2+m^2}}\: e^{-ipr\cos\theta} \nonumber \\
	&= (i\slashed{\partial}_x+m) \;\frac{1}{r}\int_0^\infty \frac{dp}{(2\pi)^3}\:
	\frac{p}{\sqrt{p^2+m^2}}\: e^{-(\varepsilon+it) \sqrt{p^2+m^2}}\: \sin(pr) \nonumber \\
	&= (i\slashed{\partial}_x+m) \frac{m^2}{(2\pi)^3} \frac{K_1 \big(m\sqrt{r^2+(\varepsilon+it)^2}
		\,\big)}{m\sqrt{r^2+(\varepsilon+it)^2}}, \label{Pepssing}
\end{align}
where the last integral was calculated using~\cite[formula (3.961.1)]{gradstein}. Here
the square root and the Bessel functions $K_0$, $K_1$ are defined using a branch cut along
the negative real axis. Carrying out the derivatives, we obtain
$$ P^\varepsilon(x,y)=\alpha_\varepsilon(\xi)(\xis-i\varepsilon\gamma^0)+\beta_\varepsilon(\xi)\1 $$
with the smooth functions
\beq \label{abeps}
\alpha_\varepsilon(\xi)=-i\frac{m^4}{(2\pi)^3}\Big(\frac{K_0(z)}{z^2}+2\:\frac{K_1(z)}{z^3}\Big)\qquad\text{and}\qquad \beta_\varepsilon(\xi)=\frac{m^3}{(2\pi)^3}\frac{K_1(z)}{z} \:,
\eeq
where we set
\[ z=m\sqrt{r^2+(\varepsilon+it)^2} \:. \]
Due to the regularization, $P^\varepsilon(x,y)$ is a smooth function.
However, in the limit~$\varepsilon \searrow 0$, singularities appear on the light cone~$\{\xi^2 = 0\}$
(for details see~\cite[\S4.4]{sector}). This can be understood from the fact that the
Bessel functions~$K_0(z)$ and~$K_1(z)$ have poles at~$z=0$, leading to singularities
on the light cone if~$\varepsilon \searrow0$.
But using that the Bessel functions are smooth for~$z \neq 0$, one also sees
that away from the light cone, $P^\varepsilon$ converges pointwise (even locally uniformly)
to a smooth function. We conclude that
\beq \label{Pconverge}
P^\varepsilon(x,y) \overset{\varepsilon \searrow 0}{\longrightarrow}
P(x,y)  \qquad \text{if~$\xi^2 \neq 0$}
\eeq
and
\beq P(x,y)=\alpha(\xi)\, \xis+\beta(\xi)\,\1 \label{P-unreg}\eeq
where the functions~$\alpha$ and~$\beta$ can be written in terms of real-valued Bessel functions as
\beq\begin{split}		\beta(\xi)&=\theta(\xi^2)\frac{m^3}{16\pi^2}\frac{Y_1(m\sqrt{\xi^2})+i\epsilon(\xi^0)J_1(m\sqrt{\xi^2})}{m\sqrt{\xi^2}}+\theta(-\xi^2)\frac{m^3}{8\pi^3}\frac{K_1(m\sqrt{-\xi^2})}{m\sqrt{-\xi^2}}\\
\alpha(\xi)&=-\frac{2i}{m}\frac{d}{d (\xi^2)}\beta(\xi)
\end{split}\label{coeff-P}
\eeq
(and~$\epsilon$ denotes the step function~$\epsilon(x)=1$ if~$x>1$ and~$\epsilon(x)=-1$ otherwise).
These functions have the expansion
\beq \label{coeffex1}
\alpha(\xi) = -\frac{i}{4 \pi^3\, \xi^4} + \O \Big(\frac{1}{\xi^2} \Big) \qquad \text{and} \qquad
\beta(\xi) = -\frac{m}{8 \pi^3\, \xi^2} + \O \big( \log(\xi^2) \big) \:.
\eeq

\subsection{The Geometry without Regularization} \label{secvac42}
We now enter the analysis of the geometric objects introduced in Section~\ref{secconstruct}
for given space-time points~$x, y \in M$. We restrict attention to the case~$\xi^2 \neq 0$
when the space-time points are not lightlike separated. This has the advantage that,
in view of the convergence~\eqref{Pconverge}, we can first consider the 
unregularized kernel~$P(x,y)$ in the form~\eqref{P-unreg}.
In Section~\ref{secreg} we can then use a continuity argument to extend the results to
small~$\varepsilon>0$. 

We first point out that, although we are working without regularization,
the fact that we took the limit~$\varepsilon \searrow 0$ of regularized objects
is still apparent because the Euclidean sign operator~\eqref{Euklid}
distinguishes a specific sign operator. This fact will enter the construction,
but of course the resulting spin connection will be Lorentz invariant.
Taking the adjoint of~\eqref{P-unreg},
\[ P(y,x)=P(x,y)^*=\overline{\alpha(\xi)}\,\xis+\overline{\beta(\xi)}\,\1, \]
we obtain for the closed chain
\beq A_{xy}=a(\xi)\, \xis+b(\xi) \,\1=A_{yx}\label{A-unreg}\eeq
with the real-valued functions $a=2\Rea(\alpha\bar{\beta})$ and $b=|\alpha|^2\xi^2+|\beta|^2$.
Subtracting the trace and taking the square, the eigenvalues of~$A_{xy}$ are computed by
\beq \label{evals}
\lambda_+=b+\sqrt{a^2\xi^2}\quad\text{and}\quad \lambda_-=b-\sqrt{a^2\xi^2} \:.
\eeq
It follows that the eigenvalues of $A_{xy}$ are real if $\xi^2>0$, whereas they form a complex conjugate pair
if $\xi^2<0$. This shows that the causal structure of Definition~\ref{def2} agrees with
the usual causal structure in Minkowski space. We next show that in the case
of timelike separation,
the space-time points are even properly timelike separated.

\begin{Lemma}\label{vac-proptime} Let $x,y\in M$ with $\xi^2 \neq 0$. Then $x$ and $y$ are
properly timelike separated (see Definition \ref{defproptl}) if and only if~$\xi^2 > 0$.
The directional sign operator of $A_{xy}$ is given by
\beq
v_{xy}=\epsilon(\xi^0)\frac{\xis}{\sqrt{\xi^2}}\,.\label{dirsignA}
\eeq
\end{Lemma}
\Proof In the case~$\xi^2<0$, the two eigenvalues~$\lambda_\pm$ in~\eqref{evals}
form a complex conjugate pair. If they are distinct, the spectrum is not real.
On the other hand, if they coincide, the corresponding eigenspace is not definite.
Thus~$x$ and~$y$ are not properly timelike separated.

In the case~$\xi^2>0$, we obtain a simple expression for~$a$,
\beq
a = 2\Rea(\alpha\bar{\beta})(m\sqrt{\xi^2})=\epsilon(\xi^0)\frac{2m^7}{(4\pi)^4}\frac{(Y_1J_0-Y_0J_1)(m\sqrt{\xi^2})}{(m\sqrt{\xi^2})^3}
=-\epsilon(\xi^0) \frac{m^3}{64 \pi^5}\: \frac{1}{\xi^4} \:, \label{a-positive}
\eeq
where we used~\cite[formula (9.1.16)]{AS} for the Wronskian of the Bessel functions~$J_1$ and~$Y_1$.
In particular, one sees that~$a \neq 0$, so that according to~\eqref{evals},
the matrix~$A_{xy}$ has two distinct eigenvalues.

\begin{figure}
\includegraphics[width=8cm]{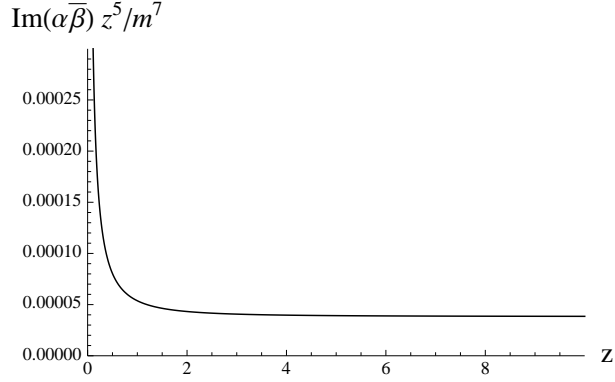}
\caption{The Bessel functions in~\eqref{Imab}.}
 \label{figbessel}
\end{figure}

Next, the calculation
\begin{align*}	\lambda_+\lambda_-&=b^2-a^2\xi^2=|\alpha|^4\, \xi^4+|\beta|^4+2|\alpha|^2\, \xi^2\, |\beta|^2
-4\xi^2 \Rea(\alpha\bar{\beta})^2 \\
&\overset{(\ast)}{\geq}|\alpha^4\,|\xi^4+|\beta|^4-2|\alpha|^2 \,\xi^2|\,\beta|^2
= \big( |\alpha^2|\xi^2-|\beta|^2 \big)^2 \geq 0
\end{align*}
shows that the spectrum of $A_{xy}$ is non-negative. In order to obtain a strict inequality,
it suffices to show that~$\Ima(\alpha\bar{\beta}) \neq 0$ (because then the inequality
in~$(\ast)$ becomes strict). After the transformation
\begin{align}	\Ima(\alpha\bar{\beta})&=-\frac{m^7}{(4\pi)^4}\left(\frac{(Y_0Y_1+J_0J_1)(z)}{z^3}
-2\: \frac{(J_1^2+Y_1^2)(z)}{z^4}\right) \nonumber \\
&=-\frac{m^7}{(4\pi)^4}\frac{1}{2z}\frac{d}{dz} \left(\frac{Y_1(z)^2+J_1(z)^2}{z^2}\right) \:,
\label{Imab}
\end{align}
where we set~$z=m\sqrt{\xi^2}>0$, asymptotic expansions of the Bessel functions
yield that the function~$\Ima(\alpha\bar{\beta})$ is positive for~$z$ near zero and near infinity.
The plot in Figure~\ref{figbessel} shows that this function is also positive in the intermediate range.

We now prove that the eigenspaces of $A_{xy}$ are definite with respect to the inner product $\Sl.|.\Sr$ on $V$. First, from~\eqref{A-unreg} it is obvious that the eigenvectors of~$A_{xy}$ coincide with those
of the operator~$\xis$.
Thus let $v\in V$ be an eigenvector of $\xis$, i.e. $\xis v=\pm\sqrt{\xi^2}\, v$. We choose a proper orthochronous Lorentz-transformation~$\Lambda$ which transforms~$\xi$ to the
vector~$\Lambda(\xi)=(t,\vec{0})$ with~$t \neq 0$.
In view of the Lorentz invariance of the Dirac equation,
there is a unitary transformation $U\in\U(V)$ with $U\gamma^lU^{-1}=\Lambda^l_j\gamma^j$. Then
the calculation
\begin{align}
\pm \sqrt{\xi^2}\: \Sl v\,|\,v\Sr&=\Sl v\,|\,\xis v\Sr=\Sl v\,|\,\eta_{ij}\xi^i\gamma^j\, v\Sr
=\eta_{kl}\,\Sl v\,|\,(\Lambda^k_i\xi^i)(\Lambda^l_j\gamma^j) \,v\Sr \nonumber \\
&=\eta_{kl}\,\Sl v\,|\,\Lambda(\xi)^k U\gamma^lU^{-1}v\Sr
=t\, \Sl v\,|\,U\gamma^0U^{-1}v\Sr \nonumber \\
&= t\, \Sl U^{-1}v\,|\,\gamma^0U^{-1}v\Sr =t\, \la U^{-1}v\,|\,U^{-1}v\ra_{\C^4}\neq0 \label{calcdefinite}
\end{align}
shows that $\Sl v\,|\,v\Sr\neq0$, and thus $v$ is a definite vector. We conclude that $x$ and $y$ are properly timelike separated.

We next show that the directional sign operator of $A_{xy}$ is given by \eqref{dirsignA}.
The calculation \eqref{calcdefinite} shows that the inner product $\Sl.|v_{xy}.\Sr$ 
with~$v_{xy}$ according to~\eqref{dirsignA} is positive definite.
Furthermore, the square of~$v_{xy}$ is given by
$$v_{xy}^2=\left(\epsilon(\xi^0)\frac{\xis}{\sqrt{\xi^2}}\right)^2=\1\,, $$
showing that~$v_{xy}$ is indeed a sign operator. Since~$v_{xy}$ obviously commutes with~$A_{xy}$,
it is the directional sign operator of $A_{xy}$.
\QED

\begin{figure}%
\includegraphics[width=6.8cm]{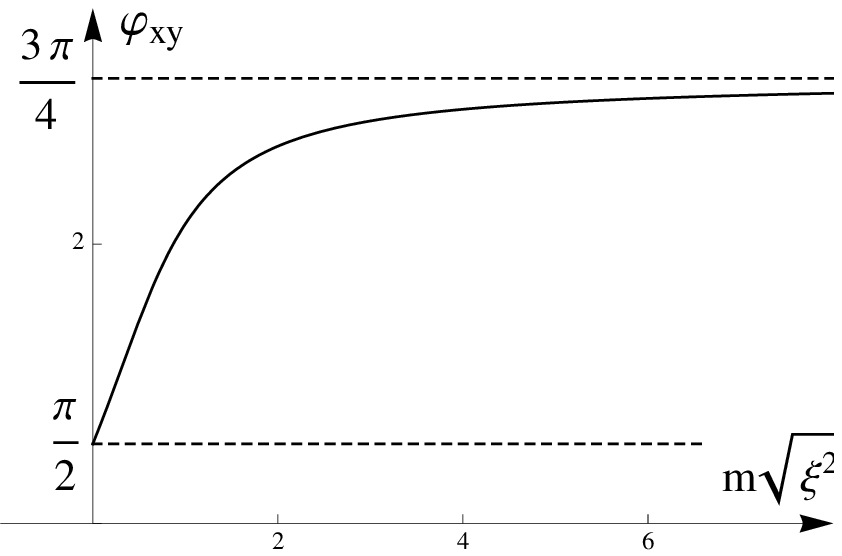} \hspace*{0.5cm}%
\includegraphics[width=6.8cm]{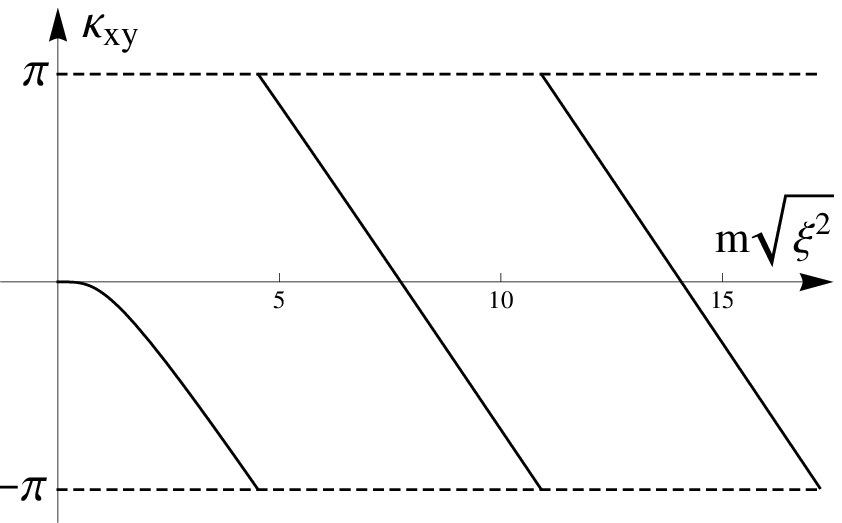}%
\caption{The phases in the spin connection in the case~$\xi^0>0$.}%
 \label{figphases}%
\end{figure}

Let us go through the construction of the spin connection in Section~\ref{sec33}.
Computing the commutator of the Euclidean sign operator~$s_x$ (see~\eqref{Euklid})
and the directional sign operator $v_{xy}$ (see~\eqref{dirsignA}),
$$[v_{xy},s_x]=\left[\epsilon(\xi^0)\frac{\xis}{\sqrt{\xi^2}}\,,\,\gamma^0\right]=2\epsilon(\xi^0)\frac{\vec{\xi}\cdot\vec{\gamma}\gamma^0}{\sqrt{\xi^2}} \:, $$
one sees that these operators are generically separated (see Definition \ref{defgensep}),
provided that we are not in the exceptional case~$\vec{\xi}\neq0$ (for which the spin connection
could be defined later by continuous continuation).
Since these two sign operators lie in the Clifford subspace~$K$ spanned by~$(\gamma^0, \ldots, \gamma^3,
i \gamma^5)$ (again in the usual Dirac representation), it follows that all the Clifford
subspaces used in the construction of the spin connection are equal to~$K$, i.e.\
\[ K_{xy} = K_{yx} = K_x^{(y)} = K_y^{(x)} = K \:. \]
All synchronization and identification maps are trivial
(see Definition~\ref{defsync} and~\eqref{phidef}).
In particular, the system is parity preserving (see Definition~\ref{defparity})
and Clifford-parallel (see Definition~\ref{def-cliffpara}).
Choosing again the basis~$(e_0 = v_{xy}, e_1, \ldots, e_4)$ of~$K$ and the spinor basis of
Corollary~\ref{corollary1}, one sees from~\eqref{P-unreg} and~\eqref{dirsignA} that~$P(x,y)$
is diagonal,
\[ P(x,y)=\begin{pmatrix} \Big(\beta+\alpha\sqrt{\xi^2}\Big)\1 & 0 \\ 0 & \Big(\beta-\alpha\sqrt{\xi^2}\Big)\1 \end{pmatrix} . \]
Thus in the polar decomposition~\eqref{polar} we get
\[ R^\pm_{xy} = \left| \beta \pm \alpha\sqrt{\xi^2} \right| \:,\qquad
\vartheta_{xy}^\pm = \arg \Big( \beta \pm \alpha\sqrt{\xi^2} \Big) \!\!\!\!\mod \pi \:,\qquad
V^\pm_{x,y} \in \{\1, -\1\}\:. \]
Computing~$\varphi_{xy}$ according to~\eqref{phival} and our convention~\eqref{phasefix2},
in the case~$\xi^0>0$ we obtain the left plot of Figure~\ref{figphases},
whereas in the case~$\xi^0<0$ one gets the same with the opposite sign.
We conclude that~$\varphi_{xy}$ is never a multiple of~$\frac{\pi}{4}$, meaning that~$x$ and~$y$
are time-directed (see Definition~\ref{deftimedir2}). Moreover, the time direction of
Definition~\ref{deftime} indeed agrees with the time orientation of Minkowski space.
Having uniquely fixed~$\varphi_{xy}$, the spin connection is given by~\eqref{Dxyansatz}
or by~\eqref{Dsecond}. A short calculation yields that~$D_{xy}$ is trivial up to a phase factor,
\beq \label{Dkappa}
D_{x,y} = e^{i \kappa_{xy}}\: \1 \:,
\eeq
where the phase~$\kappa_{xy}$ is given by
\beq \label{kxydef}
\kappa_{xy} = \arg \left( e^{i \varphi_{xy}} \: \Big(\beta+\alpha\sqrt{\xi^2}\Big) \right) 
= \arg \left( e^{-i \varphi_{xy}} \: \Big(\beta-\alpha\sqrt{\xi^2}\Big) \right)  \:.
\eeq
The function~$\kappa_{xy}$ is shown in the right plot of Figure~\ref{figphases}.
One sees that the phase factor in~$D_{x,y}$ oscillates on the length scale~$m^{-1}$.
We postpone the discussion of this phase to Section~\ref{sec44}.

Let us consider the corresponding metric connection of Definition~\ref{defmetric}.
We clearly identify the tangent space~$T_x$ with the vector space~$K$.
As the synchronization maps are trivial and
the phases in~$D_{x,y}$ drop out of~\eqref{metric2}, it is obvious
that~$\nabla_{x,y}$ reduces to the trivial connection in Minkowski space.
Finally, choosing~$u(x) = i \gamma^5$, the causal fermion system is obviously
chirally symmetric (see Definition~\ref{defreduce}).

Our results are summarized as follows.
\begin{Prp} \label{prpunreg} Let~$x, y \in M$ with~$\xi^2 \neq 0$ and~$\vec{\xi} \neq 0$.
Consider the spin connection corresponding to the Euclidean signature operator~\eqref{Euklid}
and the unregularized Dirac sea vacuum~\eqref{P-unreg}.
Then~$x$ and~$y$ are spin-connectable if and only if~$\xi^2>0$.
The spin connection~$D_{x,y}$ is trivial up to a phase
factor~\eqref{Dkappa}.
The time direction of Definition~\ref{deftime} agrees with the usual time orientation of
Minkowksi space. The corresponding metric connection~$\nabla_{x,y}$ is trivial.

Restricting attention to pairs~$(x,y) \in M \times M$ with~$\xi^2 \neq 0$ and~$\vec{\xi} \neq 0$,
the resulting causal fermion system is parity preserving, chirally symmetric and Clifford-parallel.
\end{Prp}

\subsection{The Geometry with Regularization} \label{secreg}
We now use a perturbation argument to extend some of the results of Proposition~\ref{prpunreg}
to the case with regularization.

\begin{Prp} \label{prpreg} Consider the causal fermion systems of Proposition~\ref{prp41}.
For any $x, y\in M$ with~$\xi^2 > 0$ and~$\vec{\xi} \neq 0$,
there is~$\varepsilon_0>0$ such that for all~$\varepsilon \in (0, \varepsilon_0)$ the following statements
hold: The points~$x$ and~$y$ are spin-connectable.
The time direction of Definition~\ref{deftime} agrees with the usual time orientation of
Minkowksi space. In the limit~$\varepsilon \searrow 0$,
the corresponding connections~$D^\varepsilon_{x,y}$ and~$\nabla^\varepsilon_{x,y}$ converge
to the connections of Proposition~\ref{prpunreg},
\beq \label{Dlimit}
\lim_{\varepsilon \searrow 0} D_{x,y}^\varepsilon = D_{x,y}\:, \qquad
\lim_{\varepsilon \searrow 0} \nabla_{x,y}^\varepsilon = \1 \:.
\eeq
\end{Prp}
\Proof Let $x,y\in M$ with $\xi^2>0$ and~$\vec{\xi} \neq 0$.
Using the pointwise convergence~\eqref{Pconverge}, 
a simple continuity argument shows that for sufficiently small~$\varepsilon$, the
spectrum of $A_{xy}^\varepsilon$ is strictly positive and the eigenspaces are definite.
Thus~$x$ and~$y$ are properly timelike separated. From \eqref{polar} and \eqref{phival} we conclude that in a small interval $(0,\varepsilon_0)$, the phase $\varphi^\varepsilon_{xy}$ depends continuously on $\varepsilon$  and lies in the same subinterval~\eqref{phasefix2}
as the phase~$\varphi_{xy}$ without regularization.
We conclude that for all~$\varepsilon \in (0, \varepsilon_0)$, the points~$x$ and~$y$ are spin-connectable and have the same time orientation as without regularization.
The continuity of the connections is obvious from~\eqref{Dsecond}.
\QED

We point out that this proposition makes no statement on whether the causal fermion systems
are parity preserving, chirally symmetric or Clifford-parallel.
The difficulty is that these definitions are either not stable under perturbations, or 
else they would make it necessary to choose~$\varepsilon$ independent of~$x$ and~$y$.
To be more specific, the closed chain with regularization takes the form
\[ A^\varepsilon_{xy} = a_\varepsilon\xis+b_\varepsilon\1+c_\varepsilon\gamma^0-id_\varepsilon\vec{\xi}\cdot\vec{\gamma}\gamma^0 \:, \]
where the coefficients involve the regularized Bessel functions in~\eqref{abeps},
\begin{align*}
  a_\varepsilon&=2\Rea(\alpha_\varepsilon\overline{\beta_\varepsilon})\:, &
	b_\varepsilon&=|\alpha_\varepsilon|^2(\xi^2+\varepsilon^2)+|\beta_\varepsilon|^2 \:, \\
	c_\varepsilon&=2\varepsilon\Ima(\alpha_\varepsilon\overline{\beta_\varepsilon})\:, &
	d_\varepsilon&=2\varepsilon|\alpha_\varepsilon|^2\,.
\end{align*}
A short calculation shows that for properly timelike separated points~$x$ and~$y$,
the directional sign operator is given by
\begin{align*}	v_{xy}^\varepsilon=\frac{a_\varepsilon\xis+c_\varepsilon\gamma^0-id_\varepsilon\vec{\xi}\cdot\vec{\gamma}\gamma^0}{\sqrt{a_\varepsilon^2\xi^2+2a_\varepsilon c_\varepsilon\xi^0+c_\varepsilon^2-d_\varepsilon^2\,|\vec{\xi}|^2}} \:,
\end{align*}
and the argument of the square root is positive.
A direct computation shows that the signature operators~$s_x$ and~$v^\varepsilon_{xy}$ 
span a Clifford subspace of signature~$(1,1)$.
According to Lemma~\ref{lemmaextend}, this Clifford subspace has a unique extension~$K$,
implying that~$K_{xy} = K_x^{(y)} = K$. This shows that the synchronization maps are all trivial.
However, as~$v_{xy}^\varepsilon$ involves a bilinear component which depends on~$\vec{\xi}$,
the Clifford subspaces~$K_{xy}$ and~$K_{xz}$ will in general be different, so that the
identification maps~\eqref{phidef} are in general non-trivial. Due to this complication, the system
is no longer Clifford-parallel, and it is not obvious whether the system is 
parity preserving or chirally symmetric.

\subsection{Parallel Transport Along Timelike Curves} \label{sec44}
The phase factor in~\eqref{Dkappa} resembles the $\U(1)$-phase in electrodynamics.
This phase is unphysical as no electromagnetic field is present.
In order to understand this seeming problem, one should note that in differential geometry,
the parallel transport is always performed along a continuous curve,
whereas the spin connection~$D_{x,y}$ directly connects distant points.
The correct interpretation is that the spin connection only gives the physically correct
result if the points~$x$ and~$y$ are sufficiently close to each other.
Thus in order to connect distant points~$x$ and~$y$, one should choose intermediate points
$x_1, \ldots x_N$ and compose the spin connections along neighboring points.
In this way, the unphysical phase indeed disappears, as the following construction shows.

Assume that~$\gamma(t)$
is a future-directed timelike curve, for simplicity parametrized by arc length, which is defined
on the interval~$[0, T]$ with~$\gamma(0) = y$ and~$\gamma(T) = x$.
The Levi-Civita parallel transport of spinors along~$\gamma$ is trivial.
In order to compare with the spin connection~$D^\varepsilon$, we subdivide~$\gamma$
(for simplicity with equal spacing, although a non-uniform spacing would work just
as well). Thus for any given~$N$, we define the points~$x_0, \ldots, x_N$ by
\beq \label{xNdef}
x_n = \gamma(t_n) \qquad \text{with} \qquad t_n = \frac{n T}{N}\:.
\eeq
\begin{Def} \label{defadmissible}
The curve~$\gamma$ is called {\bf{admissible}} if for all sufficiently large~$N \in \N$
there is a parameter~$\varepsilon_0>0$
such that for all~$\varepsilon \in (0, \varepsilon_0)$ and all~$n=1,\ldots, N$,
the points~$x_{n}$ and~$x_{n-1}$ are spin-connectable.
\end{Def} \noindent
If~$\gamma$ is admissible, we define the parallel transport~$D_{x,y}^{N,\varepsilon}$ by
successively composing the parallel transports between neighboring points,
\[ D^{N,\varepsilon}_{x,y} := D^\varepsilon_{x_N ,x_{N-1}} D^\varepsilon_{x_{N-1} ,x_{N-2}} \cdots D^\varepsilon_{x_1 ,x_0} \::\: V\rightarrow V\:. \]
Then the following theorem holds.
\begin{Thm} \label{thmminkowski}
Considering the family of
causal fermion systems of Proposition~\ref{prp41},
the admissible curves are generic in the sense that they are dense in the $C^\infty$-topology
(meaning that for any smooth~$\gamma$ and every~$K \in \N$,
there is a series~$\gamma_\ell$ of admissible curves such that~$D^k \gamma_\ell \rightarrow D^k \gamma$
uniformly for all~$k =0, \ldots, K$). 
Choosing~$N \in \N$ and~$\varepsilon>0$ such that
the points~$x_{n}$ and~$x_{n-1}$ are spin-connectable for all~$n=1,\ldots, N$, every point
$x_n$ lies in the future of $x_{n-1}$.
Moreover,
$$\lim_{N \rightarrow \infty} \:\lim_{\varepsilon \searrow 0} D^{N,\varepsilon}_{x,y}= \DLC_{x,y} \:, $$
where we use the identification~\eqref{Jdef}, and~$\DLC_{x,y} : S_yM \rightarrow S_xM$ denotes the trivial
parallel transport along~$\gamma$.
\end{Thm}
\Proof For any given~$N$, we know from Proposition~\ref{prpreg} that by choosing~$\varepsilon_0$
small enough, we can arrange that for all~$\epsilon\in\big(0,\epsilon_0)$ and all~$n=1, \ldots, N$,
the points~$x_{n}$ and~$x_{n-1}$ are spin-connectable and $x_n$ lies in the future of $x_{n-1}$,
provided that the vectors~$\vec{x}_n - \vec{x}_{n-1}$ do not vanish (which is obviously satisfied
for a generic curve). Using~\eqref{Dlimit} and~\eqref{Dkappa}, we obtain
\beq \label{Dprod}
\lim_{\varepsilon \searrow 0} D^{N,\varepsilon}_{x,y}=D_{x_N ,x_{N-1}} D_{x_{N-1} ,x_{N-2}} \cdots D_{x_1 ,x_0}=\exp \bigg( i\sum_{n=1}^N \kappa_{x_{n},x_{n-1}} \bigg) \1 \:.
\eeq

Combining the two equations in~\eqref{kxydef}, one finds
\[ \kappa_{xy} = \frac{1}{2}\: \arg \left( \beta^2 - \alpha^2\:\xi^2 \right) \mod \pi \:. \]
Expanding the Bessel functions in~\eqref{coeff-P} gives
\[ \beta^2 - \alpha^2\:\xi^2 = \frac{1}{16 \pi^6}\: \frac{1}{\xi^6} + \O \Big(\frac{1}{\xi^4} \Big)\:, \]
As~$\kappa_{xy}$ is smooth and vanishes in the limit~$y \rightarrow x$, we conclude that
\[ \kappa_{xy} = \O(\xi^2)\:. \]
Using this estimate in~\eqref{Dprod}, we obtain
\[ \lim_{\varepsilon \searrow 0} D^{N,\varepsilon}_{x,y}= \exp \left( i\: N\: \O(N^{-2}) \right) \1\:. \]
Taking the limit~$N \rightarrow \infty$ gives the result.
\QED

\section{Example: The Fermionic Operator in a Globally Hyperbolic Space-Time} \label{secglobhyp}
In this section we shall explore the connection between the notions of the quantum geometry introduced
in Section~\ref{secconstruct} and the common objects of Lorentzian spin geometry.
To this end, we consider Dirac spinors on a globally hyperbolic Lorentzian manifold~$(M,g)$
(for basic definitions see~\cite{baer+ginoux, baum}).
For technical simplicity, we make the following assumptions:
\begin{itemize}
\item[(A)] The manifold~$(M,g)$ is flat Minkowski space in the past
of a Cauchy hypersurface $\nn$.
\item[(B)] The causal fermion systems are introduced as the Cauchy development
of the fermion systems in Minkowski space as considered in Section~\ref{sec41}.
\end{itemize}
These causal fermion systems are constructed in Section~\ref{secPepsgrav}.
We proceed by analyzing the fermionic operator in the limit without regularization
using its Hadamard expansion (see Section~\ref{sechadamard} and Section~\ref{secPcurve}).
We then consider the spin connection along a timelike curve~$\gamma$ (see Section~\ref{secversus}).
We need to assume that in a neighborhood~${\mathscr{U}}$ of the curve~$\gamma$, the
Riemann curvature tensor~$R$ is bounded
pointwise in the sense that
\beq \label{curvcond}
\frac{\| R(x) \|}{m^2} + \frac{\|\nabla R(x)\|}{m^3} + \frac{\|\nabla^2 R(x)\|}{m^4} < c
\qquad \text{for all~$x \in {\mathscr{U}}$}\:,
\eeq
where~$\| . \|$ is a norm (for example induced by the scalar product~$\Sl .| \dot{\gamma}(t)\, . \Sr$
on~$S_{\gamma(t)} M$), and~$c<1$ is a numerical constant (which could be computed explicitly).
This conditions means that curvature should be small on the Compton scale.
It is a physically necessary assumption because otherwise the gravitational field
would be so strong that pair creation would occur, making it impossible to
speak of ``classical gravity''.
Our main result is that the spliced spin connection goes over to the Levi-Civita connection
on the spinor bundle, up to errors of the order~$\|\nabla R\|/m^3$  (see Theorem~\ref{thmgrav}).
We conclude with a brief outlook (see Section~\ref{secoutlook}).

\subsection{The Regularized Fermionic Operator} \label{secPepsgrav}
Let $(M,g)$ be a globally hyperbolic Lorentzian manifold which coincides with Minkowski space in the past of a Cauchy hypersurface $\nn$. Choosing a global time function~$\mathfrak{t}$ (see~\cite{bernal+sanchez}),
$M$ has a smooth splitting~$M \cong \R \times \nn$~with~$\nn = \mathfrak{t}^{-1}(\{0\})$.
For consistency with Section~\ref{secvac}, we use the conventions that the signature
of~$g$ is $(+\,-\,-\,-)$, and that Clifford multiplication satisfies the relation $X\cliff Y+Y\cliff X=2g(X,Y)$.
We denote the volume measure by~$d\mu(x) = \sqrt{|\det g|}\, d^4x$.
As a consequence of the smooth splitting, the manifold~$M$ is spin.
The spinor bundle~$SM$ is endowed with a Hermitian inner product of signature~$(2,2)$, which we
denote by~$\Sl .|. \Sr$. We let $\dd$ be the Dirac operator on $M$, acting on sections~$\psi \in \Gamma(M, SM)$ of the spinor bundle. For a given mass $m>0$, we consider the Dirac equation on $M$,
\beq\label{deq-globhyp}
(\dd-m)\,\psi =0 \:.
\eeq

The simplest method for constructing causal fermion systems is to replace the plane-wave
solutions used in Minkowski space (see Section~\ref{sec41}) by corresponding solutions
of~\eqref{deq-globhyp} obtained by solving a Cauchy problem. More precisely,
in the past of~$\nn$ where our space-time is isometric to Minkowski space, we again introduce the
plane-wave solution~$\psi_{\vec{k} a -}$ (see~\eqref{planewave}).
Using that the Cauchy problem has a unique solution (see~\cite{baer+ginoux} and
the integral representation~\eqref{duhamel} below),
we can extend them to smooth solutions~$\tilde{\psi}_{\vec{k} a -}$ on~$M$ by
\beq \label{psitilde}
(\dd - m)\, \tilde{\psi}_{\vec{k} a -} = 0 \:,\qquad \tilde{\psi}_{\vec{k} a -}|_{\nn} = \psi_{\vec{k} a -}\:.
\eeq
In obvious generalization of~\eqref{negen} and~\eqref{pip},
we can form superpositions of these solutions,
on which we introduce the scalar product
\beq \label{pip2}
\la \tilde{\psi} | \tilde{\phi} \ra_\H = 2 \pi \int_{{\mathfrak{t}}=\text{const}} \Sl
\tilde{\psi}({\mathfrak{t}}, x) \,|\, \nu \cliff \tilde{\phi}({\mathfrak{t}}, x) \Sr
d\mu_{\nn({\mathfrak{t}})}(x)\:,
\eeq
where~$\nu$ is the future-directed unit normal on~$\nn$
(note that this scalar product is independent of~${\mathfrak{t}}$ due to current conservation).
We again denote the corresponding Hilbert space by~$\H$.

In order to introduce a corresponding causal fermion system, we 
introduce the operators~$\ie_x$ and~$F^\varepsilon(x)$
by adapting~\eqref{iotaepsdef} and~\eqref{Fepsdef},
\begin{align*}
\ie_x \::\: S_xM &\rightarrow \H \,,\;
u \mapsto -\frac{m}{\pi} \sum_{a=1,2}
\int \frac{d^3k}{2 \omega}\: e^{-\frac{\varepsilon \omega}{2}} \:
\tilde{\psi}_{\vec{k}a -} \Sl \tilde{\psi}_{\vec{k}a -}(x) | u \Sr \\
F^\varepsilon(x) &= -\ie_x\, (\ie_x)^* \::\:
\H \rightarrow \H\:.
\end{align*}
Let us verify that~$\ie_x$ is injective: For a given non-zero spinor~$\chi \in S_xM$,
we choose a wave function~$\psi \in \H$ which is well-approximated by
a WKB wave packet of large negative energy (by decreasing the energy,
we can make the error of the approximations arbitrarily small).
Consider the operator
\[ L \::\: {\mathcal{D}}(L) \subset \H \rightarrow \H \:,\quad
(L \psi)(x) = 
-\frac{m}{\pi} \sum_{a=1,2}
\int \frac{d^3k}{2 \omega}\: e^{-\frac{\varepsilon \omega}{2}} \:
\tilde{\psi}_{\vec{k}a -}(x) \: \la \tilde{\psi}_{\vec{k}a -} | \psi \ra_\H \:, \]
where~${\mathcal{D}}(L)$ is a suitable dense domain of definition (for example
the smooth Dirac solutions with spatially compact support).
As the image of~$L$ is obviously dense in~$\H$, there is a vector~$\phi \in {\mathcal{D}}$
such that~$L \phi$ approximates~$\psi$ (again, we can make the error of this
approximation arbitrarily small). Then~$\la \phi | \ie_x \chi \ra \approx \Sl \psi(x) | \chi \Sr_x$.
By modifying the polarization and direction of the wave packet~$\psi$, we can
arrange that~$\Sl \psi(x) | \chi \Sr_x \neq 0$.

According to~\eqref{Sxdef}, the spin space~$S_x^\varepsilon$ is defined as the image of~$F_x^\varepsilon$.
We now choose a convenient basis
of~$S_x^\varepsilon$ which will at the same time give a canonical identification
of~$S_x^\varepsilon$ with the differential geometric spinor space~$S_xM$.
We first choose an eigenvector basis~$(\f^\varepsilon_\alpha(x))_{\alpha=1,\ldots,4}$
of~$S_x^\varepsilon=F^\varepsilon(x)(\H)$ with corresponding
eigenvalues
\beq \label{nuev}
\nu^\varepsilon_1(x), \nu^\varepsilon_2(x) < 0 \qquad \text{and} \qquad
\nu^\varepsilon_3(x), \nu^\varepsilon_4(x) > 0 \:.
\eeq
We normalize the eigenvectors according to
\[ \la \f^\varepsilon_\alpha(x) \,|\, \f^\varepsilon_\beta(x) \ra_\H = \frac{1}{|\nu^\varepsilon_\alpha(x)|}\:
\delta_{\alpha \beta}\:. \]
Then, according to~\eqref{ssp}, the~$(\f^\varepsilon_\alpha(x))$ are a pseudo-orthonormal
basis of~$(S_x^\varepsilon, \Sl .|. \Sr_x)$.
Next, we introduce the vectors
\[ \e^\varepsilon_\alpha(x) = (\ie_x)^* \:\f^\varepsilon_\alpha(x) \in S_xM\:. \]
A short calculation shows that these vectors form a pseudo-orthonormal
eigenvector basis of the operator
\[ (\ie_x)^* \ie_x \::\: S_x M \rightarrow S_x M \:, \]
corresponding to the eigenvalues~$\nu^\varepsilon_\alpha(x)$. In analogy to~\eqref{Jdef},
we always identify the spaces~$S_xM$ and~$S^\varepsilon_x$
via the mapping~${\mathfrak{J}}^\varepsilon_x$ defined by
\beq \label{Jepshyp}
{\mathfrak{J}}^\varepsilon_x \::\: S_xM \rightarrow S^\varepsilon_x \,,\;
\e^\varepsilon_\alpha(x) \mapsto \f^\varepsilon_\alpha(x) \:.
\eeq

Again identifying~$x$ with~$F^\varepsilon(x)$, the kernel of the
fermionic operator~\eqref{Pxydef} takes the form~\eqref{PepsF}.
Exactly as in the proof of Lemma~\ref{lemmapeps}, we find that
\beq \label{Pepsgrav}
P^\varepsilon(x,y) = -(\iota^\varepsilon_x)^* \iota^\varepsilon_y =
-\frac{m}{\pi} \sum_{a=1,2} \int \frac{d^3k}{2 \omega}\: e^{-\varepsilon \omega} 
\:|\tilde{\psi}_{\vec{k}a -}(x) \Sr \Sl \tilde{\psi}_{\vec{k} a -}(y) | \:.
\eeq
From this formula we can read off the following characterization of~$P^\varepsilon$.

\begin{Prp} \label{prp51}
The kernel of the fermionic operator~$P^\varepsilon(x,y)$ is the unique smooth
bi-solution of~\eqref{deq-globhyp}, i.e.\ in a distributional formulation
$$P^\varepsilon\big((\dd-m)\psi,\phi\big)=0=P^\varepsilon\big(\psi,(\dd-m)\phi\big)\quad\text{for all }\psi,\phi\in\Gamma_0(M, S M)\,,$$
with the following properties:
\begin{itemize}
\item[(i)] $P^\varepsilon$ coincides with the regularized Dirac sea vacuum~\eqref{Pepsdef} if~$\psi$ and~$\phi$ are both supported in the past of $\nn$.
\item[(ii)] $P^\varepsilon$ is symmetric in the sense that $P^\varepsilon(\psi,\phi)=P^\varepsilon(\phi,\psi)$.
\end{itemize}
\end{Prp}

In order to keep the analysis simple, our strategy is to take the limit~$\varepsilon \searrow 0$ at
an early stage. In the remainder of this section, we analyze this limit for~$P^\varepsilon$ and for
the Euclidean sign operator.
In preparation, we recall the relation between the Dirac Green's functions and the solution
of the Cauchy problem, adapting the methods in~\cite{baer+ginoux} to the first order Dirac system.
On a globally hyperbolic Lorentzian manifold, one can introduce the
retarded Dirac Green's function, which we denote by~$s^\wedge(x,y)$.
It is defined as a distribution on~$M \times M$, meaning that we can evaluate it with compactly supported
test functions~$\phi, \psi \in \Gamma_0(M, SM)$,
\[ s^\wedge(\phi, \psi) = \iint_{M \times M} \Sl \phi(x) | s^\wedge(x,y)\: \psi(y) \Sr\: d\mu(x)\: d\mu(y) \:. \]
We can also regard it as an operator on the test functions. Thus for~$\psi
\in \Gamma_0(M, SM)$, we set
\[ s^\wedge(x, \psi) = \int_M s^\wedge(x,y)\: \psi(y)\: d\mu(y) \in \Gamma(M, SM)\:. \]
The retarded Green's function is uniquely determined as a solution of the inhomogeneous Dirac equation
\beq \label{inhomsolve}
(\dd_x - m) \,s^\wedge(x, \psi) = \psi(x) = s^\wedge \big( x, (\dd - m) \psi \big)
\eeq
subject to the support condition
\[ \supp s^\wedge(x,.) \subset J^\wedge(x) \:, \]
where~$J^\wedge(x)$ denotes the causal past of~$x$.
The advanced Dirac Green's function~$s^\vee(x,y)$ is defined similarly. It can be obtained
from the retarded Green's function by conjugation,
\beq \label{sadjoint}
s^\wedge(x,y)^* = s^\vee(y,x) \:,
\eeq
where the star denotes the adjoint with respect to the Hermitian inner product
on the spinor bundle.

For the construction of the Dirac Green's functions, it is useful to also consider the second-order equation
\beq \label{kg-globhyp}
(\dd^2-m^2)\,\psi=0\:.
\eeq
Using the Lichnerowicz-Weitzenb\"ock identity (see~\cite{baum}), we can rewrite this equation as 
$$\Big(\Box^\nabla+\frac{\s}{4}-m^2\Big) \psi=0\,,$$
where $\Box^\nabla$ denotes the Bochner Laplacian corresponding to the spinorial Levi-Civita connection.
This shows that the operator in~\eqref{kg-globhyp} is normally hyperbolic,
ensuring the existence of the corresponding Green's function~$S^\wedge$ 
as the unique distribution on~$M \times M$ which satisfies the equation
\[ (\dd^2-m^2) \,S^\wedge(x,\phi) = \phi(x) \:. \]
and the support condition
\[ \supp S^\wedge(x,.) \subset J^\wedge(x) \]
(see~\cite[Section~3.4]{baer+ginoux}).
Then the Dirac Green's function can be obtained by the identities
\beq \label{sSrel}
s^\wedge(\psi , \phi) = S^\wedge \big( (\dd+m) \psi , \phi \big) = S^\wedge \big( \psi , (\dd+m) \phi \big) \:.
\eeq

\begin{figure}
\begin{center}
\begin{picture}(0,0)%
\includegraphics{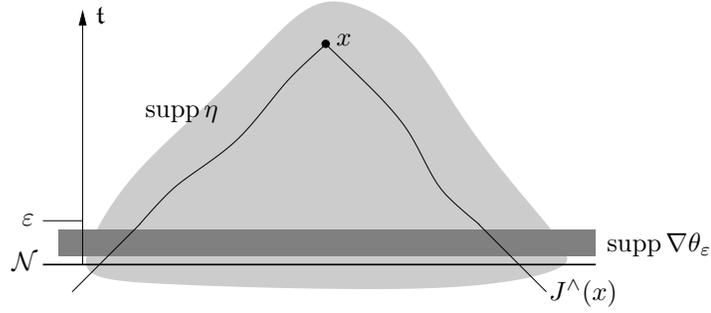}%
\end{picture}%
\setlength{\unitlength}{1367sp}%
\begingroup\makeatletter\ifx\SetFigFontNFSS\undefined%
\gdef\SetFigFontNFSS#1#2#3#4#5{%
  \reset@font\fontsize{#1}{#2pt}%
  \fontfamily{#3}\fontseries{#4}\fontshape{#5}%
  \selectfont}%
\fi\endgroup%
\begin{picture}(10815,5574)(-3449,-3965)
\put(2446,809){\makebox(0,0)[lb]{\smash{{\SetFigFontNFSS{10}{12.0}{\rmdefault}{\mddefault}{\updefault}$x$}}}}
\put(6301,-3811){\makebox(0,0)[lb]{\smash{{\SetFigFontNFSS{10}{12.0}{\rmdefault}{\mddefault}{\updefault}$J^{\wedge}(x)$}}}}
\put(-1919,1199){\makebox(0,0)[lb]{\smash{{\SetFigFontNFSS{10}{12.0}{\rmdefault}{\mddefault}{\updefault}${\mathfrak{t}}$}}}}
\put(-3434,-3271){\makebox(0,0)[lb]{\smash{{\SetFigFontNFSS{10}{12.0}{\rmdefault}{\mddefault}{\updefault}${\mathcal{N}}$}}}}
\put(-3254,-2446){\makebox(0,0)[lb]{\smash{{\SetFigFontNFSS{10}{12.0}{\rmdefault}{\mddefault}{\updefault}$\varepsilon$}}}}
\put(7351,-2926){\makebox(0,0)[lb]{\smash{{\SetFigFontNFSS{10}{12.0}{\rmdefault}{\mddefault}{\updefault}$\text{supp}\, \nabla \theta_\varepsilon$}}}}
\put(-1019,-496){\makebox(0,0)[lb]{\smash{{\SetFigFontNFSS{10}{12.0}{\rmdefault}{\mddefault}{\updefault}$\text{supp}\, \eta$}}}}
\end{picture}%
\end{center}
\caption{The cutoff functions~$\eta$ and~$\theta_\varepsilon$.} \label{figsupport}
\end{figure}

The existence of the retarded Green's function implies that the Cauchy problem
\[ (\dd - m)\, \tilde{\psi} = 0 \:,\qquad \tilde{\psi}|_{\nn} = \psi \in C^\infty(\nn) \]
has a unique smooth solution, as we now recall. To show uniqueness, assume that~$\tilde{\psi}$
is a smooth solution of the Cauchy problem. For given~$x$ in the future of~$\nn$, we choose
a test function~$\eta \in C^\infty_0(M)$ which is identically equal to one in a neighborhood
of the set~$J^\wedge(x) \cap J^\vee(\nn)$.
Moreover, for a given non-negative function~$\theta \in C^\infty(\R)$ with~$\theta|_{(-\infty, 0]} \equiv 0$
and~$\theta|_{[1, \infty)} \equiv 1$ and sufficiently small~$\varepsilon > 0$, we introduce the smooth cutoff
function~$\theta_\varepsilon(y) = \theta( {\mathfrak{t}}(y)/\varepsilon)$. Then the
product~$\phi := \theta_\varepsilon \,\eta\, \tilde{\psi}$ has compact support (see Figure~\ref{figsupport}),
and by~\eqref{inhomsolve} we obtain
\[ \tilde{\psi}(x) = \phi(x) = s^\wedge \big( x, (\dd - m) \phi \big) \\
= s^\wedge \big( x,i \eta\, (d \theta_\varepsilon)\cliff \tilde{\psi} \big) \:. \]
Taking the limit~$\varepsilon \searrow 0$, we obtain the formula
\beq \label{duhamel}
\tilde{\psi}(x) = i\,\int_{\nn} s^\wedge(x,y)\: \nu(y) \cliff \psi(y)\: d\mu_{\nn}(y)\:,
\eeq
where~$\nu$ is the normal of~$\nn$.
This formula is an explicit integral representation of the solution in terms of the Green's function and the
initial data, proving uniqueness. On the other hand, this integral representation can be used to
define~$\tilde{\psi}$, proving existence.

We next express~$P^\varepsilon(x,y)$
in terms of Green's functions and the regularized fermionic operator of Minkowski space.
\begin{Lemma} The regularized fermionic operator has the representation
\beq \label{Pepsrep}
P^\varepsilon(x,y) = \iint_{\nn \times \nn} s^\wedge(x,z_1) \,\nu(z_1) \cliff P^\varepsilon(z_1, z_2)\,
\nu(z_2)\cliff s^\vee(z_2, y)\: d\mu_\nn(z_1)\: d\mu_\nn(z_2) \:,
\eeq
with~$P^\varepsilon(z_1, z_2)$ as given by~\eqref{Pepsdef}.
\end{Lemma}
\Proof
We use~\eqref{duhamel} in~\eqref{Pepsgrav} and apply~\eqref{sadjoint}.
\QED

Setting~$\varepsilon$ to zero, we can use the statement of Proposition~\ref{prp51}
as the definition of a distributional solution of the Dirac equation.
\begin{Def}\label{Pdef-globhyp}
The distribution~$P(x,y)$ is defined as the unique distributional
bi-solution of~\eqref{deq-globhyp},
\beq \label{Peqn}
P\big((\dd-m)\psi,\phi\big)=0=P\big(\psi,(\dd-m)\phi\big)\qquad\text{for all }\psi,\phi\in\Gamma_0(M, S M)\,,
\eeq
with the following properties:
\begin{itemize}
\item[(i)] $P$ coincides with the regularized Dirac sea vacuum~\eqref{Pepsdef} if~$\psi$ and~$\phi$ are both supported in the past of $\nn$.
\item[(ii)] $P$ is symmetric in the sense that $P(\psi,\phi)=P(\phi,\psi)$.
\end{itemize}
\end{Def} \noindent
If the regularization is removed, $P^\varepsilon$ goes over to~$P$ in the following sense.
\begin{Prp} $\quad$
\begin{itemize}
\item[(a)] If~$\varepsilon \searrow 0$, $P^\varepsilon(x,y) \rightarrow P(x,y)$ as
a distribution on~$M \times M$.
\item[(b)] If~$x$ and~$y$ are timelike separated, $P(x,y)$ is a continuous function.
In the limit~$\varepsilon \searrow 0$, the function~$P^\varepsilon(x,y)$ converges to~$P(x,y)$
pointwise, locally uniformly in~$x$ and~$y$.
\end{itemize}
\end{Prp}
\Proof Part~(a) is a consequence of the uniqueness of the time evolution of distributions.
More specifically, suppose that~$\psi$ is a smooth solution of the Dirac equation. We choose
a smooth function~$\eta \in C^\infty(\R)$ with~$\eta|_{[0, \infty)} \equiv 1$ and~$\eta|_{(-\infty, -1]} \equiv 0$.
Then
\[ (\dd-m) \big( \eta(\mathfrak{t}(x))\, \psi(x) \big) = \big( \dd \eta(\mathfrak{t}(x)) \big) \cliff \psi(x) =: \phi(x) \:, \]
and the function~$\phi$ is supported in the past of~$\nn$.
Using~\eqref{inhomsolve} we obtain for any~$x$ in the future of~$\nn$ that
\beq \label{psirep}
\psi(x) = \eta(\mathfrak{t}(x))\, \psi(x) = s^\wedge(x, \phi) = (s^\wedge \ast \phi)(x) \:.
\eeq
Regarding the star as a convolution of distributions, this relations even holds if~$\psi$ is a distributional
solution of the Dirac equation. Suppose that in the past of~$\nn$, the distribution~$\psi$
converges to zero (meaning that~$\psi(\varphi) \rightarrow 0$ for every test function~$\varphi$
supported in the past of~$\nn$). Then, as the function~$\phi$ is supported in the past of~$\nn$,
it converges to zero as a distribution in the whole space-time.
The relation~\eqref{psirep} shows that~$\psi$ also converges to zero in the whole space-time.
In order to prove~(a), we first choose~$z$ in the past of~$\nn$ and apply the above argument to
the distribution~$\psi_+(x) = (P^\varepsilon-P)(x,z)$. Then according to the explicit formulas in Minkowski
space (see~\eqref{Pepsdef} and~\eqref{Pepssing}),
$\psi_+$ converges to zero in the past of~$\nn$, and thus in the whole space-time.
By symmetry, it follows that for any fixed~$x$, the distribution~$\psi_-(z) := (P^\varepsilon-P)(z,x)$
converges to zero in the past of~$\nn$.
As~$\psi_-$ is again a distributional solution of the Dirac equation, we conclude that~$\psi_-$
converges to zero in the whole space-time.

\begin{figure}
\begin{center}
\begin{picture}(0,0)%
\includegraphics{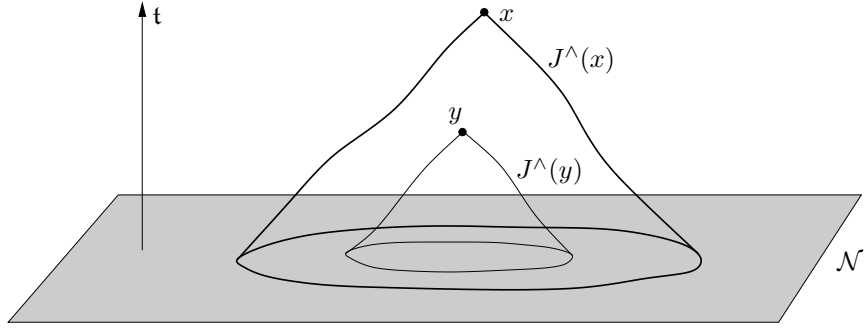}%
\end{picture}%
\setlength{\unitlength}{1367sp}%
\begingroup\makeatletter\ifx\SetFigFontNFSS\undefined%
\gdef\SetFigFontNFSS#1#2#3#4#5{%
  \reset@font\fontsize{#1}{#2pt}%
  \fontfamily{#3}\fontseries{#4}\fontshape{#5}%
  \selectfont}%
\fi\endgroup%
\begin{picture}(15509,5864)(-6405,-4808)
\put(3391,-151){\makebox(0,0)[lb]{\smash{{\SetFigFontNFSS{10}{12.0}{\rmdefault}{\mddefault}{\updefault}$J^{\wedge}(x)$}}}}
\put(2821,-2176){\makebox(0,0)[lb]{\smash{{\SetFigFontNFSS{10}{12.0}{\rmdefault}{\mddefault}{\updefault}$J^{\wedge}(y)$}}}}
\put(8656,-3841){\makebox(0,0)[lb]{\smash{{\SetFigFontNFSS{10}{12.0}{\rmdefault}{\mddefault}{\updefault}${\mathcal{N}}$}}}}
\put(2521,674){\makebox(0,0)[lb]{\smash{{\SetFigFontNFSS{10}{12.0}{\rmdefault}{\mddefault}{\updefault}$x$}}}}
\put(1606,-1096){\makebox(0,0)[lb]{\smash{{\SetFigFontNFSS{10}{12.0}{\rmdefault}{\mddefault}{\updefault}$y$}}}}
\put(-3764,629){\makebox(0,0)[lb]{\smash{{\SetFigFontNFSS{10}{12.0}{\rmdefault}{\mddefault}{\updefault}${\mathfrak{t}}$}}}}
\end{picture}%
\end{center}
\caption{The singular supports of~$s^\wedge(x,.)$ and~$s^\vee(.,y)$} \label{figsupport2}
\end{figure}

In order to prove~(b), we first note that the singular support of the causal Green's functions~$s^\wedge(x,.)$
and~$s^\vee(.,x)$ lies on the light cone~$\partial J^\wedge(x)$ centered
at~$x$ (see~\cite[Proposition~2.4.6]{baer+ginoux}
and~\eqref{sSrel}). Thus if~$x$ and~$y$ are timelike separated,
the singular supports of~$s^\wedge(x,.)$ and~$s^\vee(.,y)$ do not intersect (see Figure~\ref{figsupport2}).
Moreover, we know from~\eqref{Pconverge}
that $P^\varepsilon(z_1, z_2)$ converges as a distribution and locally uniformly away from the diagonal.
Using these facts in~\eqref{Pepsrep}, we conclude that~$P^\varepsilon(x,y)$ converges locally uniformly
to~$P(x,y)$. This also implies that~$P(x,y)$ is continuous.
\QED
We remark that~$P(x,y)$ is even a smooth function away from the light cone; for a proof
for general bi-solutions we refer to~\cite{radzikowski, kratzert}.

\begin{Prp} \label{prp55} There is a future-directed timelike unit vector field~$s$ such that
for every~$x \in M$,
\[ \lim_{\varepsilon \searrow 0} s^\varepsilon_x = s(x)\:, \]
where by~$s(x)$ we mean the operator on~$S_xM$ acting by Clifford multiplication.
\label{Lemma-euclidean}
\end{Prp}
\Proof A short calculation gives
\begin{align}
\la {\f}^\varepsilon_\alpha(x) \,|\, {\f}^\varepsilon_\beta(x) \ra_\H
&= \frac{1}{\nu^\varepsilon_\alpha \nu^\varepsilon_\beta}\:
\la \ie_x \e^\varepsilon_\alpha | \ie_x \e^\varepsilon_\beta \ra_\H
= \frac{1}{\nu^\varepsilon_\alpha \nu^\varepsilon_\beta}\:
\Sl \e^\varepsilon_\alpha | (\ie_x)^* \ie_x \e^\varepsilon_\beta \Sr \nonumber \\
&= -\frac{1}{\nu^\varepsilon_\alpha \nu^\varepsilon_\beta}\:
\Sl \e^\varepsilon_\alpha | P^\varepsilon(x,x)\, \e^\varepsilon_\beta \Sr \label{inner1} \\
{F}^\varepsilon(x)\, {\f}^\varepsilon_\alpha(x) &=
\frac{1}{\nu^\varepsilon_\alpha} \big(- \ie_x (\ie_x)^* \: \ie_x(\e^\varepsilon_\alpha) \big)
= \frac{1}{\nu^\varepsilon_\alpha}\: \ie_x\, P^\varepsilon(x,x)\, \e^\varepsilon_\alpha \nonumber \\
\la {\f}^\varepsilon_\alpha(x) \,|\, {F}^\varepsilon(x)\,{\f}^\varepsilon_\beta(x) \ra_\H
&=-\frac{1}{\nu^\varepsilon_\alpha \nu^\varepsilon_\beta}\:
\Sl \e^\varepsilon_\alpha | P^\varepsilon(x,x)^2\, \e^\varepsilon_\beta \Sr \:. \label{inner2}
\end{align}
Comparing~\eqref{inner1} with~\eqref{inner2}, one sees that in our chosen basis,
\[ {F}^\varepsilon(x) = P^\varepsilon(x,x)\:. \]
Moreover, we know from~\eqref{nuev} that~$F^\varepsilon(x)$ has two positive and two
negative eigenvalues. Therefore, it remains to prove that, after a suitable rescaling,
$P^\varepsilon(x,x)$ converges to the operator of Clifford multiplication
by a future-directed timelike unit vector~$s(x) \in T_x M$, i.e.
\beq \label{Pepsgoal}
\lim_{\varepsilon \searrow 0} \varepsilon^p P^\varepsilon(x,x) = c\, s(x)
\eeq
for suitable constants~$p$ and~$c$.

In order to prove this claim, in the past of~$\nn$ we choose a chart where the metric is
the Minkowski metric. Moreover, we choose the standard spinor frame and use the notation
of Section~\ref{secvac}. Then we can combine~\eqref{Pepsrep} with~\eqref{sSrel} to obtain
\begin{align*}
P^\varepsilon(x,x) &= \iint_{\R^3 \times \R^3} s^\wedge(x,z_1) \,\gamma^0\, P^\varepsilon(z_1, z_2)\,
\gamma^0 \, s^\vee(z_2, x)\: d^3z_1\: d^3z_2 \\
&= \iint_{\R^3 \times \R^3} \!\!\!\!\!\!\!\!S^\wedge(x,z_1) \,(-i \overset{\leftarrow}{\Pdd}_{z_1} \!+ m)\,\gamma^0
P^\varepsilon(z_1, z_2) \gamma^0 \, (i \Pdd_{z_2} \!+ m)\, S^\vee(z_2, x)\: d^3z_1\: d^3z_2\:,
\end{align*}
where~$z_{1\!/\!2} = (0, \vec{z}_{1\!/\!2})$, and the arrow indicates that the derivatives act to the left,
\[ S^\wedge(x,z) \overset{\leftarrow}{\Pdd}_{z} \equiv
\frac{\partial}{\partial z^j}\: S^\wedge(x,z)\: \gamma^j \:. \]
We now integrate by parts the spatial derivatives of~$z_1$ and~$z_2$. Using the identity
\[ (i \vec{\gamma} \vec{\nabla}_{z} + m) \: P^\varepsilon(z, y) =
-i \gamma^0 \frac{\partial}{\partial z^0} P^\varepsilon(z, y) + 2m\, P^\varepsilon(z, y) \]
and its adjoint, we obtain
\beq \label{Pepsorigin}
\begin{split}
P^\varepsilon(x,x) &=
\iint_{\R^3 \times \R^3} S^\wedge(x,z_1) \,\Big(-i (\overset{\leftarrow}{\partial}_{t_1}
+ \partial_{t_1}) + 2 m \gamma^0 \Big)\,
P^\varepsilon(z_1, z_2) \\
& \qquad \qquad \qquad \times \Big( i (\overset{\leftarrow}{\partial}_{t_2} + \partial_{t_2}) + 2 m \gamma^0 \Big)
\, S^\vee(z_2, x)\: d^3z_1\: d^3z_2\:,
\end{split}
\eeq
where~$t_{1\!/\!2} \equiv z_{1\!/\!2}^0$ are the time components of~$z_{1\!/\!2}$.

In the limit~$\varepsilon \searrow 0$, the function~$P^\varepsilon(z_1, z_2)$ becomes singular if~$z_1 = z_2$
(see~\eqref{Pepssing} in the case~$t=r=0$). Moreover, the singular supports of the
distributions~$S^\wedge(x,.)$ and~$S^\vee(.,x)$ coincide (see Figure~\ref{figsupport2} in the case~$x=y$).
As a consequence, the integral in~\eqref{Pepsorigin} diverges as~$\varepsilon \searrow 0$, having
poles in~$\varepsilon$. The orders of these poles can be obtained by a simple power counting.
In order to analyze the structure of these poles in more detail, one performs the
Hadamard expansion of the distributions~$S^\wedge$ and~$S^\vee$
(see~\cite[Section~2]{baer+ginoux} or the next section of the present paper for similar calculations
for the fermionic operator). Substituting the resulting formulas into~\eqref{Pepsorigin},
one finds that the higher orders in the Hadamard expansion give rise to lower order
poles in~$\varepsilon$. In particular, the most singular contribution to~\eqref{Pepsorigin} is
obtained simply by taking the first term of the Hadamard expansion of the Green's function~$S^\wedge(x,z_1)$,
which is a scalar multiple of the parallel transport
with respect to the spinorial Levi-Civita connection along the unique null geodesic joining~$z_1$ and~$x$.
Similarly, the Green's function~$S^\vee(z_2,x)$ may be replaced by a multiple of the parallel transport
along the null geodesic joining~$z_2$ with~$x$. Moreover, for the most singular
contribution to~\eqref{Pepsorigin} it suffices to consider the lowest order in~$m$, which means that
we may disregard the factors~$2 m \gamma^0$ in~\eqref{Pepsorigin}.
Finally, we know that in the limit~$z_1 \rightarrow z_2$, the leading contribution to~$P^\varepsilon(z_1, z_2)$
is proportional to~$\gamma^0$ (see~\eqref{Pepsgrav} and~\eqref{iis}).
Putting these facts together, the most singular contribution to~$P^\varepsilon(x,x)$ is
obtained simply by taking the operator~$\gamma^0$ at~$z_1=z_2=z$ and to parallel transport it
along the null geodesic joining~$z$ with~$x$. Integrating~$z$ over the set~$\partial J^\wedge(x) \cap \nn$,
we obtain the desired operator of Clifford multiplication in~\eqref{Pepsgoal}.
\QED

\subsection{The Hadamard Expansion of the Fermionic Operator} \label{sechadamard}
In this section we shall analyze the singularity structure of the distribution~$P$ introduced in
Definition~\ref{Pdef-globhyp} by performing the so-called Hadamard expansion.
In order to be able to apply the methods worked out in~\cite[Section~2]{baer+ginoux},
it is preferable to first consider the second-order equation~\eqref{kg-globhyp}.
The following lemma relates~$P$ to a solution of~\eqref{kg-globhyp}.
\begin{Lemma}\label{lemma-Tsymm}
Let $T$ be the unique symmetric distributional bi-solution of the Klein-Gordon equation~\eqref{kg-globhyp} which coincides with the Fourier transform of the lower mass shell
\beq \label{Tdef}
T(x,y)=\int\frac{d^4k}{(2\pi)^4}\: \delta(k^2-m^2)\, \Theta(-k^0)\: e^{-ik(x-y)}
\eeq
for $x$ and $y$ in the past of $\nn$. Then 
\beq\label{relPT}
P(\psi,\phi)=T\big((\dd+m)\psi,\phi\big).
\eeq
\end{Lemma}
\Proof
We introduce the distribution~$P_m$ by
$$P_m(\psi,\phi)=\frac{1}{2m}\: T\big((\dd+m)\psi,(\dd+m)\phi\big) \:. $$
Obviously, $P_m$ is symmetric and satisfies the Dirac equation~\eqref{Peqn}. Moreover,
a short calculation using~\eqref{Tdef} and~\eqref{Pdef} shows that~$P_m$ coincides with the regularized
Dirac sea vacuum~\eqref{Pepsdef} if~$\psi$ and~$\phi$ are both supported in the past of $\nn$.
We conclude that~$P_m$ coincides with the distribution~$P$ of Definition~\ref{Pdef-globhyp}.
Obviously, $P_m$ is also a bi-solution of the Klein-Gordon equation. Flipping the sign of~$m$,
we get another bi-solution~$P_{-m}$ of the Klein-Gordon equation.
Again using~\eqref{Tdef} and~\eqref{Pdef}, we find that the following combination of~$P_m$ and~$P_{-m}$
coincides with~$T$,
\beq \label{TPrel}
T=\frac{1}{2m}(P_m-P_{-m})\,.
\eeq

The fact that the operator~$P_m$ is a bi-solution of the Dirac equation implies that
it commutes with~$\dd$,
\beq \label{Pcommute}
P_m(\dd\psi,\phi)=mP_m(\psi,\phi)=P_m(\psi,\dd\phi)
\eeq
and similarly for $P_{-m}$. We thus obtain
\begin{align*}
	P(\psi,\phi)&\;\:=\: \frac{1}{2m}T\big((\dd+m)\psi,(\dd+m)\phi\big)\\
	&\overset{\eqref{TPrel}}{=} \frac{1}{(2m)^2}\Big(P_m\big((\dd+m)\psi,(\dd+m)\phi\big)-P_{-m}\big((\dd+m)\psi,(\dd+m)\phi\big)\Big)\\
	&\overset{\eqref{Pcommute}}{=} \frac{1}{(2m)^2}\Big(P_m\big((\dd+m)^2\psi,\phi\big)-P_{-m}\big((\dd+m)^2\psi,\phi\big)\Big) \\
	&\;\:=\: \frac{1}{2m}T\big((\dd+m)^2\psi,\phi\big)=T\big((\dd+m)\psi,\phi\big)\,,
\end{align*}
giving the result.
\QED

We now perform the Hadamard expansion of the distribution~$T$ using the methods of~\cite{deWitt,
fulling+sweeny+wald, radzikowski, moretti, baer+ginoux}. Assume that~$\Omega \subset M$ is a geodesically
convex subset (see~\cite[Definition~1.3.2]{baer+ginoux}). Then for any~$x, y \in \Omega$, there is a
unique geodesic~$c$ in~$\Omega$ joining~$y$ and~$x$.
We denote the squared length of this geodesic by
\[ \Gamma(x,y) = g \big( \exp_y^{-1}(x), \exp_y^{-1}(x) \big) \]
(note that~$\Gamma$ is positive in timelike directions and negative in spacelike directions) and remark that the identity
\beq \label{gradG}
g(\grad_x\Gamma,\grad_x\Gamma)=4\Gamma
\eeq
holds.
In order to prescribe the behavior of the singularities on the light cone, we set
\[ \Gamma_\varepsilon(x,y) = \Gamma + i \varepsilon \big(\mathfrak{t}(x) - \mathfrak{t}(y) \big) \]
and introduce the short notation
\beq \label{rule}
\frac{1}{\Gamma^p} = \lim_{\varepsilon \searrow 0} \frac{1}{(\Gamma_\varepsilon)^p} \qquad \text{and} \qquad
\log \Gamma = \lim_{\varepsilon \searrow 0} \log \Gamma_\varepsilon 
= \log |\Gamma| - i \pi\, \epsilon\big(\mathfrak{t}(x) - \mathfrak{t}(y) \big)
\eeq
(where~$\epsilon$ is again the step function),
with convergence in the distributional sense. Here the logarithm is cut along the positive real axis, with
the convention
\[ \lim_{\varepsilon \searrow 0} \log(1+i \varepsilon) = - i \pi\:. \]
In the past of~$\nn$, this prescription gives the correct singular behavior of the distribution~\eqref{Pdef}
on the light cone (for details see~\cite[eqns~(2.5.39)-(2.5.41)]{PFP}). Using the
methods of~\cite{fulling+sweeny+wald}, it follows that this prescription holds globally.
We remark that the rule~\eqref{rule} also implements the
local spectral condition in~\cite{radzikowski}.

In~\cite[Section~2]{baer+ginoux}, the Hadamard expansion is worked out in detail for the
causal Green's functions of a normally hyperbolic operator. Adapting the methods and results in a
straightforward way to the distribution~$T$, we obtain the Hadamard expansion
\beq \label{Hadaexp-T}
(- 8 \pi^3)\, T(x,y)=\frac{\vleck}{\Gamma}\:\Pi^y_x+\frac{\log(\Gamma)}{4}\:V^y_x+ \Gamma \log(\Gamma)\:
W^y_x + \Gamma H^y_x + \O(\Gamma^2 \log\Gamma)
\eeq
(the normalization constant~$(- 8 \pi^3)$ can be read off from~\eqref{Pepssing} and~\eqref{coeffex1},
because $T(x,y)$ coincides with~$\beta/m$ if~$x$ and~$y$ are in the past of~$\nn$).
Here~$\vleck(x,y)$ is the square root of the van Vleck-Morette determinant
(see for example~\cite{moretti}), which in normal coordinates around $y$ is given by 
\beq
\vleck(x,y)=|\det(g(x))|^{-\frac{1}{4}}\,.\label{vleck-formula}
\eeq
Moreover, $\Pi^y_x: S_yM\rightarrow S_xM$ denotes the spinorial Levi-Civita parallel transport
along~$c$. The linear mappings $V^y_x, W^y_x, H^y_x:S_yM\rightarrow S_xM$ are called Hadamard coefficients. They depend smoothly on $x$ and $y$, and can be determined via the Hadamard recurrence
relations~\cite{deWitt}.
The Hadamard coefficient $V^y_x$ is given explicitly by formula \eqref{hadamardcoeff} in the appendix.
Writing the result of Lemma~\ref{lemma-Tsymm} with distributional derivatives as~$P(x,y)=(\dd_x+m)T(x,y)$,
we obtain the Hadamard expansion of~$P$ by differentiation.
\begin{Corollary} \label{coroll57}
The distribution~$P(x,y)$ has the Hadamard expansion
\begin{align}\label{Hadaexp-P}
(- 8 \pi^3)&\, P(x,y)=-\frac{i \vleck}{\Gamma^2}\grad_x\Gamma\cliff \Pi^y_x+\frac{i}{\Gamma}\grad_x\vleck\cliff\: \Pi^y_x+\frac{\vleck}{\Gamma}(\dd_x+m)\Pi^y_x \nonumber\\ &+\frac{i}{4\Gamma}\grad_x\Gamma\cliff V^y_x 
+\frac{\log(\Gamma)}{4}\, (\dd_x+m)V^y_x+
i (1+\log(\Gamma))\,	\grad_x \Gamma \cliff W^y_x  \nonumber\\
& + i \grad_x\Gamma\cliff H^y_x + \O(\Gamma\log\Gamma)\,.
\end{align}
\end{Corollary}\noindent

\subsection{The Fermionic Operator Along Timelike Curves} \label{secPcurve}
Assume that~$\gamma(t)$ is a future-directed, timelike curve which joins two space-time points~$p, q \in M$.
For simplicity, we parametrize the curve by arc length on the interval~$[0, t_\text{max}]$ such
that~$\gamma(0) = q$ and~$\gamma(t_\text{max}) = p$. For any given~$N$, we define the points~$x_0, \ldots, x_N$ by
\beq \label{xinter}
x_n = \gamma(t_n) \qquad \text{with} \qquad t_n = \frac{n}{N}\: t_\text{max}\:.
\eeq
Note that these points are all timelike separated, and that the geodesic distance of
neighboring points is of the order~$1/N$.
In this section we want to compute~$P(x_{n+1}, x_n)$ in powers of~$1/N$.
To this end, we consider the Hadamard expansion of Corollary~\ref{coroll57}
and use that~$0 < \Gamma(x_{n+1}, x_n) \in \O(1/N)$. Thus our main task is to expand
the Hadamard coefficients in~\eqref{Hadaexp-P} in powers of~$1/N$. For ease in notation, we set
\[ x = x_{n+1} \qquad \text{and} \qquad y = x_n\:. \]
Possibly by increasing~$N$, we can arrange that~$x$ and~$y$ lie in a geodesically convex
subset~$\Omega \subset M$. We let~$c$ be the unique geodesic in~$\Omega$ joining~$y$ and~$x$,
\beq \label{def-c} 
c:[0,1]\rightarrow M\,,\quad c(\tau):=\exp_y(\tau T)\quad\text{with }T:=\exp^{-1}_y(x) \:. 
\eeq
We also introduce the expansion parameter
\[ \delta := \sqrt{\Gamma(x,y)} = \sqrt{g(T,T)} \in \O \Big( \frac{1}{N} \Big) \:. \]

Next, we let~$\{e_0=\delta^{-1} T,\,e_1,\,e_2,\,e_3\}$ be a pseudo-orthonormal basis of $T_yM$, i.e.
$$g(e_j,e_k)=\epsilon_j \:\delta_{jk}\,,$$
where the signs $\epsilon_j$ are given by
\beq
\epsilon_j:=\bca+1 & \text{if~$j=0$} \\ -1 & \text{if~$j=1,2,3$} \:. \eca \label{epsalphadef}
\eeq
We extend this basis to a local pseudo-orthonormal frame of $T\Omega$ by 
\beq \label{def-ponframe}
e_j(z)=\Lambda^y_z\, e_j\,,
\eeq
where~$\Lambda^y_z$ denotes the Levi-Civita parallel transport in $TM$ along the
unique geodesic in~$\Omega$ joining~$y$ and~$z$.
Then the following propositions hold.

\begin{Prp} \label{prp58}
The kernel of the fermionic operator has the expansion
\beq
(-8 \pi^3)\:P(x,y) =-\frac{i}{\Gamma^2}\,\grad_x\Gamma\cliff \Pi^y_x + \frac{m}{\Gamma}\,\Pi^y_x+\O(\delta^{-1})\,.  
\eeq
\end{Prp}

\begin{Prp} \label{prp59}
The closed chain has the expansion
\begin{align}
(-8 \pi^3)^2\: A_{xy} =\:& c(x,y)\: \1_{ S_xM} \label{Ayx4} \\
&+ m \Big( m^2 - \frac{\s}{12} \Big)\,\frac{\Ima(\log\Gamma)}{2\Gamma^2}\,\grad_x\Gamma
\label{Ayx1} \\
&+ i \big[\grad_x \Gamma, X_{xy} \big] + \big\{ \grad_x \Gamma, Y_{xy} \big\} \label{Ayx3}\\
&+\O(\delta^{-1} \log \delta) \:, \nonumber
\end{align}
where all operators act on~$S_xM$.
Here~$X_{xy}$ and~$Y_{xy}$ are symmetric linear operators and
\[ X_{xy} = \O(\delta^{-3})\:,\qquad Y_{xy} = \O(\delta^{-3} \, \log \delta)\:. \]
\end{Prp}\noindent
The proof of Propositions \ref{prp58} and \ref{prp59} is given in Appendix~\ref{AppA}, where we also compute some of the
Hadamard coefficients explicitly in terms of curvature expressions.

Note that the contribution~\eqref{Ayx1} is of the
order~$\delta^{-3}$, whereas~\eqref{Ayx3} is of the order~$\O(\delta^{-2} \log \delta)$.
The term~\eqref{Ayx1} amounts to Clifford multiplication with~$\grad_x \Gamma$
and is thus analogous to the term~$a(\xi) \, \xi\slsh$ in the closed chain~\eqref{A-unreg} of
Minkowski space. The contributions~\eqref{Ayx3} will be discussed in detail in the next section.

\subsection{The Unspliced versus the Spliced Spin Connection} \label{secversus}
In this section, we compute the unspliced and spliced spin connections and compare them.
We write the results of Corollary~\ref{coroll57} and Proposition~\ref{prp59} as
\begin{align} 
(-8 \pi^3)\: P(x,y) =\:& -\frac{i}{\Gamma^2}\grad_x\Gamma\cliff \Pi^y_x
+\frac{m}{\Gamma} \:\Pi^y_x + \O(\delta^{-1}) \label{Pexp} \\
(-8 \pi^3)^2\: A_{xy} =\:& c_{xy} + a_{xy}\: \grad_x \Gamma \nonumber \\
&+ i \big[\grad_x \Gamma, X_{xy} \big] + \big\{ \grad_x \Gamma, Y_{xy} \big\} 
+ \O \big( \delta^{-1} \log \delta \big) \:, \label{A0rep}
\end{align}
where
\[ a_{xy} \sim \delta^{-4} \qquad \text{and} \qquad X_{xy} \sim \delta^{-3} \:. \]
We want to compute the directional sign operator~$v_{xy}$
(see Definition~\ref{defdirsig}) in an expansion in powers of~$\delta$.
To this end, we first remove the commutator term in~\eqref{A0rep} by a unitary transformation,
\beq \label{UtA}
(-8 \pi^3)^2\: e^{-i Z_{xy}} \,A_{xy} \,e^{i Z_{xy}} = c_{xy} + 
a_{xy}\: \grad_x \Gamma + \big\{ \grad_x \Gamma, Y_{xy} \big\} 
+ \O \big( \delta^{-1} \log \delta \big) \:,
\eeq
where we set
\beq \label{Zdef} 
Z_{xy} = -\frac{X_{xy}}{a_{xy}} \sim \delta\:. 
\eeq
We let~$u \in T_xM$ be a future-directed timelike unit vector pointing in the direction of~$\grad_x \Gamma$.
Then the operator~$u$ (acting by Clifford multiplication) is a sign operator
(see Definition~\ref{defsign}), which obviously commutes with the right side of~\eqref{UtA}.
Hence the directional sign operator (see Definition~\ref{defdirsig}) is obtained from~$u$ by
unitarily transforming backwards,
\beq \label{vrep3} v_{xy} = e^{i Z_{xy}} \:u\: e^{-i Z_{xy}}
= u + i \big[Z_{xy}, u \big] + \O \big( \delta^{2} \log^2 \delta \big) .
\eeq

In order to construct the synchronization map at~$x$, it is convenient to work with
the distinguished subspace~${\mathfrak{K}}(x)$ of~$\Symm(S_xM)$ spanned by the
operators of Clifford multiplication
with the vectors~$e_0, \ldots, e_3 \in T_xM$ and the pseudoscalar
operator~$e_4 =-e_0 \cdots e_3$ (thus in the usual Dirac representation,
${\mathfrak{K}} = \la \gamma^0, \ldots, \gamma^3, i \gamma^5 \ra$).
The subspace~${\mathfrak{K}}$ is a distinguished Clifford subspace
(see Definition~\ref{defcliffsubspace} and~Definition~\ref{deftangrep}).
The inner product~\eqref{anticommute2} extends the Lorentzian metric on~$T_xM$
to~${\mathfrak{K}}(x)$.
The space~$\Symm(S_xM)$ is spanned by the 16 operators~$\1$, $e_j$ and~$\sigma_{jk} = \frac{i}{2}[e_j, e_k]$
(where~$j,k \in \{0, \ldots, 4\}$), giving the basis representation
\[ Z_{xy} = c + \sum_{j,k=0}^4 B^{jk} \sigma_{jk} + \sum_{j=0}^4 w^j e_j \:. \]
The first summand is irrelevant as it drops out of the commutator in~\eqref{vrep3}.
The second summand gives a contribution to~$v_{xy}$ which lies in the distinguished Clifford subspace,
\beq \label{def-Delu} 
\Delta u := i \sum_{j,k=0}^4 \big[B^{jk} \sigma_{jk}, u \big] = 4 \sum_{j,k=0}^4 B^{jk} u_j\: e_k \in {\mathfrak{K}}\:, 
\eeq
whereas the last summand gives a bilinear contribution
\[ i \big[w, u \big] \qquad \text{with}\qquad w := \sum_{j=0}^4 w^j e_j\:. \]
We thus obtain the representation
\beq \label{vrep1}
v_{xy} =  u + \Delta u +  i \big[w, u \big] + \O \big( \delta^{2} \log^2 \delta \big) .
\eeq
We next decompose~$w \in {\mathfrak{K}}$ as the linear combination
\beq \label{wrep}
w = \alpha\, u + \beta\, s(x) + \rho \qquad \text{with} \qquad
\rho \perp u \quad \text{and} \quad \rho \perp s(x)\:.
\eeq
If~$u$ and~$s(x)$ are linearly dependent, we choose~$\beta=0$.
Otherwise, the coefficients~$\alpha$ and~$\beta$
are uniquely determined by the orthogonality conditions.
Substituting this decomposition into~\eqref{vrep1}, we obtain
\beq \label{vrep2}
v_{xy} = e^{i \rho} \:e^{i \beta s(x)} \:\big( u + \Delta u \big) \:  e^{-i \beta s(x)} \: e^{-i \rho}
+ \O \big( \delta^{2} \log^2 \delta \big)\:.
\eeq
Comparing with Lemma~\ref{lemma3} and Definition~\ref{defsync}, one finds that~$e^{i \rho}$
is the synchronization map~$U^{u, s(x)} $ at~$x$. The mapping~${e^{i \beta s(x)}}$, on the
other hand, identifies the representatives~${\mathfrak{K}}, K_x^{(y)}
\in \T^{s_x}$ of the tangent space~$T_x$ (see Definition~\ref{deftangent}).
Using the notation introduced after Definition~\ref{defdirsig} and at the beginning
of Section~\ref{secmetricconn}, we have~$U_{xy} = e^{i \rho}$ and
\beq \label{xysync}
K_{xy} = e^{i \rho}\,K_x^{(y)} \, e^{-i \rho} \qquad \text{and} \qquad
K_x^{(y)} = e^{i \beta s(x)}\,{\mathfrak{K}}(x)\, e^{-i \beta s(x)} \:.
\eeq

We next compute the synchronization map at the point~$y$.
Since the matrices~$A_{xy}$ and~$A_{yx}$ have the same characteristic polynomial, we know that
\[ v_{xy} \,P(x,y) = P(x,y)\, v_{yx} \:. \]
Multiplying by
\[ (- 8 \pi^3)^{-1}\: P(x,y)^{-1} = \frac{i}{4}\: \Gamma\:\Pi^x_y \grad_x\Gamma
-\frac{m}{4}\:\Gamma^2 \:\Pi^x_y + \O(\delta^{5}) \]
(where we used~\eqref{Pexp} and~\eqref{gradG}),
a direct calculation using~\eqref{vrep1} gives
\[ v_{yx} =  \Pi^x_y \,\Big( \,u - \Delta u - i \big[w, u \big] \,\Big)\, \Pi^y_x 
+ \O \big( \delta^{2} \log^2 \delta \big) . \]
Using that~$s(y) = \Pi^x_y \,s(x)\, \Pi^y_x + \O(\delta)$, we obtain similar to~\eqref{xysync}
\beq \label{yxsync2}
K_{yx} = \Pi^x_y e^{-i \rho} \Pi^y_x \,K_y^{(x)} \, \Pi^x_y e^{i \rho} \Pi^y_x \quad \text{and} \quad
K_y^{(x)} = \Pi^x_y e^{-i \beta s(x)} \Pi^y_x \,{\mathfrak{K}}(y)\,\Pi^x_y e^{i \beta s(x)} \Pi^y_x \:.
\eeq
We are now ready to compute the spin connections introduced
in Definitions~\ref{spinconable} and~\ref{defsplicedD}.
\begin{Prp} \label{prpDspliced}
The unspliced and spliced spin connections are given by
\begin{align}
D_{x,y} &= \Big( \1 + (\Delta u ) \cliff u + 2 i\, (\beta \,s(x) + \rho) \Big)\, \Pi^y_x + \O(\delta^2 \log^2 \delta) 
\label{Dunspliced} \\
D_{(x,y)} &= \Big( \1 + (\Delta u ) \cliff u \Big)\, \Pi^y_x + \O(\delta^2 \log^2 \delta)\:. \label{Dspliced}
\end{align}
\end{Prp}
\Proof We first compute the unspliced spin connection using the characterization
of Theorem~\ref{thmspinconnect}. A short calculation using~\eqref{Pexp} and~\eqref{vrep1} gives
\begin{align*}
(- 8 \pi^3)^2\, A_{xy} &= \frac{4}{\Gamma^3} + \O(\delta^{-4}) \\
A_{xy}^{-\frac{1}{2}}\: P(x,y) &= -\Big( \big|\! -8 \pi^3 \big|^{-1} A_{xy}^{-\frac{1}{2}} \Big) \: \Big( (- 8 \pi^3)\, P(x,y) \Big) \\
&= \frac{i}{2} \:\Gamma^{-\frac{1}{2}} \grad_x\Gamma \cliff \Pi^y_x
- \frac{m}{2} \: \Gamma^{\frac{1}{2}}\:\Pi^y_x + \O(\delta^2) \\
&= i u \cliff \Pi^y_x - \frac{m}{2} \: \Gamma^{\frac{1}{2}}\:\Pi^y_x + \O(\delta^2) \:.
\end{align*}
In order to evaluate the condition~(ii) of Theorem~\ref{thmspinconnect}, it is easiest to
transform the Clifford subspaces~$K_{xy}$ and~$K_{yx}$ to the distinguished Clifford
subspace~${\mathfrak{K}}(x)$ and~${\mathfrak{K}}(y)$, respectively.
In view of~\eqref{xysync} and~\eqref{yxsync2}, we can thus rewrite
the condition~(ii) of Theorem~\ref{thmspinconnect} by demanding that the
unitary transformation
\[ V := e^{-i \beta s(x)}\: e^{-i \rho} \:\Big( e^{i \varphi\, v_{xy}}\: A_{xy}^{-\frac{1}{2}}\: P(x,y) \Big)\:
\Pi^x_y\, e^{-i \rho} e^{-i \beta s(x)} \,\Pi^y_x \]
transforms the distinguished Clifford subspaces to each other,
\beq \label{Vcond}
V \,{\mathfrak{K}}(y)\, V^{-1} = {\mathfrak{K}}(x)\:.
\eeq
The operator~$V$ is computed by
\begin{align}
V &= e^{-i \beta s(x)}\: e^{-i \rho} \: e^{i \varphi \: v_{xy}}
\left( i u - \frac{m}{2} \: \Gamma^{\frac{1}{2}}  \right)
e ^{-i \rho} e^{-i \beta s(x)} \,\Pi^y_x + \O(\delta^2) \nonumber \\
&\!\!\!\overset{\eqref{vrep2}}{=} e^{i \varphi \,(u + \Delta u)}\:
e^{-i \beta s(x)}\: e^{-i \rho} \left( i u - \frac{m}{2} \: \Gamma^{\frac{1}{2}}  \right)
e^{-i \rho} e^{-i \beta s(x)} \,\Pi^y_x + \O(\delta^2) \nonumber \\
&= \left\{ e^{i \varphi \,(u + \Delta u)} \big( i u - \frac{m}{2} \: \Gamma^{\frac{1}{2}} 
+ 2 \beta \,\la u, s(x) \ra \big) \right\} \Pi^y_x + \O(\delta^2 \log^2 \delta) . \label{curly}
\end{align}
Now the condition~\eqref{Vcond} means that the curly brackets in~\eqref{curly}
describe an infinitesimal Lorentz transformation on~${\mathfrak{K}}(x)$.
Thus the brackets must only have a scalar and a bilinear contribution, but no vector
contribution. This leads us to choose~$\varphi$ such that
\beq \label{phiphase}
\sin \varphi = -1 + \O(\delta^2 \log^2 \delta) \:,\quad
\cos \varphi = -\frac{m}{2} \: \Gamma^{\frac{1}{2}} 
+ 2 \beta \,\la u, s(x) \ra + \O(\delta^2 \log^2 \delta) < 0
\eeq
(note that this choice of~$\varphi$ is compatible with our convention~\eqref{phasefix2}).
It follows that
\begin{align*}
V &= \Big( \1 + (\Delta u ) \cliff u \Big) \Pi^y_x + \O(\delta^2 \log^2 \delta) \\
D_{x,y} \, \Pi^x_y &= e^{i \varphi\, v_{xy}}\: A_{xy}^{-\frac{1}{2}}\: P(x,y)\, \Pi^x_y \\
&= \1 + (\Delta u ) \cliff u + i [w,u]\, u + 2 i \beta\, \la u,s(x) \ra \, u + \O(\delta^2 \log^2 \delta) \\
&\!\!\!\overset{\eqref{wrep}}{=} \1 + (\Delta u ) \cliff u + 2 i\, (\beta \,s(x) + \rho) + \O(\delta^2 \log^2 \delta) \:.
\end{align*}
Finally, using the notions of Definition~\ref{defsplicedD}, we obtain
\begin{align}
U_x^{|y)} &= e^{-i \beta s(x)}\, e^{-i \rho} \:,\qquad
U_y^{(x|} =  e^{-i \rho}\, e^{-i \beta s(x)} \label{Uxyextra} \\
D_{(x,y)} &= U_x^{|y)}\, D_{x,y}\, U_y^{(x|} = \1 + (\Delta u ) \cliff u+ \O(\delta^2 \log^2 \delta)\:,
\end{align}
completing the proof.
\QED

The terms in the statement of the above proposition are quantified
in the next lemma, which is again proven in the appendix.
\begin{Lemma} \label{lemma511}
The linear operators~$\Delta u$ and~$\beta s + \rho$ in~\eqref{Dspliced}
and~\eqref{Dunspliced} have the expansions
\begin{align}
\Delta u &= \frac{1}{6\, \delta}\:\Big(m^2 - \frac{\s}{12} \Big)^{-1} \epsilon_j\,(\nabla_{e_j} R)(T, e_j)\, T + \O(\delta^2 \log^2 \delta)
\label{Deluex} \\
\beta s + \rho &= \Bigg[ \O \Big( \frac{1}{m}\, \|\epsilon_j\,\Ric(T, e_j)\, e_j\| \Big) 
+ \O \bigg( \frac{\delta}{m^3}\: \big(\|R\|^2 + \|\nabla^2 R\| \big) \bigg) \Bigg] \Big(1 + \O \Big( \frac{\s}{m^2} \Big) \Big) \nonumber\\
 &\quad+ \O(\delta^2 \log^2 \delta)\:.  \label{betaex}
\end{align}
\end{Lemma}

Let us discuss these formulas. 
We first point out that all the terms in~\eqref{Deluex}
and~\eqref{betaex} are of the order~$\O(\delta)$. Thus the corresponding
correction terms in~\eqref{Dunspliced} and~\eqref{Dspliced} are also of the order~$\O(\delta)$.
In the next section, we shall see that adding up all these correction terms along a timelike curve will
give a finite deviation from the spinorial Levi-Civita parallel transport.
If we assume furthermore that the Compton scale is much smaller than the length scale where
curvature effects are relevant, 
\beq \label{Rscale}
\frac{\|\nabla^2 R\|}{m^4} \ll \frac{\|\nabla R\|}{m^3} \ll \frac{\|R\|}{m^2} \ll 1 \:,
\eeq
then this deviation will even be small. More specifically, the term involving the Ricci tensor
in~\eqref{betaex} is the leading correction term. As shown in Proposition~\ref{prpDspliced},
this leading correction enters the unspliced spin connection, but drops out of the spliced
spin connection. This explains why it is preferable to work with the spliced spin connection.

The above calculations also reveal another advantage of splicing:
The corrections in the spliced spin connection are bilinear contributions
(see~\eqref{Dspliced} and~\eqref{Deluex}) and can thus be interpreted as describing
an infinitesimal Lorentz transformation. However, the corrections in the unspliced spin connection
(see~\eqref{Dunspliced} and~\eqref{wex}) involve vector contributions, which have the unpleasant
feature that they do not leave the distinguished Clifford subspaces~${\mathfrak{K}}(x)$ invariant.

\subsection{Parallel Transport Along Timelike Curves} \label{sec55}
We are now in the position to prove the main theorem of this section.
We return to the setting of the beginning of Section~\ref{secPcurve} and consider
a future-directed, timelike curve~$\gamma$ which joins two space-time points~$p, q \in M$.
For any given~$N$, we again define the intermediate points~$x_0, \ldots, x_N$ by~\eqref{xinter}.
We then define the parallel transport~$D_{xy}^{N,\varepsilon}$ by successively composing
the spliced spin connection between neighboring points,
\beq \label{Dprod2}
D^{N, \varepsilon}_{(p,q)} := D^\varepsilon_{(x_N, x_{N-1})} \:D^\varepsilon_{(x_{N-1}, x_{N-2})}
\:\cdots\: D^\varepsilon_{(x_1, x_0)} \::\:
S_q \rightarrow S_p \:,
\eeq
where $D^\varepsilon$ is the spliced spin connection induced from the regularized fermionic
operator~$P^\varepsilon$.
Substituting the formulas~\eqref{Dspliced} and~\eqref{Deluex}, one gets~$N$ correction
terms $(\Delta u) \cliff u$, each of which is of the order~$\delta \sim N^{-1}$. Thus in the
limit~$N \rightarrow \infty$, we get a finite correction, which we now compute.

\begin{Thm} \label{thmgrav}
Let $(M,g)$ be a globally hyperbolic manifold which is isometric to Minkowski space
in the past of a Cauchy-hypersurface $\nn$.
Then the admissible curves (see Definition~\ref{defadmissible}) are dense in the
$C^\infty$-topology. Choosing~$N \in \N$ and~$\varepsilon>0$ such that
the points~$x_{n}$ and~$x_{n-1}$ are spin-connectable for all~$n=1,\ldots, N$, every point
$x_n$ lies in the future of $x_{n-1}$. Moreover,
\begin{align*}
\lim_{N \rightarrow \infty} \:\lim_{\varepsilon \searrow 0} D^{N,\varepsilon}_{(p,q)}
& =  \DLC_{p,q} \, \Texp \bigg( \frac{1}{6} \int_\gamma \Big(m^2 - \frac{\s}{12} \Big)^{-1} \\
&\qquad \times \DLC_{q,\gamma(t)}
\Big[\epsilon_j\, (\nabla_{e_j} R)\big(\dot{\gamma}(t), e_j \big)\, \dot{\gamma}(t) \Big]
\cliff \,\dot{\gamma}(t) \cliff \DLC_{\gamma(t), q} \: dt \bigg)\,,
\end{align*}
where~$\gamma(t)$ is a parametrization by arc length, and~$\DLC_{p,q}$ denotes the
parallel transport along~$\gamma$ with respect to the spinorial Levi-Civita connection,
and~$\Texp$ is the time-ordered exponential
(we here again identify~$S_x^\varepsilon$ and~$S_xM$ via~\eqref{Jepshyp}).
\end{Thm}
\Proof Substituting the formula~\eqref{Dspliced} into~\eqref{Dprod2}, one gets a product
of~$N$ linear operators. Taking the limit~$N \rightarrow \infty$ and using that differential
quotients go over to differentials, one obtains a solution of the linear ordinary differential equation
\[ \frac{d}{dt} D_{(\gamma(t), q)} =
\Big( \lim_{\delta \searrow 0} \frac{1}{\delta} \:(\Delta u) \cliff u \Big) \cliff\: D_{(\gamma(t), q)} \:. \]
Here the limit~$\delta \searrow 0$ can be computed explicitly using~\eqref{Deluex}.
Then the differential equation can be solved in terms of the time-ordered exponential (also called Dyson
series; see~\cite[Section~1.2.1 and~7.17.4]{zeidler1}). This gives the result.
\QED
This theorem shows that in the limit~$\varepsilon \searrow 0$ and locally in the neighborhood
of a given space-time point, the spliced spin connection reduces to the
spinorial Levi-Civita connection, up to a correction term which involves line integrals of derivatives
of the Riemann tensor along~$\gamma$. Computing the holonomy of a closed curve, one sees that
the corresponding spliced spin curvature equals the Riemann curvature, up to higher order
curvature corrections.

For clarity, we point out that the above theorem does not rely on the fact that we are working
with distinguished representatives of the tangent spaces.
Namely, replacing~\eqref{Dprod2} by the products of the unspliced spin connection
with intermediate splice maps,
\[ D^{N, \varepsilon}_{p,q} := D^\varepsilon_{x_N, x_{N-1}} \,U_{x_{N-1}}^{(x_N | x_{N-2})}
\,D^\varepsilon_{x_{N-1}, x_{N-2}} \,U_{x_{N-2}}^{(x_{N-1} | x_{N-3})} \:\cdots\:
U_{x_{N-1}}^{(x_{2} | x_{0})}\, D^\varepsilon_{x_1, x_0} \:, \]
the above theorem remains true (to see this, we note that in view of~\eqref{Dsplicedef} and~\eqref{Uharmony},
the parallel transpors~$D^{N, \varepsilon}_{p,q}$ and~$D^{N, \varepsilon}_{(p,q)}$
differ only by the two factors~$U_{x_N}^{|x_{N-1})}$ and~$U_{x_0}^{(x_1|}$,
which according to~\eqref{Uxyextra} and Lemma~\ref{lemma511} converge to the identity matrix).

We now apply the above theorem to the metric connection.
\begin{Corollary} Under the assumptions of Theorem~\ref{thmgrav}, the
metric connection and the Levi-civita connection are related by
\beq \label{eqLorentz}
\lim_{N \rightarrow \infty} \:\lim_{\varepsilon \searrow 0} \nabla^N_{x,y} - \nablaLC_{x,y} 
= \O \!\left( L(\gamma)\: \frac{\| \nabla R\|}{m^2} \right)
\Big(1 + \O \Big( \frac{\s}{m^2} \Big) \Big) \:,
\eeq
where~$L(\gamma)$ is the length of the curve~$\gamma$, and
\[ \nabla^{N, \varepsilon}_{p,q} := \nabla_{x_N, x_{N-1}} \:\nabla_{x_{N-1}, x_{N-2}}
\:\cdots\: \nabla_{x_1, x_0} \::\:
T_q \rightarrow T_p \:. \]
\end{Corollary}
\Proof This follows immediately from Theorem~\ref{thmgrav} and the identity
\[ \nabla^{N, \varepsilon}_{p,q} u_q =  D^{N, \varepsilon}_{(p,q)} u_q \cliff  D^{N, \varepsilon}_{(q,p)} \:, \]
where we again identify the tangent space~$T_xM$ with the distinguished
Clifford subspace~${\mathfrak{K}}(x)$ of~$\Symm(S_xM)$ (see after~\eqref{vrep3}).
\QED
We finally discuss the notions of parity-preserving, chirally symmetric
and future-transitive fermion systems (see Definitions~\ref{def-parpreserv}, \ref{def-chirsymm}
and~\ref{locft}).
Since our expansion in powers of~$\delta$ only gives us information on~$P(x,y)$
for nearby points~$x$ and~$y$, we can only analyze local versions of these definitions.
Then the expansion~\eqref{phiphase} shows that the fermion system without regularization
is locally future-transitive and locally parity-preserving. Moreover, as the formula~\eqref{Dspliced}
only involves an even number of Clifford multiplications, the fermion system is
locally chirally symmetric (with the vector field~$u(x)$ in Definition~\ref{def-chirsymm} chosen as
$i$ times the pseudoscalar matrix), up to the error term specified in~\eqref{Dspliced}.

\section{Outlook} \label{secoutlook}
We conclude by putting the previous constructions into a broader context and
by mentioning possible directions of future research.
We first point out that the assumptions~(A) and~(B) at the beginning of Section~\ref{secglobhyp}
should be considered only as a technical simplification.
More generally, the fermionic operator can be introduced using a causality argument
which gives a canonical splitting of the solution space of the Dirac equation into two subspaces.
One of these subspaces extends the notion of the Dirac sea to interacting systems
(see~\cite[Section 2.4]{PFP}). Apart from the recent construction in a space-time of finite life-time~\cite{finite},
this method has been worked out only perturbatively in terms of the
so-called causal perturbation expansion (see~\cite{grotz} and for linearized gravity~\cite[Appendix B]{firstorder}).
This shortcoming was our motivation for the above assumptions~(A) and~(B),
which made it possible to carry out all constructions non-perturbatively.
To avoid confusion, we note that the fermionic operator constructed by solving
the Cauchy problem (see Proposition~\ref{Pepsgrav}) does in general {\em{not}} coincide with
the physical fermionic operator obtained by the causal perturbation expansion
in the same space-time (because solving the Cauchy problem for vacuum initial data
is usually not compatible with the global construction in~\cite[eqns~(2.2.16) and~(2.2.17)]{PFP}).
However, these two fermionic operators have the same singularity structure on the light cone,
meaning that after removing the regularization, both fermionic operators have the
same Hadamard expansion. Since in the constructions of Sections~\ref{sechadamard}-\ref{sec55},
we worked exclusively with the formulas of the Hadamard expansion, all the results in these sections
immediately carry over to the physical fermionic operator.

We also point out that throughout this paper, we worked with the simplest possible
regularization by a convergence generating factor~$e^{-\varepsilon |k^0|}$
(see Lemma~\ref{lemmapeps}). More generally, one could consider a broader class of
regularizations as introduced in~\cite[\S4.1]{PFP}. All our results will carry over, provided
that the Euclidean operator has a suitable limit as~$\varepsilon \searrow 0$
(similar to~\eqref{Euklid} and Lemma~\ref{prp55}).

Our constructions could also be generalized to systems with
several families of elementary particles (see~\cite[\S2.3]{PFP}).
In this setting, only the largest mass will enter the conditions~\eqref{curvcond}
and~\eqref{Rscale}, so that it is indeed possible to describe physical systems
involving fermions with an arbitrarily small or vanishing rest mass. Working with several generations
also gives the freedom to perform local transformations before taking the partial trace (as is worked
out in~\cite[Section~7.6]{sector} for axial potentials).
This freedom can be used to modify the logarithmic poles of the fermionic operator on the light cone.
In this context, an interesting future project is to study causal fermion systems in the presence
of an electromagnetic field. We expect that the spin connection will then also
include the $\U(1)$-gauge connection of electrodynamics.

Another direction of future research would be to study the geometry of causal fermion systems
with regularization (i.e.\ without taking the limit~$\varepsilon \searrow 0$). It seems an
interesting program to study the ``quantum structure'' of the resulting space-times.

From the mathematical point of view, the constructions in this paper extend the basic
notions of Lorentzian spin geometry to causal fermion systems. However,
most of the classical problems in geometric analysis and differential geometry have not yet
been analyzed in our setting.
For example, it has not yet been studied how ``geodesics'' are introduced
in causal fermion systems, and whether such geodesics can be obtained by minimizing the ``length
of curves'' (similar as in~\eqref{Dprod2}, such a ``curve'' could be a finite sequence of space-time points).
Maybe the most important analytic problem is to get a connection between the
geometric objects defined here and the causal action principle (see~\cite[\S3.5]{PFP} and~\cite{continuum}).
From the geometric point of view, our notions of connection and curvature
describe the local geometry of space-time. It is a challenging open problem to explore
how these local notions are related to the global geometry and topology of space-time.

\appendix
\section{The Expansion of the Hadamard Coefficients} \label{AppA}
In this section we will derive an expansion of the Hadamard coefficients in~\eqref{Hadaexp-P} in powers
of~$\delta$. Using these expansions, we will then prove Proposition~\ref{prp58},
Proposition~\ref{prp59} and Lemma~\ref{lemma511}.
In terms of the pseudo-orthonormal frame $e_j$ (see~\eqref{def-ponframe}),
the Dirac operator on~$SM$ is given by
\beq
\dd\psi=i\,\epsilon_j\, e_j\cliff \nabla_{e_j}\psi\,,\qquad\text{with }\psi\in\Gamma(M,SM)\,. \label{diroprep}
\eeq
Here~$\nabla$ denotes the spinorial Levi-Civita connection, the dot denotes Clifford multiplication, and the
signs~$\epsilon_j$ are defined in~\eqref{epsalphadef}.
We denote space-time indices by Latin letters $j,k,\ldots\in\{0,1,2,3\}$, and spatial indices by Greek letters $\alpha,\beta,\ldots\in\{1,2,3\}$. Furthermore, we use Einstein's summation convention.
In order to calculate the derivatives of the spinorial parallel transport $\Pi^y_x$ with respect to the vectors~$e_j$,
we introduce suitable local coordinates. To this end, we consider the family of geodesics
\beq
c_s(t):=c(t,s_1,s_2,s_3):=\exp_y\big(tu+ts_\alpha e_\alpha\big)\,,
\label{def-cs}
\eeq
where $u=\exp_y^{-1}(x)=\delta\, e_0$.
The curve $c_0$ obviously coincides with the curve $c$ defined in \eqref{def-c}.
The exponential map~\eqref{def-cs} also gives rise to local coordinates $(t,s_\alpha)$ around $y$, with
corresponding local coordinate vector fields
\beq
T:=\frac{\partial c_s}{\partial t}\quad\text{ and }\quad Y_\alpha:=\frac{\partial c_s}{\partial s_\alpha}\,. 
\label{def-TY}
\eeq
The vector field $T$ is the tangent field of the curves $c_s$, and in terms of~$e_j$ it is given by
\beq
T=\delta \,e_0+s_\alpha \,e_\alpha\,.
\label{Tealpha}
\eeq
Since in this appendix we always consider variations of the curve~$c_0$, we can assume
that $s_\alpha=\O(\delta)$, and thus
$$T=\O(\delta)\,.$$
Moreover, the vector field~$T$ is timelike and the fields~$Y_\alpha$ are spacelike. By definition of the vector
fields~$e_j$ and of the spinorial parallel transport~$\Pi^y_{c_s(t)}$, it also follows that
\begin{align}
	\nabla_T \,\Pi^y_{c_s(t)}=\:&0 \label{TPi}\\
	\nabla_T \,e_j|_{c_s(t)}=\:&0 \label{Te}\\
	\nabla_{e_j}\Pi^y_{c_s(t)}=\:&\O(\delta) \label{eaPi}\\
	\nabla_{e_j}e_k|_{c_s(t)}=\:&\O(\delta) \label{eaeb}\,.
\end{align}
The vector fields $Y_\alpha$ are Jacobi fields, i.e. they are solutions of the Jacobi equation
\beq \label{Jacobi}
\nabla_T\nabla_T Y_\alpha = R(T,Y_\alpha)\,T
\eeq
with initial conditions
\beq \label{Jbcond}
Y_\alpha|_{t=0}=0 \qquad \text{and} \qquad \nabla_TY_\alpha|_{t=0} =e_\alpha \,,
\eeq
where $R$ denotes the Riemann tensor on $TM$. This initial value problem
can be solved perturbatively along each curve $c_s$, giving the expansion
\beq \label{Y-eexp}
Y_\alpha|_{c_s(t)} = t\, e_\alpha+\Lambda^y_{c_s(t)}\int_0^t\!\! d\tau\:\Lambda^{c_s(\tau)}_y\int_0^\tau\!\!
d\sigma\:\sigma\,\Lambda^{c_s(\sigma)}_{c_s(\tau)} R(T,e_\alpha) T|_{c_s(\sigma)}+\O(\delta^4) \:.
\eeq

The spinorial curvature tensor $\rr$ on $ S M$ is defined by the relation
$$\rr(X,Y)\psi:=\nabla_X\nabla_Y\psi-\nabla_Y\nabla_X\psi-\nabla_{[X,Y]}\psi \:,$$
valid for any $X,Y\in T_pM$ and $\psi\in\Gamma(M,SM)$. In the local pseudo-orthonormal frame~$(e_j)$, it
takes the form
\beq
\rr(X,Y)\, \psi = \frac{1}{4}\: \epsilon_j\epsilon_k\, g\big(R(X,Y)e_j,e_k\big) \:e_j\cliff e_k\cliff \psi\,.
\label{rrexplicit}
\eeq
Using~\eqref{TPi} and the fact that the local coordinate vector fields $T$ and $Y_\alpha$ commute, we
conclude that
\beq \label{TYpi}
\nabla_T\nabla_{Y_\alpha} \,\Pi^y_{c(t)}=\rr(T,Y_\alpha) \,\Pi^y_{c(t)}.
\eeq
Integrating this equation gives
\beq\nabla_{Y_\alpha}\Pi^y_{c(t)}=\Pi^y_{c(t)}\int_0^t\Pi_y^{c(\tau)}\rr(T,Y_\alpha)|_{c(\tau)}\Pi^y_{c(\tau)}d\tau\,.
\label{YjPi}
\eeq
Using this formula, we can now derive the expansion of the Hadamard coefficient~$\dd_x \Pi^y_x$.
\begin{Lemma} \label{lemma-DPi}
The Hadamard coefficient~$\dd_x \Pi^y_x$ has the expansion
\begin{align*}
	\dd_x\Pi^y_x =&- \frac{i}{2}\:\epsilon_j\Big(\int_0^1\: t \Ric(T,e_j)|_{c(t)}\:dt\Big)\,
	e_j\cliff \Pi^y_x \\
	&+\frac{i}{24}\,\epsilon_j \epsilon_p\epsilon_q \Big(\epsilon_k\, g\big(R(e_j,T)T,e_k\big)\big|_x 
	\int_0^1\: t\,  g\big(R(T,e_k)e_p,e_q\big)|_{c(t)}\:dt\Big)e_j\cliff e_p\cliff e_q\cliff \Pi^y_x \\
	&+\O(\delta^4)\,.
\end{align*}
\end{Lemma}
\Proof
From \eqref{Y-eexp} we conclude that 
\begin{align*}
	Y_\alpha|_{c(t)}&=te_\alpha|_{c(t)} + \frac{t^3}{6}\,R(T,e_\alpha)T|_{c(t)}+\O(\delta^3) \\
	&=te_\alpha|_{c(t)}+ \frac{t^3}{6}\,\epsilon_k g\big(R(T,e_\alpha)T,e_k\big)e_k|_{c(t)}+\O(\delta^3)\,,
\end{align*}
where we performed a Taylor expansion of the integrand in \eqref{Y-eexp} around $c(t)$. Thus
$$e_\alpha|_x=Y_\alpha-\frac{1}{6}\,g\big(R(e_\alpha,T)T,e_\beta)\big)Y_\beta+\O(\delta^3)\,.$$
Next, from~\eqref{YjPi} we conclude that
\begin{align}
	\nabla_{e_\alpha}\Pi^y_x=\:&\Pi^y_x\int_0^1 d\tau\:\tau\Pi^{c(\tau)}_y\rr(T,e_\alpha)|_{c(\tau)}\Pi^y_{c(\tau)} \nonumber\\	&+\frac{1}{6}\,\epsilon_k\,g\big(R(e_\alpha,T)T,e_k)\big)\big|_x\Pi^y_x\int_0^1 d\tau\:\tau\Pi^{c(\tau)}_y\rr(T,e_k)|_{c(\tau)}\, \Pi^y_{c(\tau)} \nonumber\\
	&+\O(\delta^4)\,.
	\label{ejPi}
\end{align}
Representing the Dirac operator as in \eqref{diroprep}, we find 
\begin{align*}
	\dd_x \Pi^y_x=\,&i \Pi^y_x\int_0^1 d\tau\:\tau\Pi^{c(\tau)}_y \epsilon_j e_j\cliff\rr(T,e_j)|_{c(\tau)}\Pi^y_{c(\tau)} \\
	&+\frac{i}{24}\,\epsilon_j \epsilon_p\epsilon_q \Big(\epsilon_k\, g\big(R(e_j,T)T,e_k\big)\big|_x 
	\int_0^1\: t\,  g\big(R(T,e_k)e_p,e_q\big)|_{c(t)}\:dt\Big)e_j\cliff e_p\cliff e_q\cliff \Pi^y_x \\
	&+\O(\delta^4)\:,
\end{align*}
where we used \eqref{rrexplicit} and the fact that the vector fields~$e_j$ are parallel along~$c$. 
The result now follows from the identity
\beq
\epsilon_j e_j \cliff  \rr(e_j,X)\psi=\frac{1}{2}\epsilon_j\Ric(e_j,X)e_j\cliff \psi \qquad\text{for $X\in T_pM$ and $\psi\in \Gamma(M,SM)$}\,, 
\label{rrRic}
\eeq
which is easily verified by applying~\eqref{rrexplicit} as well as the first Bianchi identities.
\QED

We now compute the expansion of the coefficients $V^y_x$ and $\dd_x V^y_x$. The Hadamard recursion relations in~\cite{deWitt, baer+ginoux} yield that the first Hadamard coefficient is given by the formula
\beq \label{hadamardcoeff}
\begin{split}
V^y_{c(t)} &= -\frac{1}{t}\:\vleck(c(t),y)\:\Pi^y_{c(t)} \int_0^t d\tau\:\vleck^{-1}(c(\tau),y) \\
& \qquad \times\:\Pi^{c(\tau)}_y\,
\Big(\Box^\nabla_z +\frac{\s(z)}{4}-m^2\Big) \Big(\vleck(z,y)\: \Pi^y_{z} \Big) \Big|_{z=c(\tau)} \:.
\end{split}
\eeq
Note that this formula remains true if we replace the curve $c$ by the curve $c_s$ as
defined in~\eqref{def-cs}.
For the computation of the term in \eqref{hadamardcoeff} which contains the Bochner Laplacian, it is most convenient to work in local normal coordinates around $y$,
\beq
\Omega\ni p=\exp_y(x_je_j)\,.\label{def-normcoord}
\eeq
The corresponding coordinate vector fields are given by
\beq
X_j:=\frac{\partial}{\partial x_j}\,. 
\label{def-Xa}
\eeq
In these coordinates, the Bochner Laplacian is given by
\beq
\Box^\nabla\Pi^y_{c_s(t)}=-g^{jk}\nabla_{X_j}\nabla_{X_k}\Pi^y_{c_s(t)}+g^{jk}\nabla_{\nabla_{X_j}X_k}\Pi^y_{c_s(t)}\,,
\label{bochner-def}
\eeq
where $g^{jk}$ is the inverse matrix of $g_{jk}=g(X_j,X_k)$.
Moreover, the vector fields~$X_j$ transform according to
\beq X_0=\frac{1}{\delta}T-\frac{s_\alpha}{t\delta}Y_\alpha\qquad\text{and}\qquad X_\alpha=\frac{1}{t}Y_\alpha\,,\label{XYtrafo}
\eeq
where~$T$ and~$Y_\alpha$ are the coordinate vector fields in ~\eqref{def-TY}. Also, from~\eqref{Tealpha}, \eqref{Y-eexp} and \eqref{XYtrafo}, it follows that
\beq
X_j=e_j+\O(\delta^2)\,.\label{Xaea}
\eeq
More precisely, we have the following lemma for the expansion of the metric.
\begin{Lemma}
In the local normal coordinates \eqref{def-normcoord}, the metric~$g$ has the expansion
\beq \label{g-expansion} \begin{split}
g(X_j,X_k)|_{c_s(t)} =\:& \epsilon_j\delta_{jk}-\frac{t^2}{3}\,g\big(R(e_j,T)T,e_k\big)|_{c_s(t)} \\
&+\frac{t^3}{6}\,g\big((\nabla_{T}R)(e_j,T)T,e_k\big)|_{c_s(t)}+\O(\delta^4)\,.
\end{split}
\eeq
\end{Lemma}
\Proof
Inserting \eqref{Y-eexp} into \eqref{XYtrafo}, we find
\begin{align}
	X_\alpha|_{c_s(t)} &= e_\alpha|_{c_s(t)}+\frac{1}{t}\Lambda^y_{c_s(t)}\int_0^t d\tau\:\Lambda^{c_s(\tau)}_y\int_0^\tau d\sigma\:\sigma\,\Lambda^{c_s(\sigma)}_{c_s(\tau)} R(T,e_\alpha)T|_{c_s(\sigma)}+\O(\delta^4) \nonumber\\
	&= e_\alpha|_{c_s(t)}+\frac{1}{t}\Lambda^y_{c_s(t)}\int_0^t d\tau\:\Lambda^{c_s(\tau)}_y\int_0^\tau d\sigma\:\sigma\,\Big(R(T,e_\alpha) T|_{c_s(\tau)} \nonumber\\	&\quad\qquad\qquad+(\sigma-\tau)\nabla_TR(T,e_\alpha)T|_{c_s(\tau)}+\O(T^4)\Big)+\O(\delta^4) \nonumber\\
	&= e_\alpha|_{c_s(t)}+\frac{1}{t}\,\int_0^t d\tau\:\Big(\frac{\tau^2}{2}\,R(T,e_\alpha) T|_{c_s(\tau)} -\frac{\tau^3}{6}\nabla_TR(T,e_\alpha)T|_{c_s(\tau)}\Big)+\O(\delta^4) \nonumber\\
	&=e_\alpha|_{c_s(t)}+\frac{1}{t}\,\int_0^t d\tau\:\Big(\frac{\tau^2}{2}\,R(T,e_\alpha)T|_{c_s(t)} \nonumber\\
	&\quad\qquad\qquad+\frac{\tau^2}{2}(\tau-t)\nabla_TR(T,e_\alpha) T|_{c_s(t)}-\frac{\tau^3}{6}\nabla_TR(T,e_\alpha)T|_{c_s(\tau)}\Big)+\O(\delta^4) \nonumber \\
	&= e_\alpha|_{c_s(t)}+\frac{t^2}{6}\,R(T,e_\alpha)T|_{c_s(t)}-\frac{t^3}{12}(\nabla_TR)(T,e_\alpha)T|_{c_s(t)}+\O(\delta^4)\,,\label{showcase}
\end{align}
where we expanded the integrands in a Taylor series around $c_s(t)$. Moreover, we used
that~$T=\O(\delta)$ and that~$T$ and~$e_j$ are parallel along the curve $c_s$.
Substituting~\eqref{Tealpha} and~\eqref{showcase} into~\eqref{XYtrafo}, we then find
\begin{align}	X_0|_{c_s(t)} &= \frac{1}{\delta}T-\frac{s_\alpha}{t\delta}Y_\alpha=e_0+\frac{s_\alpha}{\delta}e_\alpha|_{c_s(t)}-\frac{s_\alpha}{\delta}X_\alpha|_{c_s(t)} \nonumber\\
&= e_0|_{c_s(t)}-\frac{t^2}{6\delta}R(T,s_\alpha e_\alpha)T|_{c_s(t)}+\frac{t^3}{12\delta}(\nabla_TR)(T,s_\alpha e_\alpha)T|_{c_s(t)}+\O(\delta^4) \nonumber\\
&= e_0-\frac{t^2}{6\delta}R(T,T-\delta e_0)T|_{c_s(t)}+\frac{t^3}{12\delta}(\nabla_TR)(T,T-\delta e_0)|_{c_s(t)}+\O(\delta^4) \nonumber\\
&= e_0|_{c_s(t)}+\frac{t^2}{6}R(T,e_0)T|_{c_s(t)}-\frac{t^3}{12}(\nabla_TR)(T,e_0)T|_{c_s(t)}+\O(\delta^4)\,. \label{X0exp}
\end{align}
Thus, inserting \eqref{showcase} and \eqref{X0exp} into the metric, we obtain
\begin{align*}
g(X_j,X_k) =\:& g(e_j,e_k)+\frac{t^2}{6}\,g\big(e_j,R(T,e_k)T\big)+\frac{1}{6}\,g\big(R(T,e_j)T,e_k\big) \\
&-\frac{t^3}{12}\,g\big(e_j,(\nabla_{T}R)(T,e_k)T\big)-\frac{t^3}{12}\,g\big((\nabla_{T}R)(T,e_j)T,e_k\big)+\O(\delta^4) \\
=\:&\epsilon_j\delta_{jk}-\frac{t^2}{3}\,g\big(R(e_j,T)T,e_k\big)+\frac{t^3}{6}\,g\big((\nabla_{T}R)(e_j,T)T,e_k\big)+\O(\delta^4)\,,
\end{align*}
where the first Bianchi identities were used in the last step.
\QED
We now expand the function~$\vleck$ and related terms in powers of $\delta$.
\begin{Lemma}\label{vleck-lemma} For the square root of the van Vleck-Morette determinant $\vleck$, the following expansions hold:
\begin{align}	\vleck(c_s(t),y) &=1+\frac{t^2}{12}\Ric(T,T)-\frac{t^3}{24}(\nabla_T\Ric)(T,T)
+\O(\delta^4) \label{Vexp} \\
\partial_{Y_\alpha}\vleck|_{c_s(t)} &=\frac{t^2}{6}\Ric(T,e_\alpha)+\O(\delta^2) \label{YVexp} \\
\grad\vleck|_{c_s(t)} &=\frac{t}{6}\epsilon_j\Ric(T,e_j)X_j +\O(\delta^2) \label{gradVexp} \\
\Box\vleck|_{c_s(t)} &=-\frac{\s}{6}+\O(\delta^2) \label{BoxVexp} \,.
\end{align}
\end{Lemma}
\Proof
We first recall the expansion for the matrix determinant
$$\det(\1+A)=1+\tr(A)+\O(A^2)\,.$$
From this identity and \eqref{g-expansion}, we obtain
\begin{align*}	
|\det(g)| =\:& -\det(g) \\
=\:& \det\Big[\1-\frac{t^2}{3}\,\epsilon_j\, g\big(R(e_j,T)T,e_k\big)|_{c_s(t)} \nonumber \\
&+\frac{t^3}{6}\,\epsilon_j\, g\big((\nabla_{T}R)(e_j,T)T,e_k\big)|_{c_s(t)}+\O(\delta^4)\Big] \\
=\:& 1+\tr\Big[-\frac{t^2}{3}\,\epsilon_{j}\,g\big(R(e_j,T)T,e_k\big)+\frac{t^3}{6}\,\epsilon_{j}
\,g\big((\nabla_{T}R)(e_j,T)T,e_k\big)\Big]+\O(\delta^4) \\
=\:& 1-\frac{t^2}{3}\,\Ric(T,T)+\frac{t^3}{6}\,(\nabla_T\Ric)(T,T)+\O(\delta^4)\,.
\end{align*}
Hence
$$\vleck=|\det(g)|^{-\frac{1}{4}}=1+\frac{t^2}{12}\,\Ric(T,T)-\frac{t^3}{24}\,(\nabla_T\Ric)(T,T)+\O(\delta^4)\,,$$
giving \eqref{Vexp}. 
Next, we calculate
$$\partial_{Y_\alpha}\vleck=\frac{t^2}{6}\,\Ric(T,\nabla_{Y_\alpha}T)+\O(\delta^2)=\frac{t^2}{6}\,\Ric(T,\nabla_{T}Y_\alpha)+\O(\delta^2)=\frac{t^2}{6}\,\Ric(T,e_\alpha)+\O(\delta^2)\,,$$
proving~\eqref{YVexp}. Using \eqref{g-expansion} and \eqref{Xaea}, the gradient of $\vleck$ is given by
$$	\grad\vleck=g^{jk}(\partial_{X_j}\vleck)X_k=\epsilon_j (\partial_{X_j}\vleck)X_j+\O(\delta^2)=\epsilon_j \Big(\partial_{X_j}\frac{t^2}{12}\Ric(T,T)\Big)e_j+\O(\delta^2)\,.$$ 
The derivatives with respect to $X_j$ are computed to be
$$\partial_{X_\alpha}\frac{t^2}{12}\Ric(T,T)=\frac{1}{t}\partial_{Y_\alpha}\frac{t^2}{12}\Ric(T,T)=\frac{t}{6}\Ric(T,e_\alpha)+\O(\delta^2)\,,$$
and 
\begin{align}
\partial_{X_0}\frac{t^2}{12}\Ric(T,T)=\:&\Big(\frac{1}{\delta}\nabla_T-\frac{s_\alpha}{t\delta}\nabla_{Y_\alpha}\Big)\frac{t^2}{12}\Ric(T,T) \nonumber\\
=\:&\frac{1}{\delta}\frac{2t}{12}\Ric(T,T)-\frac{s_\alpha}{t\delta}\frac{t^2}{6}\Ric(T,\nabla_{Y_\alpha}T)+\O(\delta^2) \nonumber\\
=\:&\frac{t}{6}\Ric\Big(T,e_0+\frac{s_\alpha}{\delta}e_\alpha\Big)-\frac{t}{6}\Ric\Big(T,\frac{s_\alpha}{\delta}e_\alpha\Big)+\O(\delta^2) \nonumber\\
=\:&\frac{t}{6}\Ric(T,e_0)+\O(\delta^2)\,,
\label{X0Ric}
\end{align}
where we used \eqref{Tealpha}, \eqref{XYtrafo} and \eqref{YVexp}. Thus
$$\grad\vleck=\frac{t}{6}\,\epsilon_j\, \Ric(T,e_j)X_j+\O(\delta^2)\,,$$
which shows~\eqref{gradVexp}.
Using that $g_{jk}=\epsilon_j \delta_{jk}+\O(\delta^2)$, we find
\begin{align*}	\Box\vleck=&-\frac{1}{\sqrt{|\det(g)|}}\,\partial_{X_j}\Big(\sqrt{|\det(g)|}\,g^{jk}\partial_{X_k}\vleck\Big) \\
=&-\epsilon_j\partial_{X_j}\partial_{X_j}\Big(1+\frac{t^2}{12}\,\Ric(T,T)-\frac{t^3}{24}\,(\nabla_T\Ric)(T,T)\Big)+\O(\delta^2) \:.
\end{align*}
The spatial derivatives in this formula are calculated by
\begin{align*}	
\partial_{X_\alpha}\partial_{X_\alpha}\vleck=&\frac{1}{t^2}\partial_{Y_\alpha}\partial_{Y_\alpha}\Big(1+\frac{t^2}{12}\,\Ric(T,T)-\frac{t^3}{24}\,(\nabla_T\Ric)(T,T)\Big)+\O(\delta^2) \\
=\:&\partial_{Y_\alpha}\Big(\frac{1}{6}\,\Ric(T,e_\alpha)+
\frac{t}{12}\,(\nabla_{e_\alpha}\Ric)(T,T)
-\frac{t}{24}\,\nabla_{Y_\alpha}\nabla_T\Ric(T,T)\Big)+\O(\delta^2) \\
=\:&\partial_{Y_\alpha}\Big(\frac{1}{6}\,\Ric(T,e_\alpha)+
\frac{t}{12}\,(\nabla_{e_\alpha}\Ric)(T,T)
-\frac{t}{24}\,\nabla_{T}\nabla_{Y_\alpha}\Ric(T,T)\Big)+\O(\delta^2) \\
=\:&\partial_{Y_\alpha}\Big(\frac{1}{6}\,\Ric(T,e_\alpha)+
\frac{t}{24}\,(\nabla_{e_\alpha}\Ric)(T,T)
-\frac{t}{12}\,(\nabla_{T}\Ric)(T,e_\alpha)\Big)+\O(\delta^2) \\
=\:&\frac{1}{6}\,\Ric(e_\alpha,e_\alpha)+
\frac{t}{4}\,(\nabla_{e_\alpha}\Ric)(T,e_\alpha)
-\frac{t}{12}\,\nabla_{Y_\alpha}\nabla_{T}\Ric(T,e_\alpha)+\O(\delta^2) \\
=\:&\frac{1}{6}\,\Ric(e_\alpha,e_\alpha)+
\frac{t}{6}\,(\nabla_{e_\alpha}\Ric)(T,e_\alpha)
-\frac{t}{12}\,(\nabla_{T}\Ric)(e_\alpha,e_\alpha)+\O(\delta^2) \,.
\end{align*}
The derivatives with respect to $X_0$ are calculated similar to \eqref{X0Ric} and give
$$
 \partial_{X_0}\partial_{X_0}\vleck=\frac{1}{6}\,\Ric(e_0,e_0)+
\frac{t}{6}\,(\nabla_{e_0}\Ric)(T,e_0)
-\frac{t}{12}\,(\nabla_{T}\Ric)(e_0,e_0)+\O(\delta^2)\,.
$$
We thus obtain
\begin{align*}
\Box\vleck
=\:&-\epsilon_j\Big(\frac{1}{6}\,\Ric(e_j,e_j)+
\frac{t}{6}\,(\nabla_{e_j}\Ric)(T,e_j)
-\frac{t}{12}\,(\nabla_{T}\Ric)(e_j,e_j) \Big)+\O(\delta^2) \\
=\:&-\frac{\s}{6}-\frac{t}{6}\,\dvg(\Ric)(T)+\frac{t}{12}\partial_T\s+\O(\delta^2)
= -\frac{\s}{6}+\O(\delta^2)\:,
\end{align*}
where in the last step we used the second Bianchi identities. 
\QED

We next derive the expansion of the Hadamard coefficient~$V^y_x$.
\begin{Lemma}\label{lemma-V} 
The Hadamard coefficient $V^y_x$ has the expansion 
\begin{align}
V^y_x =\:& m^2\,\Pi^y_x-\frac{\s}{12}\,\Pi^y_x+\frac{\partial_T \s}{24}\,\Pi^y_x \label{Vyx1} \\
&+\frac{\epsilon_j\epsilon_k\epsilon_l}{24}\,g\big((\nabla_{e_j}R)(T,e_j)e_k,e_l\big)e_k\cliff e_l\cliff \Pi^y_x \label{Vyx2}\\
&+\delta^2\, v^s\,\Pi^y_x+\delta^2\, v^b_{jk}\,e_j\cliff e_k\cliff \Pi^y_x+\delta^2\,v^p\,e_0\cliff e_1\cliff e_2\cliff e_3\cliff \Pi^y_x+\O(\delta^3)\,, \label{Vyx3}
\end{align}
where the coefficients $v^s$, $v^b_{jk}$ and $v^p$ are real-valued functions.
\end{Lemma}
\Proof
As $V^y_x$ is a Hadamard coefficient of the second-order equation \eqref{kg-globhyp}, all contributions to $V^y_x$ involve an even number of Clifford multiplications and only real-valued functions. As a consequence, the
higher order terms can be written in the general form \eqref{Vyx3}. In order to calculate the leading terms, we
note that inserting the expansion $g^{jk}=\epsilon_j\delta_{jk}+\O(\delta^2)$ into the definition of the Bochner Laplacian \eqref{bochner-def} yields
\begin{align}
\Box^\nabla\Pi^y_{c(t)} &=-\nabla_{X_0}\nabla_{X_0}\Pi^y_{x}+\nabla_{X_\alpha}\nabla_{X_\alpha}\Pi^y_{x}+\O(\delta^2)=\frac{1}{t^2}\nabla_{Y_\alpha}\nabla_{Y_\alpha}\Pi^y_{c(t)}+\O(\delta^2) \nonumber\\
&= -\frac{1}{4}\:\Pi^y_{c(t)}\int_0^t d\tau\:\frac{\tau^2}{t^2}\,\Pi^{c(\tau)}_y \epsilon_j\epsilon_k\epsilon_l\,g\big((\nabla_{e_j}R)(T,e_j) \,e_k,e_l\big)e_k\cliff e_l\cliff \Pi^y_{c(\tau)}
	+\O(\delta^2) \nonumber\\
&=-\frac{t}{12}\:\epsilon_j\epsilon_k\epsilon_l\,g\big((\nabla_{e_j}R)(T,e_j)e_k,e_l\big)e_k\cliff e_l\cliff \Pi^y_{c(t)}+\O(\delta^2)\,, \label{BoxPi}
\end{align}
where we used formulas~\eqref{eaPi}, \eqref{eaeb}, \eqref{rrexplicit}, \eqref{YjPi} and a Taylor expansion of the integrand around $c(t)$.
Inserting into the definition of $V^y_x$ (see formula \eqref{hadamardcoeff}), we obtain
\begin{align*}
	V^y_{x} =\:& -\vleck(x,y)\:\Pi^y_x \int_0^1 d\tau\:\vleck^{-1}(c(\tau),y) \:\Pi^{c(\tau)}_y\,
\Big(\Box^\nabla +\frac{\s}{4}-m^2\Big) \big(\vleck\Pi^y_{c(\tau)} \big) \\
=\:& -\Pi^y_x \int_0^1 d\tau\:\Pi^{c(\tau)}_y\,
\Big(\Box\vleck +\frac{\s}{4}-m^2\Big) \Pi^y_{c(\tau)} \\
&+\Pi^y_x \int_0^1 d\tau\:\Pi^{c(\tau)}_y\,
\Big(2\nabla_{\grad\vleck}-\Box^\nabla\Big)\Pi^y_{c(\tau)}+\O(\delta^2) \\
=\:&-\Pi^y_x \int_0^1 d\tau\:\Pi^{c(\tau)}_y\,
\Big(\frac{\s}{12}-m^2\Big) \Pi^y_{c(\tau)} \\
&+\Pi^y_x \int_0^1 d\tau\:\Pi^{c(\tau)}_y\,
\frac{\tau}{12}\,\epsilon_j\epsilon_k\epsilon_l\,g\big((\nabla_{e_j}R)(T,e_j)e_k,e_l\big)e_k\cliff e_l\cliff \Pi^y_{c(\tau)}+\O(\delta^2) \\
=\:&\Big(m^2-\frac{\s}{12}+\frac{\partial_T\s}{24}\Big) \Pi^y_x \\
&+\frac{\epsilon_j\epsilon_k\epsilon_l}{24}\,g\big((\nabla_{e_j}R)(T,e_j)e_k,e_l\big)e_k\cliff e_l\cliff \Pi^y_{x}+\O(\delta^2)\,,
\end{align*}
where we used \eqref{Vexp}, \eqref{BoxVexp}, \eqref{eaPi}, \eqref{gradVexp}, \eqref{BoxPi} and
again performed a Taylor expansion of the integrands around $x$.
\QED

The expansion of the Hadamard coefficient $\dd_x V^y_x$ is given in the next lemma.
\begin{Lemma}\label{lemma-DV}
The Hadamard coefficient~$\dd_x V^y_x$ has the expansion
\[ \dd_x V^y_x = i\,\delta\,d^v_j\,e_j\cliff \Pi^y_x+i\,\delta\,d^a_{jkl}\,e_j\cliff e_k\cliff e_l\cliff \Pi^y_x + \O(\delta^2)\,, \]
where the coefficients $d^v_j$ and $d^a_{jkl}$ are real-valued functions.
\end{Lemma}
\Proof We apply the Dirac operator
to the expansion of Lemma~\ref{lemma-V}. The derivatives of the factors~$\Pi^y_x$
can be computed using Lemma~\ref{lemma-DPi} and~\eqref{ejPi}, giving contributions of the form
\beq \label{term1}
\dd_x V^y_x \asymp
i\,\delta\,d^v_j\,e_j\cliff \Pi^y_x+i\,\delta\,d^a_{jkl}e_j\cliff e_k\cliff e_l\cliff \Pi^y_x + \O(\delta^2) \:.
\eeq
If the derivative acts on the scalar curvature in the second summand in~\eqref{Vyx1}
or on the factor~$T$ in the last summand in~\eqref{Vyx1},
the resulting terms can be rewritten using the second Bianchi identities as
\beq \label{term2}
\dd_x V^y_x \asymp
-i\frac{\epsilon_j}{12}\,\dvg(\Ric)(e_j)e_j\cliff \Pi^y_x \:.
\eeq
On the other hand, if the derivative acts on the scalar curvature in the last summand in~\eqref{Vyx1},
we get terms of the form~\eqref{term1}.
Similarly, if the Riemann tensor in~\eqref{Vyx2} is differentiated, we again get terms of the form~\eqref{term1}.
Moreover, differentiating the factor~$T$ in~\eqref{Vyx2} gives the contribution
\beq \label{term3}
\dd_x V^y_x \asymp
i\frac{\epsilon_j}{12}\,\dvg(\Ric)(e_j)e_j\cliff \Pi^y_x \:,
\eeq
which cancels against the term \eqref{term2}.
Finally, we need to be concerned about differentiating the error term in~\eqref{Vyx2}.
Noting that all the contributions to~$V^y_x$ involve an even number of Clifford multiplications
and only real-valued functions,
applying the Dirac operator obviously gives terms of the form~\eqref{term1}.
\QED
The last relevant Hadamard coefficients~$W^y_x$ and~$H^y_x$ can be expanded as follows.
\begin{Lemma}\label{lemma-WH}
The Hadamard coefficients $W^y_x$ and $H^y_x$ have the expansion
\begin{align}
W^y_x&=w^s\,\Pi^y_x + w^b_{jk}\,e_j\cliff e_k\cliff \Pi^y_x + w^p\,e_0\cliff e_1\cliff e_2\cliff e_3\cliff \Pi^y_x + \O(\delta)
\label{W-exp} \\
H^y_x&=h^s\,\Pi^y_x + h^b_{jk}\,e_j\cliff e_k\cliff \Pi^y_x + h^p\,e_0\cliff e_1\cliff e_2\cliff e_3\cliff \Pi^y_x + \O(\delta)\,,
\label{H-exp}
\end{align}
where all coefficients are real-valued functions.
\end{Lemma}
\Proof
As $W^y_x$ and $H^y_x$ are Hadamard coefficients of the second-order equation \eqref{kg-globhyp}, all contributions to $W^y_x$ and $H^y_x$ involve an even number of Clifford multiplications and only real-valued functions. Thus, $W^y_x$ and $H^y_x$ can be written in the form \eqref{W-exp} and \eqref{H-exp}, respectively.
\QED

We now come to the proof of the propositions stated in Section~\ref{secPcurve}.
\Proof[Proof of Proposition~\ref{prp58}]
We rewrite the results of Lemmas \ref{lemma-DPi}, \ref{lemma-V}, \ref{lemma-DV} and \ref{lemma-WH} in the form
\begin{align}
	\dd_x\Pi^y_x=\:&i\,\delta\,c^s_j\,e_j\cliff \Pi^y_x+i\,\delta^3\,c^a_{jkl}\,e_j\cliff e_k\cliff e_l\cliff \Pi^y_x+\O(\delta^4) \nonumber\\
	V^y_x=\:&v^s\,\Pi^y_x +\delta\,\tilde{v}_{kl}\,e_k\cliff e_l\cliff \Pi^y_x +\delta^2\, v^b_{jk}\,e_j\cliff e_k\cliff \Pi^y_x+\delta^2\,v^p\,e_0\cliff e_1\cliff e_2\cliff e_3\cliff \Pi^y_x+\O(\delta^3) \nonumber\\
	\dd_x V^y_x=\:&i\,\delta\,d^v_j\,e_j\cliff \Pi^y_x+i\,\delta\,d^a_{jkl}\,e_j\cliff e_k\cliff e_l\cliff \Pi^y_x+\O(\delta^2) \nonumber\\
	W^y_x=\:&w^s\,\Pi^y_x + w^b_{jk}\,e_j\cliff e_k\cliff \Pi^y_x + w^p\,e_0\cliff e_1\cliff e_2\cliff e_3\cliff \Pi^y_x +\O(\delta) \nonumber\\
	H^y_x=\:&h^s\,\Pi^y_x + h^b_{jk}\,e_j\cliff e_k\cliff \Pi^y_x + h^p\,e_0\cliff e_1\cliff e_2\cliff e_3\cliff \Pi^y_x +\O(\delta) \,,
	\label{coeff-scheme}
\end{align}
where all coefficients are real-valued functions. 
Here each factor~$\delta$ corresponds to a factor~$T$ in the resulting explicit formulas (for details see~\cite{drgrotz}).
The coefficients $v^s$ and $\tilde{v}_{jk}$ are given by
\beq \label{def-vtilde}
v^s=m^2-\frac{\s}{12}+\delta\,\tilde{v}^s\qquad\text{and}\qquad \tilde{v}_{kl}=\frac{1}{\delta} \frac{\epsilon_j\epsilon_k\epsilon_l}{24}\,g\big((\nabla_{e_j}R)(T,e_j)e_k,e_l\big)\,,
\eeq
where $\tilde{v}^s$ is a real-valued function.
Inserting this formulas into the Hadamard expansion \eqref{Hadaexp-P}, we find
\begin{align}
	(-8&\pi^3)\,P(x,y)=-\frac{i\vleck}{\Gamma^2}\grad_x\Gamma\cliff \Pi^y_x+\frac{i}{\Gamma}\grad_x\vleck\cliff\: \Pi^y_x \nonumber\\	&+\frac{\vleck}{\Gamma}\Big(m+i\,\delta\,c^s_j\,e_j+i\,\delta^3\,c^a_{jkl}\,e_j\cliff e_k\cliff e_l\Big)\cliff \Pi^y_x \nonumber\\
&+\frac{i}{4\Gamma}\grad_x\Gamma\cliff\Big(v^s +\delta\,\tilde{v}_{kl}\,e_k\cliff e_l +\delta^2\, v^b_{jk}\,e_j\cliff e_k+\delta^2\,v^p\,e_0\cliff e_1\cliff e_2\cliff e_3\Big)\cliff \Pi^y_x \nonumber\\
&+\frac{m}{4}\big[\log|\Gamma|-i\pi\,\epsilon\big(\mathfrak{t}(x)-\mathfrak{t}(y) \big)\big]\Big(v^s +\delta\,\tilde{v}_{kl}\,e_k\cliff e_l\Big)\cliff \Pi^y_x \nonumber\\
&+\frac{1}{4}\big[\log|\Gamma|-i\pi\,\epsilon\big(\mathfrak{t}(x)-\mathfrak{t}(y) \big)\big]\Big(i\,\delta\,d^v_j\,e_j+i\,\delta\,d^a_{jkl}\,e_j\cliff e_k\cliff e_l\Big)\cliff \Pi^y_x \nonumber\\
&+i\big[1+\log|\Gamma|-i\pi\,\epsilon\big(\mathfrak{t}(x)-\mathfrak{t}(y) \big)\big]\grad_x\Gamma\cliff\Big(w^s + w^b_{jk}\,e_j\cliff e_k + w^p\,e_0\cliff e_1\cliff e_2\cliff e_3\Big)\cliff \Pi^y_x	\nonumber\\	&+i\,\grad_x\Gamma\cliff\Big(h^s + h^b_{jk}\,e_j\cliff e_k + h^p\,e_0\cliff e_1\cliff e_2\cliff e_3\Big)\cliff \Pi^y_x+\O(\delta^2\log\delta)\,. \label{Pexp1}
\end{align}
The Clifford relations immediately yield the identities
\begin{align*}
	\grad_x\Gamma\cliff \big(f_{jk}\,e_j\cliff e_k\big) &=\delta\,f_j\,e_j+\delta\,f_{jkl}\,e_j\cliff e_k\cliff e_l \\
	\grad_x\Gamma\cliff \big(f\,e_0\cliff e_1\cliff e_2\cliff e_3\big) &=\delta\,f_{jkl}\,e_j\cliff e_k\cliff e_l\,,
\end{align*}
where all coefficients are real-valued functions.
Using these identities in~\eqref{Pexp1} and combining terms which are of the same order in $\delta$
and contain the same number of Clifford multiplications, we obtain
\begin{align}
	(-8\pi^3)\,&P(x,y)=-\frac{i}{\Gamma^2}\grad_x\Gamma\cliff \Pi^y_x+\frac{m}{\Gamma}\,\Pi^y_x+\frac{i}{\Gamma}\,\delta\, p^{(1)}_j\,e_j\cliff \Pi^y_x+m\log|\Gamma|\,v^s\,\Pi^y_x \label{P1}\\	
	&+\frac{i}{4\Gamma}\,\delta\,\tilde{v}_{kl}\,\grad_x\Gamma\cliff e_k\cliff e_l\cliff \Pi^y_x -\frac{i\pi}{4}\,\epsilon\big(\mathfrak{t}(x)-\mathfrak{t}(y)\big) m\,v^s\,\Pi^y_x \label{P2}\\
	&+\log|\Gamma|\,\delta \Big(i\,p^{(2)}_j\,e_j+\frac{m}{4}\,\tilde{v}_{kl}\,e_k\cliff e_l+i\,p^{(3)}_{jkl}\,e_j\cliff e_k\cliff e_l\Big)\cliff \Pi^y_x \label{P3}\\	&+\delta\,\pi\,\epsilon\big(\mathfrak{t}(x)-\mathfrak{t}(y)\big)\Big(\,p^{(4)}_j\,e_j-\frac{i\,m}{4}\,\tilde{v}_{kl}\,e_k\cliff e_l+\,p^{(5)}_{jkl}\,e_j\cliff e_k\cliff e_l\Big)\cliff \Pi^y_x \label{P4}\\
	&+\delta \Big(i\,p^{(6)}_j\,e_j+i\,p^{(7)}_{jkl}\,e_j\cliff e_k\cliff e_l\Big)\cliff \Pi^y_x +\O(\delta^2\log\delta)\,. \label{P5}
\end{align}
Here all coefficients $p^{(1)}_j,\ldots,p^{(7)}_{jkl}$ are real-valued functions of the order $\O(\delta^0)$.
Using the Clifford relations,
the composition of three Clifford multiplications can be rewritten as
as vector and axial components,
\beq \label{3gamma}
e_j\cliff e_k\cliff e_l = \big( g_{jk} e_l + 
g_{kl} e_j - g_{jl} e_k \big) + i \epsilon_n\,
\epsilon_{jkln}\: e_5 \cliff e_n 
\eeq
(where~$\epsilon_{jkln}$ is the totally anti-symmetric tensor,
and~$e_5 = i e_0 e_1 e_2 e_3$ denotes the pseudoscalar matrix;
see~\cite[Appendix~A]{bjorken}). 
Thus, in \eqref{P3}, \eqref{P4} and \eqref{P5} the resulting vector components can be combined with the corresponding vector components in these lines.
The resulting axial component in \eqref{P3} can be written as
\beq \label{Pxyv}
(-8 \pi^3)\: P(x,y) \asymp \log|\Gamma| \:\delta \: a_j \:e_5 \cliff e_j \cliff \Pi^y_x
\eeq
with real coefficients~$a_j$. Moreover, from~\eqref{Hadaexp-P} and the previous calculations
one sees that there is no other contribution to~$P(x,y)$ of this form.
As the expression~$\delta \: a_j \:e_5 \cliff\, e_j$ is linear in~$\delta$ and smooth
in~$x$ and~$y$, it is
odd under permutations of~$x$ and~$y$ (this can also be understood from the fact
that the linear factor~$\delta$ corresponds to a factor~$T$ in the resulting explicit formulas;
for details see~\cite{drgrotz}). Also using the identity~$(e_5 \cliff e_j)^* = e_5 \cliff e_j$, we obtain
\[ (-8 \pi^3)\: P(x,y) = (-8 \pi^3)\: P(y,x)^* = -\log(\Gamma) \:\delta \: a_j \:e_5 \cliff e_j \cliff \Pi^y_x \:. \]
Comparing with~\eqref{Pxyv}, we conclude that the coefficients~$a_j$ vanish.
For the same reason, the term in \eqref{P5} containing three Clifford multiplications reduces to a vectorial contribution. Finally, the axial contribution in \eqref{P4} resulting from the decomposition \eqref{3gamma} can be written in the form
\beq \label{Pxyv2}
(-8 \pi^3)\: P(x,y) \asymp i\,\delta\,\epsilon\big(\mathfrak{t}(x)-\mathfrak{t}(y)\big) \: a_j \:e_5 \cliff e_j \cliff \Pi^y_x
\eeq
with real coefficients~$a_j$, and from~\eqref{Hadaexp-P} and the previous calculations one sees that there is no other
contribution to~$P(x,y)$ of this form. However, the term \eqref{Pxyv2} is odd under conjugation but even
when interchanging~$x$ and~$y$. Therefore, we conclude that the coefficients~$a_j$ vanish.
We thus obtain the following expansion of the kernel of the fermionic operator,
\begin{align}
	(-8\pi^3)\,&P(x,y)=-\frac{i}{\Gamma^2}\grad_x\Gamma\cliff \Pi^y_x+\frac{m}{\Gamma}\,\Pi^y_x+\frac{i}{\Gamma}\,\delta\, p^{(1)}_j\,e_j\cliff \Pi^y_x+\frac{m}{4}\log|\Gamma|\,v^s\,\Pi^y_x \nonumber\\	&+\frac{i}{4\Gamma}\,\delta\,\tilde{v}_{kl}\,\grad_x\Gamma\cliff e_k\cliff e_l\cliff \Pi^y_x -\frac{i\pi}{4}\,\epsilon\big(\mathfrak{t}(x)-\mathfrak{t}(y)\big) m\,v^s\,\Pi^y_x \nonumber\\
	&+\log|\Gamma|\,\delta \Big(i\,\tilde{p}^{(2)}_j\,e_j+\frac{m}{4}\,\tilde{v}_{kl}\,e_k\cliff e_l\Big)\cliff \Pi^y_x \nonumber\\	&+\delta\,\pi\,\epsilon\big(\mathfrak{t}(x)-\mathfrak{t}(y)\big)\Big(\,\tilde{p}^{(4)}_j\,e_j-\frac{i\,m}{4}\,\tilde{v}_{kl}\,e_k\cliff e_l\Big)\cliff \Pi^y_x +i\,\delta\,\tilde{p}^{(6)}_j\,e_j\cliff \Pi^y_x \nonumber\\
	&+\O(\delta^2\log\delta)\,,
	\label{kernel-exp}
\end{align}
where $\tilde{p}^{(2)}_j$, $\tilde{p}^{(4)}_j$ and $\tilde{p}^{(6)}_j$ are real-valued functions.
The first two terms in this expansion show that Proposition \ref{prp58} holds. The other terms will be needed to calculate the expansion of the closed chain.
\QED

\Proof[Proof of Proposition~\ref{prp59}]
Using the expansion \eqref{kernel-exp}, we compute
\begin{align*}
(-8\pi^3)^2 A_{yx}=&\:(-8\pi^3)^2 P(x,y)^*\,P(x,y) \\ =&\:c(x,y)\,\1_{S_yM}+\frac{\pi\,m}{2\Gamma^2}\,v^s\,\epsilon\big(\mathfrak{t}(x)-\mathfrak{t}(y)\big)\,\Pi^y_x\,\grad_x\Gamma\cliff\Pi^y_x \\
&+\frac{im}{4\Gamma^2}\,(\delta+\delta\log|\Gamma|)\,\Pi^y_x\,\big\{\grad_x\Gamma,\tilde{v}_{kl}\,e_k\cliff e_l\big\}\cliff\Pi^y_x \\
&+\frac{i\pi}{\Gamma^2}\,\delta\,\epsilon\big(\mathfrak{t}(x)-\mathfrak{t}(y)\big)\,\Pi^y_x\,\big[\grad_x\Gamma,\tilde{p}^{(4)}_j\,e_j\big]\cliff\Pi^y_x \\
&+\frac{m\pi}{4\Gamma^2}\,\delta\,\epsilon\big(\mathfrak{t}(x)-\mathfrak{t}(y)\big)\,\Pi^y_x\,\big[\grad_x\Gamma,\tilde{v}_{kl}\,e_k\cliff e_l\big]\cliff\Pi^y_x +\O(\delta^{-1}\log\delta) \\	
=&\:c(x,y)\,\1_{S_yM} \\
&-m\Big(m^2-\frac{\s}{12}+\delta\,\tilde{v}^s\Big)\,\frac{\pi\,\epsilon\big(\mathfrak{t}(x)-\mathfrak{t}(y)\big)}{2\Gamma^2}\,\grad_y\Gamma\cliff \1_{S_yM} \\
&-\frac{m}{4\Gamma^2}\,(\delta+\delta\log|\Gamma|)\,\big\{\grad_x\Gamma,i\tilde{v}_{kl}\,e_k\cliff e_l\big\}\cliff\1_{S_yM} \\
&-\frac{i\pi}{\Gamma^2}\,\delta\,\epsilon\big(\mathfrak{t}(x)-\mathfrak{t}(y)\big)\,\big[\grad_x\Gamma,\tilde{p}^{(4)}_j\,e_j\big]\cliff\1_{S_yM} \\
&+i\frac{m\pi}{4\Gamma^2}\,\delta\,\epsilon\big(\mathfrak{t}(x)-\mathfrak{t}(y)\big)\,\big[\grad_x\Gamma,i\tilde{v}_{kl}\,e_k\cliff e_l\big]\cliff\1_{S_yM} +\O(\delta^{-1}\log\delta)\:,
\end{align*} 
where we used that~$\tilde{v}_{kl}=-\tilde{v}_{lk}$ according to~\eqref{def-vtilde}. Moreover, we used the
formula for~$v^s$ in \eqref{def-vtilde} as well as the identities
$$\Pi^x_y\,e_j\cliff \Pi^y_x=e_j\cliff \1_{S_yM}\qquad\text{and}\qquad\Pi^x_y\,\grad_x\Gamma\cliff \Pi^y_x=-\grad_y\Gamma\cliff \1_{S_yM}\,.$$
Now the operators $X_{yx}$ and $Y_{yx}$ defined by
\begin{align}
X_{yx}:=&\:\frac{\pi}{\Gamma^2}\,\delta\,\epsilon\big(\mathfrak{t}(x)-\mathfrak{t}(y)\big)\,\Big(\frac{i\,m}{4}\,\tilde{v}_{kl}\,e_k\cliff e_l-\tilde{p}^{(4)}_j\,e_j \Big) \label{Xex}\\
Y_{yx}:=&\:-\frac{m}{4\Gamma^2}\Big((\delta+\delta\log|\Gamma|)\,i\tilde{v}_{kl}\,e_k\cliff e_l 
-\delta\,\pi\,\tilde{v}^s\,\epsilon\big(\mathfrak{t}(x)-\mathfrak{t}(y)\big)\Big) \label{Yex}
\end{align}
are obviously symmetric, linear operators on $S_yM$. Moreover, $X_{yx}$ is of the
order~$\O(\delta^{-3}\log\delta)$, whereas~$Y_{yx}$ is of the order $\O(\delta^{-3})$.
Interchanging~$x$ and~$y$ completes the proof.
\QED

We finally prove the expansions \eqref{Deluex} and \eqref{betaex} in Lemma~\ref{lemma511}.
\Proof[Proof of Lemma~\ref{lemma511}]
From \eqref{Ayx1}, we conclude that the coefficient $a_{xy}$ in \eqref{A0rep} is given by
$$a_{xy}= m \Big( m^2 - \frac{\s}{12} \Big)\,\frac{\Ima(\log\Gamma)}{2\Gamma^2}\,.$$
Thus we obtain from \eqref{Xex} and \eqref{def-vtilde} that the operator $Z_{xy}$ in \eqref{Zdef} is given by
\beq \label{Zex}
Z_{xy}=\frac{1}{2}\Big(m^2-\frac{\s}{12}\Big)^{-1}\Big[\frac{\epsilon_j \epsilon_k \epsilon_l}{12}\,g\big((\nabla_{e_j}R)(T,e_j)e_k,e_l\big)\,\frac{i}{2}[e_k,e_l]-\frac{4\delta}{m}\tilde{p}^{(4)}_j\,e_j\Big]\,.
\eeq
Moreover, since $x$ lies in the future of $y$ and $\grad_x\Gamma$ is normalized according to \eqref{gradG}, the future-directed timelike unit vector $u$ introduced after \eqref{Zdef} is given by
$$u=\frac{1}{2\delta}\,\grad_x\Gamma=\frac{1}{\delta}\,T\,.$$
Therefore, the vector $\Delta u$ introduced in \eqref{def-Delu} is given by
$$\Delta u=\frac{1}{6\,\delta}\Big(m^2-\frac{\s}{12}\Big)^{-1} \epsilon_j (\nabla_{e_j}R)(T,e_j)T\,,$$
proving \eqref{Deluex}.

The operator $w$ in \eqref{vrep1} is given by the vectorial part of \eqref{Zex}, i.e.
\beq \label{wex}
w=-\frac{2}{m}\Big(m^2-\frac{\s}{12}\Big)^{-1} \delta\,\tilde{p}^{(4)}_j\,e_j\,.
\eeq
A short review of the proofs of Propositions \ref{prp58} and \ref{prp59} yields that the functions $\tilde{p}^{(4)}_j$ are combinations of the real-valued functions appearing in the expansion of the Hadamard coefficients $\dd_x V^y_x$, $W^y_x$ and $H^y_x$ in \eqref{coeff-scheme}.
These functions are calculated explicitly in \cite{drgrotz}. They are of the order
$$\O \Big( \frac{m^2}{\delta}\, \|\epsilon_j\,\Ric(T, e_j)\, e_j\| \Big)+\O \big(\|R\|^2 + \|\nabla^2 R\| \big) \,.$$
Inserting into formula \eqref{wex}, we conclude that $w$ is of the order
$$\Bigg[ \O \Big( \frac{1}{m}\, \|\epsilon_j\,\Ric(T, e_j)\, e_j\| \Big) 
+ \O \bigg( \frac{\delta}{m^3}\: \big(\|R\|^2 + \|\nabla^2 R\| \big) \bigg) \Bigg] \Big(1 + \O \Big( \frac{\s}{m^2} \Big) \Big)\,.$$
Now \eqref{betaex} follows immediately from the representation \eqref{wrep}.
\QED

%\bibliographystyle{amsplain}
%\bibliography{../felix}

\begin{thebibliography}{10}

\bibitem{AS}
M.~Abramowitz and I.A. Stegun, \emph{Handbook of {M}athematical {F}unctions
  with {F}ormulas, {G}raphs, and {M}athematical {T}ables}, National Bureau of
  Standards Applied Mathematics Series, vol.~55, U.S. Government Printing
  Office, Washington, D.C., 1964.

\bibitem{baer+ginoux}
C.~B{\"a}r, N.~Ginoux, and F.~Pf{\"a}ffle, \emph{Wave {E}quations on
  {L}orentzian {M}anifolds and {Q}uantization}, ESI Lectures in Mathematics and
  Physics, European Mathematical Society (EMS), Z\"urich, 2007.

\bibitem{baum}
H.~Baum, \emph{Spinor structures and {D}irac operators on pseudo-{R}iemannian
  manifolds}, Bull. Polish Acad. Sci. Math. \textbf{33} (1985), no.~3-4,
  165--171.

\bibitem{becker+schwarz}
K.~Becker, M.~Becker, and J.H. Schwarz, \emph{String {T}heory and
  {M}-{T}heory}, Cambridge University Press, Cambridge, 2007.

\bibitem{bernal+sanchez}
A.N. Bernal and M.~S{\'a}nchez, \emph{On smooth {C}auchy hypersurfaces and
  {G}eroch's splitting theorem}, Comm. Math. Phys. \textbf{243} (2003), no.~3,
  461--470.

\bibitem{bjorken}
J.D. Bjorken and S.D. Drell, \emph{Relativistic {Q}uantum {M}echanics},
  McGraw-Hill Book Co., New York, 1964.

\bibitem{sorkin}
L.~Bombelli, J.~Lee, D.~Meyer, and R.D. Sorkin, \emph{Space-time as a causal
  set}, Phys. Rev. Lett. \textbf{59} (1987), no.~5, 521--524.

\bibitem{connes}
A.~Connes, \emph{Noncommutative {G}eometry}, Academic Press Inc., San Diego,
  CA, 1994.

\bibitem{deWitt}
B.S. DeWitt and R.W. Brehme, \emph{Radiation damping in a gravitational field},
  Ann. Physics \textbf{9} (1960), 220--259.

\bibitem{U22}
F.~Finster, \emph{Local {$\rm U(2,2)$} symmetry in relativistic quantum
  mechanics}, arXiv:hep-th/9703083, J. Math. Phys. \textbf{39} (1998), no.~12,
  6276--6290.

\bibitem{firstorder}
\bysame, \emph{Light-cone expansion of the {D}irac sea to first order in the
  external potential}, arXiv:hep-th/9707128, Michigan Math. J. \textbf{46}
  (1999), no.~2, 377--408.

\bibitem{PFP}
\bysame, \emph{The {P}rinciple of the {F}ermionic {P}rojector}, hep-th/0001048,
  hep-th/0202059, hep-th/0210121, AMS/IP Studies in Advanced Mathematics,
  vol.~35, American Mathematical Society, Providence, RI, 2006.

\bibitem{osymm}
\bysame, \emph{Fermion systems in discrete space-time---outer symmetries and
  spontaneous symmetry breaking}, arXiv:math-ph/0601039, Adv. Theor. Math.
  Phys. \textbf{11} (2007), no.~1, 91--146.

\bibitem{discrete}
\bysame, \emph{A variational principle in discrete space-time: Existence of
  minimizers}, arXiv:math-ph/0503069, Calc. Var. Partial Differential Equations
  \textbf{29} (2007), no.~4, 431--453.

\bibitem{sector}
\bysame, \emph{An action principle for an interacting fermion system and its
  analysis in the continuum limit}, arXiv:0908.1542 [math-ph] (2009).

\bibitem{lrev}
\bysame, \emph{From discrete space-time to {M}inkowski space: Basic mechanisms,
  methods and perspectives}, arXiv:0712.0685 [math-ph], Quantum {F}ield
  {T}heory (B.~Fauser, J.~Tolksdorf, and E.~Zeidler, eds.), Birkh\"auser
  Verlag, 2009, pp.~235--259.

\bibitem{continuum}
\bysame, \emph{Causal variational principles on measure spaces},
  arXiv:0811.2666 [math-ph], J. Reine Angew. Math. \textbf{646} (2010),
  141--194.

\bibitem{srev}
\bysame, \emph{A formulation of quantum field theory realizing a sea of
  interacting {D}irac particles}, arXiv:0911.2102 [hep-th], Lett. Math. Phys.
  \textbf{97} (2011), no.~2, 165--183.

\bibitem{grotz}
F.~Finster and A.~Grotz, \emph{The causal perturbation expansion revisited:
  Rescaling the interacting {D}irac sea}, arXiv:0901.0334 [math-ph], J. Math.
  Phys. \textbf{51} (2010), 072301.

\bibitem{rrev}
F.~Finster, A.~Grotz, and D.~Schiefeneder, \emph{Causal fermion systems: A
  quantum space-time emerging from an action principle}, arXiv:1102.2585
  [math-ph], Quantum {F}ield {T}heory and {G}ravity (F.~Finster, O.~M\"uller,
  M.~Nardmann, J.~Tolksdorf, and E.~Zeidler, eds.), Birkh\"auser Verlag, Basel,
  2012, pp.~157--182.

\bibitem{QFTconf}
F.~Finster, O.~M\"uller, M.~Nardmann, J.~Tolksdorf, and E.~Zeidler,
  \emph{Quantum {F}ield {T}heory and {G}ravity. {C}onceptual and mathematical
  advances in the search for a unified framework.}, Birkh\"auser Verlag, Basel,
  2012.

\bibitem{finite}
F.~Finster and M.~Reintjes, \emph{A non-perturbative construction of the
  fermionic projector on globally hyperbolic manifolds {I} -- {S}pace-times of
  finite lifetime}, arXiv:1301.5420 [math-ph] (2013).

\bibitem{friedrich}
T.~Friedrich, \emph{Dirac {O}perators in {R}iemannian {G}eometry}, Graduate
  Studies in Mathematics, vol.~25, American Mathematical Society, Providence,
  RI, 2000.

\bibitem{fulling+sweeny+wald}
S.A. Fulling, M.~Sweeny, and R.M. Wald, \emph{Singularity structure of the
  two-point function quantum field theory in curved spacetime}, Comm. Math.
  Phys. \textbf{63} (1978), no.~3, 257--264.

\bibitem{GLR}
I.~Gohberg, P.~Lancaster, and L.~Rodman, \emph{Indefinite {L}inear {A}lgebra
  and {A}pplications}, Birkh\"auser Verlag, Basel, 2005.

\bibitem{gradstein}
I.S. Gradshteyn and I.M. Ryzhik, \emph{Table of {I}ntegrals, {S}eries, and
  {P}roducts}, Fourth edition prepared by Ju. V. Geronimus and M. Ju. Ce\u\i
  tlin., Academic Press, New York, 1965.

\bibitem{drgrotz}
A.~Grotz, \emph{A {L}orentzian quantum geometry}, Dissertation Universit\"at
  Regensburg, urn:nbn:de:bvb:355-epub-231289, 2011.

\bibitem{kiefer}
C.~Kiefer, \emph{Quantum {G}ravity}, second ed., International Series of
  Monographs on Physics, vol. 136, Oxford University Press, Oxford, 2007.

\bibitem{kratzert}
K.~Kratzert, \emph{Singularity structure of the two point function of the free
  {D}irac field on a globally hyperbolic spacetime}, Ann. Phys. (8) \textbf{9}
  (2000), no.~6, 475--498.

\bibitem{lawson+michelsohn}
H.B. Lawson, Jr. and M.-L. Michelsohn, \emph{Spin {G}eometry}, Princeton
  Mathematical Series, vol.~38, Princeton University Press, Princeton, NJ,
  1989.

\bibitem{moretti}
V.~Moretti, \emph{Proof of the symmetry of the off-diagonal heat-kernel and
  {H}adamard's expansion coefficients in general {$C^\infty$} {R}iemannian
  manifolds}, Comm. Math. Phys. \textbf{208} (1999), no.~2, 283--308.

\bibitem{paschke+verch}
M.~Paschke and R.~Verch, \emph{Local covariant quantum field theory over
  spectral geometries}, gr-qc/0405057, Classical Quantum Gravity \textbf{21}
  (2004), no.~23, 5299--5316.

\bibitem{radzikowski}
M.J. Radzikowski, \emph{Micro-local approach to the {H}adamard condition in
  quantum field theory on curved space-time}, Comm. Math. Phys. \textbf{179}
  (1996), no.~3, 529--553.

\bibitem{strohmaier}
A.~Strohmaier, \emph{On noncommutative and pseudo-{R}iemannian geometry},
  math-ph/0110001, J. Geom. Phys. \textbf{56} (2006), no.~2, 175--195.

\bibitem{thiemann}
T.~Thiemann, \emph{Modern {C}anonical {Q}uantum {G}eneral {R}elativity},
  Cambridge Monographs on Mathematical Physics, Cambridge University Press,
  Cambridge, 2007.

\bibitem{zeidler1}
E.~Zeidler, \emph{Quantum {F}ield {T}heory. {I}. {B}asics in mathematics and
  physics}, Springer-Verlag, Berlin, 2006.

\end{thebibliography}
\providecommand{\bysame}{\leavevmode\hbox to3em{\hrulefill}\thinspace}
\providecommand{\MR}{\relax\ifhmode\unskip\space\fi MR }
\providecommand{\MRhref}[2]{%
  \href{http://www.ams.org/mathscinet-getitem?mr=#1}{#2}
}
\providecommand{\href}[2]{#2}

\end{document}